\documentclass[useAMS,usenatbib,usegraphicx]{mn2e}
\usepackage{ifpdf}
\usepackage{pdfscreen}
\usepackage{pslatex}
\usepackage{amsmath,amsfonts}
\newcommand{\Hfull}{\mathcal{H}}
\newcommand{\Hsec}{\mathcal{H}_{\idm{sec}}}
\newcommand{\Hrel}{\mathcal{H}_{\idm{GR}}}
\newcommand{\Hflat}{\mathcal{H}_{\idm{QM}}}

\newcommand{\HflatS}{\mathcal{H}_{\idm{spin}}}
\newcommand{\HflatT}{\mathcal{H}_{\idm{TB}}}

\newcommand{\Hpert}{\mathcal{H}_{\idm{pert}}}
\newcommand{\Hmut}{\mathcal{H}_{\idm{NG}}}

\newcommand{\HKepl}{\mathcal{H}_{\idm{kepl}}}

\newcommand{\Xint}{\mathcal{X}}
\newcommand{\R}{\mathcal{R}}

\newcommand{\RP}{${\cal S}$}
\newcommand{\RPm}{${\cal S}_{\pi}$}
\newcommand{\RPp}{${\cal S}_0$}
\newcommand{\TSR}{TSR}
\newcommand{\UE}{UE}
\def\vec#1{{\mathbf{#1}}}
\def\idm#1{{\mbox{\scriptsize #1}}}
\def\corr#1{{ #1}}
\title[Generalized secular planetary problem]
{Secular dynamics of coplanar, non-resonant planetary system
under the general relativity and quadrupole moment perturbations}
\author[C. Migaszewski and K. Go\'zdziewski]{
Cezary Migaszewski$^{1}$\thanks{E-mail: c.migaszewski@astri.uni.torun.pl} and 
Krzysztof Go\'zdziewski$^{1}$\footnotemark[1]\thanks{E-mail: k.gozdziewski@astri.uni.torun.pl}\\
$^{1}$Toru\'n Centre for Astronomy, Nicolaus Copernicus University, 
Gagarin Str. 11, 87-100 Toru\'n, Poland}
\begin{document}

\date{Accepted 2008 September 20.  Received 2008 September 19; in original form
2008 April 14}
\pagerange{\pageref{firstpage}--\pageref{lastpage}} \pubyear{2008}

\maketitle

\label{firstpage}

\begin{abstract}
We construct a secular theory of a coplanar system of $N$-planets not involved
in strong mean motion resonances, and which are far from collision zones.
Besides the point-to-point Newtonian mutual interactions, we  consider the
general relativity corrections to the gravitational potential of the star and
the innermost planet, and also a modification of this potential by the
quadrupole moment and tidal distortion of the star. We focus on hierarchical
planetary systems.  After averaging the model Hamiltonian with a simple
algorithm making use of very basic properties of the Keplerian motion, we obtain
analytical secular theory of the high order in the semi-major axes ratio. A
great precision of the analytic approximation is demonstrated with the numerical
integrations of the equations of motion.  A survey regarding model parameters
(the masses, semi-major axes, spin rate of the star) reveals a rich and
non-trivial dynamics of the secular system. Our study is focused on its
equilibria. Such solutions predicted by the classic secular theory,  which
correspond to aligned (mode~I) or anti-aligned (mode~II) apsides, may be
strongly affected by the gravitational corrections. The so called true secular
resonance, which is a new feature of the classic two-planet problem discovered
by Michtchenko \& Malhotra (2004), may appear in other, different regions of the
phase space of the generalized model. 
\corr{
We found bifurcations of mode~II which emerge new, yet unknown in the
literature, secularly unstable equilibria and a complex structure of the phase
space. These equilibria may imply secularly unstable orbital configurations
even for initially moderate eccentricities.
} 
The point mass gravity corrections can affect the long term-stability in the
secular time scale,  which may directly depend on the age of the host star
through its spin rate. We also analyze the secular dynamics of the
$\upsilon$~Andromede system in the realm of the generalized model. Also in this
case of the three-planet system, new secular equilibria may appear.
\end{abstract}


\section{Introduction}
The currently known sample of extrasolar planets\footnote{See
www://exoplanet.eu, http://exoplanets.org, http://exoplanets.eu} comprises of
many multiple-planet configurations. Their architectures are diverse and,
usually, very different from the Solar system configuration. Some of them
consist of so called hot-Jupiters or hot-Neptunes, with  semi-major axes $\sim
0.05$~au and the orbital periods of a few days.  The multiple-planet systems may
also contain companions in relatively distant orbits. Such systems are 
non-resonant (or hierarchical). 

The discovery of multi-planet systems and their unexpected orbital diversity
initiated interest in their secular evolution. Generally, the analysis of
secular interactions, including tidal effects, general relativity (GR), and
quadrupole moment (QM) corrections to the Newtonian gravitation (NG) of point
masses, follows the theory of compact multiple stellar systems
\citep{Mardling2002,Nagasawa2005}.  The secular evolution  of planetary systems
has been already intensively studied in this context by many authors. For
instance, \cite{Adams2006a} consider the GR corrections in the sample of
detected extrasolar systems. They have shown that the GR interactions can be particularly 
important for the long-term dynamics of  short-period planets. Also
\cite{Adams2006b,Mardling2007} study the evolution of such objects taking into
account the tidal circularization of their orbits, and secular excitation of the
eccentricity of the innermost planet by more distant companions. These secular
effects imply constraints on the orbital elements of the innermost orbits,
including yet undetected Earth-like planets, which are expected to be found in
such systems. The  quadrupole moment  and relativistic effects may detectably
affect the light-curves of stars hosting transiting planets in a few years
time-scale  \citep{MiraldaEscude2002,Agol2005,Winn2005}.  In  general, the
interplay of apparently subtle perturbations with the  point mass Newtonian
gravity depends on many physical and orbital parameters, and it may lead to
non-trivial and interesting dynamical phenomena. \corr{Still, 
the problem is of particular
importance for studies of the long-term stability of the Solar system
\citep[see very recent works by][]{Benitez2008,Laskar2008}.}

Because a fully general study of such effects would be a very difficult task, we
introduce some simplification to the planetary model. We skip tidal effects
caused by dissipative interactions of extended star and planetary bodies, in
particular  the tidal friction (TF) damping the eccentricity and the tidal
distortion (TD) that modifies planetary figures,  and could influence their
gravitational interaction with the parent star. The tidal effects are typically
much smaller than the leading Newtonian, relativistic and quadrupole moment
contributions \citep{Mardling2007}. Also their time-scale is usually much longer
than of the most  significant conservative effects.    In this work, we focus on
the secular dynamics of planetary systems over an intermediate time scale
relevant to the  conservative perturbations.  Our goal is to obtain a
qualitative picture of the secular system that may be useful as the first order
approximation to more general model of the long-term planetary dynamics. 

Moreover, we show in this paper that even if we skip the non-conservative  tidal
effects then the GR and/or QM {\em corrections} to the 
point mass Newtonian gravity can be
much more significant for the secular dynamics than the mutual NG interactions
alone. These effects may change qualitatively  the view of the dynamics of
planetary systems  as predicted by the Laplace-Lagrange theory
\citep{Murray2000} or its recent versions
\citep[e.g.,][]{Lee2003,Michtchenko2004,Michtchenko2006,Libert2005,Veras2007,Migaszewski2008a}. 

The model without dissipative effects may be investigated  with the help of
conservative Hamiltonian theory. Assuming that planetary orbits are well
separated and the system is far from mean motion resonances and collision zones,
we can apply the averaging proposition \citep{Arnold1993} to derive the long
term evolution of their orbital elements. This approach can be classified among
secular theories having a long history in the context of the Solar system
\citep{Brouwer1961,Murray2000}.  

The present work is a step towards a generalization of the secular model of
coplanar 2-planet system by \cite{Michtchenko2004} and its analytical version
applied to $N$-planet system in \citep{Migaszewski2008a}.  To explore the phase
space globally without restrictions on eccentricities, we simplify the equations
of motion through the averaging of perturbations to the Keplerian motion, with
the help of semi-numerical method proposed by \cite{Michtchenko2004}. To obtain
analytical results, we also use  a very simple averaging algorithm that can be
applied to perturbations dependent on the mutual distance of interacting bodies
\citep{Migaszewski2008a}. These works demonstrate that the secular evolution can
be precisely described in wide ranges of the orbital parameters, including
eccentricities up to 0.8--0.9, for well separated (hierarchical)  orbits with
small ratio of the semi-major axes, $\alpha\sim 0.1$.  Here, the secular NG
theory is very helpful to derive 
the more general model including the GR and QM  effects. Basically, without the
averaging, the only possibility of investigating the long-term secular dynamics
relies on numerical solutions of the equations of motion \citep{Mardling2002}.
However, due to extremely different time scales which are related as days (the
orbital periods of inner planets) to $10^3$--$10^6$~years of the secular orbital
evolution, the numerical integrations are of very limited use when we want to
investigate large volumes of initial conditions rather than isolated orbits. In
such a case,  the analytical theory can be very helpful to get a deep insight
into the secular dynamics. Moreover, when necessary, particular solutions can be studied
in detail with the help of the direct numerical integrations.

The plan of this paper is the following. In Sect.~2 we formulate the generalized
model of a coplanar, $N$-planet system, accounting for the relativistic and
quadrupole moment corrections to the NG-perturbed motion of the innermost
planet. To make the paper self-consistent, we average out the perturbations to
the Keplerian motion  with the help of the averaging algorithm described in
\citep{Migaszewski2008a}. In Sect.~3 we apply the analytical and numerical tools
to study the influence of the GR and QM corrections on the secular evolution. 
In particular, we recall the concept of the so called representative  plane of
initial conditions \citep{Michtchenko2004}. We focus on the  search for
stationary solutions in the averaged and reduced systems (periodic orbits in the
full systems), and we investigate their stability and bifurcations with the help
of phase diagrams.  In this section we also derive interesting conclusions on
the stability of the unaveraged systems. In Sect.~4, we study the phase space of
two-planet systems in wide ranges of parameters governing their orbital
configurations (masses, semi-major axes ratios, eccentricities) and physical
parameters (flattening of the star). To illustrate the application of the
secular theory to multi-planet configurations, we consider the three-planet
$\upsilon$~Andr system, and we investigate the QM (stellar rotation) influence
on its secular orbital evolution.
%
\section{Generalized model of $N$-planet system}
%
The dynamics of the planetary system can be modeled by the Hamiltonian function
written with respect to canonical Poincar\'e  variables \citep[see,
e.g.,][]{Poincare1897,Laskar1995}, and    expressed by a sum of two terms,
$
\Hfull = \HKepl + \Hpert,
$
where
\begin{equation}
\HKepl = \sum_{i=1}^{N} {\bigg( \frac{\mathbf{p}_i^2}{2 \beta_i} 
- \frac{\mu_i \beta_i}{r_i} \bigg)}
\label{HKepl}
\end{equation}
stands for  integrable part comprising of the direct sum of the relative,
Keplerian motions of $N$ planets and the host star. Here, the dominant point
mass of the star is ${m_0}$, and $m_i \ll m_0$, $i=1,\ldots,N$ are the point
masses of the $N$-planets.  For each planet--star pair we define the mass
parameter ${\mu_i=k^2~(m_0+m_i)}$ where $k$ is the Gauss gravitational constant,
and ${\beta_i=(1/m_i+1/m_0)^{-1}}$ are the so called reduced masses. We consider
the perturbing Hamiltonian  $\Hpert$ as a sum of three terms, 
\begin{equation}
\Hpert = \Hmut + \Hrel + \Hflat,
\label{Hpert}
\end{equation}
where $\Hmut$ is for the mutual point-mass interactions  between planets,
$\Hrel$ is for the general (post-Newtonian) relativity corrections to the
Newtonian gravity, and $\Hflat$ takes into account the dynamical flattening and
tidal distortion of the parent star. 

We consider the secular effects of $\Hmut$, which can be expressed as follows:
\begin{equation}
\Hmut = \sum_{i=1}^{N-1} \sum_{j>i}^{N} {\bigg(
 - \underbrace{\frac{k^2 m_i m_j}{\Delta_{i,j}}}_{\textrm{\small direct part}} +
\underbrace{\frac{\mathbf{p}_i \cdot \mathbf{p}_j}{m_0}}_{\textrm{\small indirect
part}}\bigg)},
\label{Hmut}
\end{equation}
where  ${\mathbf{r}_i}$ are for the position vectors of planets relative to the
star,  ${\mathbf{p}_i}$ are for their conjugate momenta relative to the {\em
barycenter} of the whole $(N+1)$-body system, 
${\Delta_{i,j}=\|\mathbf{r}_i-\mathbf{r}_j\|}$ denote the relative distance
between planets $i$  and $j$.

For  hierarchical planetary  systems (weakly interacting binaries), the secular
time scale of Newtonian point-mass interactions may be as short as 
$10^{3}$~years, up to Myrs. In general, these interactions cause slow
circulation of the apsidal lines. The GR correction to the NG potential of the
star--planet system also leads to the circulation of pericenters.
Moreover, the time scale of this effect may be comparable to that one forced by
mutual Newtonian interactions between planets. In  such a situation, one should
necessarily include the relativistic corrections to the model of motion.
%
\subsection{General relativity corrections}
%
The GR correction will be applied only to the innermost planet, so we skip the
direct GR perturbations on the motion of the more distant companions, as well as
the mutual GR interactions caused by planetary masses. Usually, the GR 
corrections to the Newtonian potential are expressed in terms of the PPN
formalism \cite[e.g.,][]{Kidder1995}. Alternatively, we found a very clear paper
of \cite{Richardson1988} {conveniently providing explicit Hamiltonian of the two
body problem with the GR term}. Following these authors, $\Hrel' \equiv
\Hrel/\beta$ (i.e., $\Hrel$ rescaled by the  reduced mass) may be written as
follows:
\begin{equation}
\Hrel' = \gamma_1 \vec{P}^4 + \gamma_2 \frac{\vec{P}^2}{r} + \gamma_3
\frac{\left(\vec{r} \cdot \vec{P}\right)^2}{r^3} + \gamma_4 \frac{1}{r^2},
\label{Hrel}
\end{equation}
where $\gamma_1,\gamma_2,\gamma_3,\gamma_4$ are coefficients defined 
through
\[
\gamma_1 = - \frac{\left(1 - 3 \nu\right)}{8 c^2}, \quad
   \gamma_2 = - \frac{\mu \left(3 + \nu\right)}{2 c^2},\quad
\gamma_3 = \frac{\mu^2}{2 c^2}, \quad
  \gamma_4 = - \frac{\mu \nu}{2 c^2},
\]
and  $c$ is the velocity of light in a vacuum, $\mu = k^2 (m_0 + m_1)$,  $\nu
\equiv m_0 m_1 / (m_0 + m_1)^2$, $r$ is the astrocentric distance, and
$\vec{P}$ is  the astrocentric momentum of the
innermost planet (normalized through the reduced mass): 
\begin{equation} \vec{P} = \vec{v} + \frac{1}{c^2}
\left[ 4\gamma_1 (\vec{v} \cdot \vec{v}) \vec{v} + 
\frac{2\gamma_2}{r} \vec{v} + \frac{2\gamma_4}{r^3} (\vec{r} \cdot \vec{v})
\vec{r} \right],
\end{equation} 
where $ \vec{v} \equiv \dot{\vec{r}}$ stands for the  astrocentric velocity of
the innermost planet (the relativistic corrections from  other  planetary bodies
in the system are skipped). Hence, in the relativistic Hamiltonian, we put 
$\vec{P} = \vec{v}$ with the accuracy of $\vec{O}({c^{-2}})$ and then the
Hamiltonian is conserved up to the order of $O({c^{-4}})$.
%
\subsection{Rotational and tidal distortions of the star}
%
Fast rotating stars are significantly flattened and that in turn may lead to
important deviations of the NG potential. In the absence of a close planet
and  non-radial pulsations,  the star has the rotational symmetry, regarding its
mass density and shape. The gravitational potential of such a body can  be
expanded in harmonic series expressed in terms of  the Stokes coefficients. The
well known property of this expansion is that the gravitational potential of
rotationally symmetric bodies retains only terms related to the so called {\em
zonal harmonics}:
\begin{equation}
\HflatS = \frac{\mu \beta}{r} \sum_{l=2}^{\infty} 
{J_{l} \left(\frac{R_0}{r}\right)^{l}} P_l(\sin\phi),
\label{Hflat}
\end{equation}
where $R_0$ stands for the characteristic radius of a sphere encompassing the
body, and it can be fixed as the equatorial radius of the star, $P_l(\sin\phi)$
is the Legendre polynomial of the $l$-th order, $\phi$ is the astrocentric
latitude and $J_l$ are non-dimensional Stokes coefficients, $l>1$. It can be
shown that for rotationally symmetric objects, $J_l=0$ for $l$ odd.  Basically,
$J_l$ (and the leading term with $J_2$, in particular) can be determined
numerically  with the help of the theory of stellar interiors combined with a
model of rotation and helio-seismic data \citep{Godier1999,Pijpers1998}.  To
calculate these coefficients, one must know the mass density as well as the
hydrostatic figure of the star.  There are attempts to estimate $J_2$ for stars
hosting planets.  For instance, \cite{Iorio2006} determined quadrupole moment of
HD~209458~\citep{Charbonneau2000} as $J_2 \sim 3.5 \times 10^{-5}$, 
\corr{however with large error of $\sim 10^{-3}$, making the result not very
credible}.  In general, the estimates of $J_{l}$ are very  uncertain even for  the
well explored Sun. For this slowly rotating dwarf (with rotational period of
$\sim 30$~days), $J_2 \sim 10^{-7}$ with an upper bound of $10^{-6}$. Moreover,
in the literature we found approximations of current  $J_2$ of the Sun in the
range up to $10^{-5}$. 

Due to indefiniteness of higher order zonal harmonics, we consider the 
dynamical effects of the leading $J_2$ term only.  For a coplanar system, in
which planets move around the star in its equatorial plane,  the first non-zero
term in harmonic series, Eq.~\ref{Hflat}, has the form of:
\begin{equation}
\label{eq:hrf}
\HflatS \approx -\frac{1}{2} \mu \beta J_2 R_0^2 \frac{1}{r^3}.
\end{equation}
The above formulae may be conveniently expressed 
\citep{Mardling2002} in terms of
the stellar spin frequency:
\begin{equation}
\HflatS \approx -\frac{1}{6} \beta R_0^5 \left(1 + m_1/m_0\right) k_L \Omega^2
\frac{1}{r^3},
\end{equation}
where $k_L$ is the tidal Love number \citep{Murray2000} 
which can be related to $J_2$ through:
\begin{equation}
k_L = 3 \frac{J_2}{q}, \quad q = \frac{\Omega^2 R_0^3}{k^2 m_0},
\end{equation}
$q$ is the ratio of centrifugal acceleration and gravitation at the
surface of the star, and $\Omega$ is the spin rate.
Hence, we have:
\begin{equation}
J_2 = \frac{1}{3} k_L \frac{R_0^3 \Omega^2}{k^2 m_0}.
\end{equation}
In this work we adopt the standard value of $k_L=0.02$ \citep{Nagasawa2005}.
Obviously, the dynamical flattening is more significant for fast rotating stars.
Before the Zero Age Main Sequence (ZAMS) stage, the rotational periods may be as
low as a few days, down to 1~day for young ($\sim 100$~Myr) Sun-like stars. The
rotational periods of $\sim 40$~days are typical for 8~Gyrs old, evolved
objects.  According to the scaling rule of  $J_2 \sim \Omega^2$, the zonal
harmonics of young objects may be as large as $10^{-4}$.   Because the
rotational period may change by two orders of magnitude during the life-time of
the star, also the quadrupole moment may change by a few orders of magnitude. 
Hence, flattening caused by fast rotation not only can affect significantly the
orbital acceleration which can compete with the GR and  mutual NG contributions
but it may also introduce a dependence of the system dynamics on the age of the
parent star \citep[][see also Sect.~5 and Sect.~6]{Nagasawa2005}.

We also consider a contribution of the tidal bulge (TB) caused by  the presence
of point-mass innermost planet, assuming that only the extended star is
distorted due to the tidal interactions. Then the Hamiltonian is corrected by the
following term \citep{Mardling2002}:
\begin{equation}
\label{eq:htb}
\HflatT = -\beta R_0^5 k^2 m_1 \left(1 + m_1/m_0\right) k_L \frac{1}{r^6}. 
\end{equation}
Both gravitational corrections are then $\Hflat = \HflatS + \HflatT$.

We account for direct dynamical effects related to $\Hflat$ only in the motion
of the innermost planet although these perturbations can be also  quite easily
included for the remaining star-planet pairs.  In the realm of the secular
theory of non-resonant, hierarchical systems, we assume that other companions
are much more distant from the star. For the semi-major axes ratio $\sim 0.1$,
the GR+QM perturbations acting on such outer companions are by orders of
magnitude smaller than for the innermost planet. However, the dynamics of the
innermost body influences  indirectly the secular dynamics of the whole system.
This will be demonstrated in this work in the cases of two- and three-planet
configurations. We also underline that with the simplified  model of
interactions, we can study the dynamics of quite compact star--planets 
configurations  because they can be parameterized, in general, by 
individual planetary
masses, semi-major axes, and physical parameters of the star.
%
\subsection{The secular model of non-resonant planetary system}
%
To apply the canonical perturbation theory, we first transform $\Hfull$ to the
following form:
\begin{equation}
 \Hfull(\vec{I},\vec{\phi}) = \HKepl(\vec{I}) + \Hpert(\vec{I},\vec{\phi}),
\label{poincare} 
\end{equation}
where $(\vec{I},\vec{\phi})$ stand for the action-angle variables, and
$\Hpert(\vec{I},\vec{\phi}) \sim \epsilon \HKepl(\vec{I})$,  where $\epsilon \ll
1$ is a small parameter. In this work, we apply the well known 
approach to analyze the equations of motion induced by Hamiltonian,
Eq.~\ref{poincare}, that relies on the averaging proposition \citep[see,
e.g.,][]{Arnold1993}.  By averaging the perturbations with respect to the fast
angles ({\em the mean longitudes} or {\em the mean anomalies}) over their
periods,  we obtain the secular Hamiltonian which does not depend on these fast
angles. Simultaneously,  the  conjugate momenta to the fast angles become
integrals of the secular problem.  In the planetary system with a dominant
stellar mass, the orbits (apsidal lines) also slowly rotate due to mutual
interactions, hence the longitudes of periastron and the longitudes of node
become slow angles. Assuming that no strong  mean motion resonances are present,
and the system is far enough from collisions, the averaging makes it possible to
reduce the number of the degrees of freedom, and to obtain qualitative
information on the long-term changes of the slowly varying orbital elements
(i.e., on the slow angles and their conjugate momenta).

The transformation of the Hamiltonian to the required form (Eq.~\ref{poincare})
may be accomplished   by expressing this Hamiltonian with respect to the
(modified) Delaunay canonical elements. These variables can be related to the
Keplerian canonical elements \citep{Murray2000}. Actually, we use the following
set of canonical action-angle variables which can be obtained after an
appropriate canonical transformation of the Delaunay elements:
\begin{eqnarray}
\label{dvars}
{l_i \equiv \mathcal{M}_i}, & \quad {L_i=\beta_i~\sqrt{\mu_i~a_i}},\nonumber\\
{g_i \equiv \varpi_i}, & \quad  {G_i=L_i~\sqrt{1-e_i^2}},\\
{h_i \equiv \Omega_i}, & \quad{H_i=G_i~(\cos~I_i}-1),\nonumber
\end{eqnarray}
where $\mathcal{M}_i$ are the mean anomalies,  $a_i$ stand for canonical
semi-major axes,  $e_i$ are the eccentricities,   $I_i$ denote inclinations,
$\varpi_i$ are the longitudes of pericenter, and $\Omega_i$ denote the
longitudes of ascending node. The choice of $\varpi_i$ instead of $\omega_i$ is
important for further applications, because $\Omega_i$ are undefined (and
irrelevant) for the dynamics of the coplanar system. We note that the
geometrical, canonical elements ($a_i$, $e_i$, $I_i$, $\varpi_i$, $\Omega_i$)
may be derived through the formal transformation between classic (astro-centric)
Keplerian elements and the relative  Cartesian coordinates
\citep[e.g.,][]{FerrazMello2006a,Morbidelli2003}, with appropriate rescaling of the astrocentric
velocities. In the settings adopted here, the Cartesian  coordinates are
understood as Poincar\'e coordinates, i.e., {\em astrocentric} positions of
planets, and canonical momenta taken relative to the {\em barycenter} of the
system. 

The $N$-planet Hamiltonian expressed in terms of the modified  Delaunay variables
(\ref{dvars}) has the form of:
\[
\Hfull = -\sum_{i=1}^N \frac{\mu_i^2 \beta^3_i}{2 L_i^2} +
\Hpert\underbrace{(L_i,l_i,G_i,g_i,H_i,h_i)}_{i=1,\ldots,N}.
\]
In this Hamiltonian, $l_i$ play the role of the fast angles. In the absence of
strong mean motion resonances, these angles can be eliminated by the following
averaging formulae:
\begin{eqnarray}
\label{secular}
&& \Hsec=\frac{1}{(2 \pi)^N}\underbrace{\int_0^{2 \pi} \ldots 
\int_0^{2\pi}}_{i=1,\ldots,N}{\Hmut \, d\mathcal{M}_1 \ldots d\mathcal{M}_N} +\\
&& \quad + \frac{1}{2\pi} \int_0^{2\pi}{\Hrel \,d\mathcal{M}_1} 
+ \frac{1}{2\pi} \int_0^{2\pi}{\Hflat \,d\mathcal{M}_1}\nonumber.
\end{eqnarray}
Hence, we can rewrite the secular Hamiltonian in the symbolic form of:
\begin{equation}
\Hsec = \left<\Hmut\right> + \left<\Hrel\right> + \left<\Hflat\right>.
\label{Hsec}
\end{equation}
Now, we try to average out each component of this Hamiltonian expressed  with
respect to the fast angles (the
mean anomalies).
%
\subsection{Averaging Newtonian point-to-point interactions}
%
A simple averaging of  the Hamiltonian of the classic planetary
$N+1$-body model is described in our previous paper \citep{Migaszewski2008a}. 
The algorithm makes use on the very basic properties of the Keplerian motion and
it relies on appropriate change of integration variables in Eq.~\ref{secular}.
It may be also applied to  perturbations expressed through powers of the
relative distance.

To recall the main result, the secular Hamiltonian  of the $N$-planet system can be
described as a sum of two-body Hamiltonians evaluated over all pairs of planets:
\begin{equation}
\left<\Hmut\right> = \sum_{i=1}^{N-1}  \sum_{j>i}^{N}{\left<\Hmut^{(i,j)}\right>}.
\label{avHmut}
\end{equation}
The direct part of the disturbing Hamiltonian reads as follows:
\begin{eqnarray}
\label{expanssion}
& & \left<\Hmut^{(i,j)}\right> = -\frac{k^2 m_i m_j}{a_j} \times  \nonumber \\
&& \quad \times \left[1 + \sqrt{1-e_j^2} \sum_{l=2}^{\infty}
{\left(\frac{\alpha_{i,j}}{1-e_j^2}\right)^l
\mathcal{R}^{(i,j)}_l(e_i,e_j,\Delta\varpi_{i,j})}\right].
\end{eqnarray}
The explicit formulae for functions $\R^{(i,j)}_l(e_i,e_j,\Delta\varpi_{i,j})$ are
given in \citep{Migaszewski2008a}. It is well known that the indirect part
averages out to a constant and it does not contribute to the secular dynamics
\citep{Brouwer1961}. 
%
\subsection{Averaging the PPN relativistic potential}
%
Making use of the formulae for the relativistic PPN Hamiltonian by
\cite{Richardson1988}, we write down the mean relativistic potential as follows:
\begin{equation}
\left<\Hrel'\right> = \gamma_1 \left<\vec{v}^4\right> + \gamma_2
\left<\frac{\vec{v}^2}{r}\right> + \gamma_3 \left<\frac{\left(\vec{r} \cdot
\vec{v}\right)^2}{r^3}\right> + \gamma_4 \left<\frac{1}{r^2}\right>,
\label{Hrelsec}
\end{equation}
with the accuracy of $O(c^{-2})$.
We should average out each component of this Hamiltonian over the mean
anomalies. It appears to be possible with  the algorithm in
\citep{Migaszewski2008a}. To average out the whole PPN Hamiltonian, we must
calculate a  few integrals  written in the general form of:
\begin{equation}
\left<\Xint\right> \equiv \frac{1}{2\pi} \int_0^{2\pi} {\Xint \,
d\mathcal{M}}
\equiv \frac{1}{2\pi} \int_0^{2\pi} {\left[\Xint \, {\cal J}\right] \, df},
\end{equation}
where ${\cal J}$ is a scaling function defined with: 
\begin{equation}
{\cal J} \equiv \frac{d\mathcal{M}}{df} = \frac{\left(1 - e^2\right)^{3/2}}{\left(1 + e
\cos{f}\right)^2},
\end{equation}
and $f$ is the true anomaly of the innermost planet (note that we skip index
"1" of that planet). Components  of the
mean Hamiltonian can be written explicitly as follows:
\begin{eqnarray}
\label{v4}
&&\left<\vec{v}^4\right> = \frac{n^4 a^4}{\sqrt{1 - e^2}} \left[1 + \frac{3}{2}
e^2 + e^4 \mathcal{F}_1\right],\\
\label{v2r}
&&\left<\frac{\vec{v}^2}{r}\right> = \frac{n^2 a}{\sqrt{1 - e^2}} \left[1 + e^2
\mathcal{F}_2\right],\\
\label{rvr3}
&&\left<\frac{\left(\vec{r} \cdot \vec{v}\right)^2}{r^3}\right> = \frac{n^2 a}{\sqrt{1 - e^2}} e^2
\mathcal{F}_2,\\
\label{r2}
&&\left<\frac{1}{r^2}\right> = \frac{1}{a^2 \sqrt{1 - e^2}}.
\end{eqnarray}
Functions $\mathcal{F}_1$ and $\mathcal{F}_2$ are given through:
\begin{eqnarray}
&& \mathcal{F}_1 = \frac{1}{2\pi} \int_0^{2\pi} {\frac{\sin^4{f}}{\left(1 + e
\cos{f}\right)^2}~df},\\
&& \mathcal{F}_2 = \frac{1}{2\pi} \int_0^{2\pi} {\frac{\sin^2{f}}{1 + e \cos{f}}~df}.
\end{eqnarray}
These integrals can be calculated in the closed form:
\begin{equation}
\mathcal{F}_1 = \frac{3 \left(2 - e^2 - 2\sqrt{1-e^2}\right)}{2 e^4}, \quad
\mathcal{F}_2 = \frac{1-\sqrt{1-e^2}}{e^2}.
\end{equation}
Finally, the secular relativistic potential is the following:
\begin{equation}
\left<\Hrel'\right> = - \frac{3 \mu^2}{c^2 a^2 \sqrt{1 - e^2}} +
\frac{\mu^2 \left(15 - \nu\right)}{8 a^2 c^2}.
\label{eq28}
\end{equation} 
{
Because all secular Hamiltonian corrections we account for in this work (see
below) do not depend on the  mean anomalies, $L_{1,2}$ are constants of motion. 
It also means  that the second term
in Eq.~\ref{eq28} reduces to
the constant and does not contribute to the long-term dynamics.
}
Writing down the secular Hamiltonian with respect to  the action-angle
variables, we have:
\begin{equation}
\left<\Hrel'\right> = -\frac{3 \mu^4 \beta^4}{c^2 L^3 G} + \mbox{const}.
\label{avHrel}
\end{equation}
Here, the action variables $(L,G)$ are defined with usual formulae known
in the Keplerian motion theory. We recall that in the secular PPN Hamiltonian,
only the star-innermost planet interactions are considered.  The  perturbation
has been also scaled by the reduced mass. To  properly add the GR star-inner
planet  interactions to the Hamiltonian of $N$-planets, we should account for
the mass factor. Because the canonical Delaunay elements of the $N$-planet system are
defined   through Eq.~\ref{dvars}, hence, to obtain the proper scaling, we should  multiply
the two-body Hamiltonian, Eq.~\ref{avHrel}, through $\beta \equiv 1/(1/m_0 +
1/m_1)$.  Finally, we obtain the following secular Hamiltonian related to the GR
correction: 
\begin{equation} 
\left<\Hrel\right> = -\frac{3 \mu^4 \beta^5}{c^2 L^3 G} + \mbox{const}. 
\label{avbHrel}
\end{equation} 
In this Hamiltonian all slow angles are cyclic, 
hence $L$ and $G$ would be integrals of motion in the absence of 
{mutual planetary interactions}. However, the element $G$ is
no longer constant when we consider the NG contributions. 
The evolution of canonical angle $\varpi$, reads as follows: 
\begin{equation}
\dot{\varpi}_{\idm{GR}} = \frac{\partial \left<\Hrel\right>}{\partial G} =
\frac{3 \mu^4 \beta^5}{c^2 L^3 G^2} = \frac{3 \mu^{3/2}}{c^2 a^{5/2} \left(1 -
e^2\right)}. 
\label{orel}
\end{equation} 
This is the well known formulae describing the  {\em
relativistic advance of pericenter}. We derive it here to keep the paper
self-consistent.  In the above formulae,  we skip  the symbol
$\left<\ldots\right>$ of the mean, which basically should encompass the
canonical angle $\varpi$  and the conjugate actions ($L, G$) as well as the
symbols of orbital elements ($a,e$).  However, we should keep in mind that after
calculating the average over the fast angles, these actions/elements have the
sense of the {\em mean} actions/elements.
%
\subsection{The effect of quadrupole moment of the star}
%
To average out the quadrupole moment of the star, we have to calculate integrals
from the astrocentric distance of the planet taken in the power of ``-3'' for the spin
distortion, and the power of ``-6'' for the tidal distortion. To calculate these averages,
we express the astrocentric distance $r$ of the inner planet with respect to the
true anomaly and  we replace the integration variables $d\mathcal{M} =
\mathcal{J} df$. For $l>1$, we  find
\begin{equation}
\left<\frac{1}{r^l}\right> = \frac{(1 - e^2)^{1/2}}{a^l 
\left(1 - e^2\right)^{l-1}}
\sum_{s=0}^{l-2} {{{l-2}\choose{s}} e^s \left<\cos^s{f}\right>},
\end{equation}
where, for $s$ even, the average of $\left<\cos^s{f}\right>$  over the mean
anomaly reads as follows:
\begin{equation}
\left<\cos^s{f}\right> = \frac{2^s \pi }{\Gamma \left(\frac{1}{2}-\frac{s}{2}\right)^2
\Gamma (s+1)},
\end{equation}
while for $s$ odd, the average over the mean anomaly
$\left<\cos^s{f}\right>=0$.  The average of the leading term in
$\left<\HflatS\right>$ has the form of:
\begin{equation}
\left<\HflatS\right> \approx -\frac{\beta  k_L
    \left(1 + m_1/m_0\right) R_0^5 \Omega^2}{6 a^3
    \left(1-e^2\right)^{3/2}}.
\label{avHflat}
\end{equation}
Having this secular Hamiltonian, we can calculate the apsidal 
frequency forced by the spin-induced QM of the star:
\begin{equation}
\dot{\varpi}_{\idm{spin}} = 
\frac{\partial\,\left<\HflatS\right>}{\partial\,G} \approx 
\frac{\beta^7 k_L (1 + m_1/m_0) \mu ^3 R_0^5 \Omega^2}{2 G^4 L^3}.
\label{oqm}
\end{equation}

Similarly, we calculate the leading term in $\left<\HflatT\right>$:
\begin{equation}
\left<\HflatT\right> \approx -\frac{\beta k_L R_0^5 \left(\frac{3}{8} e^4+3 e^2+1\right) k^2
    m_1 \left(1+m_1/m_0\right)}{a^6 \left(1-e^2\right)^{9/2}},
\label{avHflatT}
\end{equation}
and the apsidal frequency due to this correction:
\begin{equation}
\dot{\varpi}_{\idm{TB}} = 
\frac{\partial\,\left<\HflatT\right>}{\partial\,G} \approx 
\frac{15 \beta^{13} k_L R_0^5 \left(G^4-14 L^2 G^2+21 L^4\right)
    m_1 \mu^7}{8 G^{10} L^7 m_0}.
\label{oqmT}
\end{equation}
The tidal bulge induced by a close companion may be 
very important in some  configurations.  In fact, the TB effect may be 
even dominant
over the spin-induced quadrupole moment perturbations. For instance, for $\upsilon$~Andr~b it is
$\sim 20$~times as big \citep[see][for details]{Mardling2002,Nagasawa2005}.
%
\section{Tools to analyze the secular dynamics}
%
To overview the dynamical influence of the GR and QM  corrections on
the point mass NG dynamics, we compare the {\em secular} evolution of the
eccentricity of the innermost planet in the sample of detected multiple extrasolar
systems. In this experiment, we basically follow \cite{Adams2006a}, however,
solving the {\em averaged} equations of motion by means of the numerical
integrator. Moreover, we carry out this test to illustrate the different and rich
behaviors of  multi-planet systems in the detected sample; a systematic
analysis is described in subsequent sections. The results are shown 
in Fig.~\ref{fig:fig1}. Each panel in this figure is labeled with the  parent star name.
The orbital elements of the selected systems, and the parameters of their parent
stars come  from the Jean Schneider  Encyclopedia of Extrasolar
Planets\footnote{http://exoplanet.eu}. In this experiment, we fixed 
rotational period of the star, $T_{\idm{rot}}=30$~days, 
$k_L=0.02$ (hence $J_2 \sim 10^{-7}$--$10^{-6}$).   In some cases, the differences between predictions of the
generalized theory and by the classical model are significant, in some other
cases both theories give practically the same  outcome.  Looking at the elements
of the examined systems, we may conclude that the corrections become very
important  for systems with the innermost planet  close to the star, and with
other bodies relatively distant.  

In fact, it is already well known \cite[e.g.,][]{Adams2006a} that the additional
effects are important when the strength and characteristic time scale of
perturbations stemming from the GR and QM become comparable with the secular
time-scale of mutual NG interactions.  Then the GR and QM  regarded as {\em
perturbing} effects may induce significant  contribution to the secular
evolution and the interplay of these effects with the NG interactions may lead
to very complex and rich dynamics. As we will see below, this condition is
particularly well satisfied  for strictly hierarchical  systems, e.g.,
HD~217107, HD~38529 \citep{Fischer2001},  HD~190360~\citep{Vogt2005},
HD~11964~\citep{Butler2006},  $\upsilon$ Andromedae~\citep{Butler1999}. The
secular time scales can be also comparable when the planetary masses are
relatively small because the NG-induced apsidal  frequency decreases
proportionally to the products of these masses while, in the first
approximation, the GR and spin-induced apsidal motion of the innermost orbit
does not depend directly on the planetary mass (see Eqs.~\ref{orel},~\ref{oqm}).

Still, the  results illustrated in Fig.~\ref{fig:fig1}  have limited
significance for the study of the global dynamics. Drawing one-dimensional
time--orbital  element plots, we can analyze the dynamical evolution only for a
few isolated initial conditions. This can be a serious drawback, if we recall
that the initial conditions of the discovered systems are still known with large
uncertainties. Some  critically important parameters governing the secular
evolution, like the masses, nodal lines and inclinations of orbits are poorly
constrained by the observations or undetermined at all.  There is also a
technical problem: the  direct integrations over the secular time-scale are CPU
intensive. Yet looking at  isolated configurations, we obtain only a local view of
the dynamics. Instead, following the methodology of Poincar\'e, we should try to
understand globally the perturbing effects and the resulting dynamics. We should
investigate the whole families of solutions rather than a few isolated
phase-space trajectories. In such a case, the application  of secular analytical
theories become critically important.
\ifpdf
\begin{figure*}
\centering
\includegraphics[width=17.6cm]{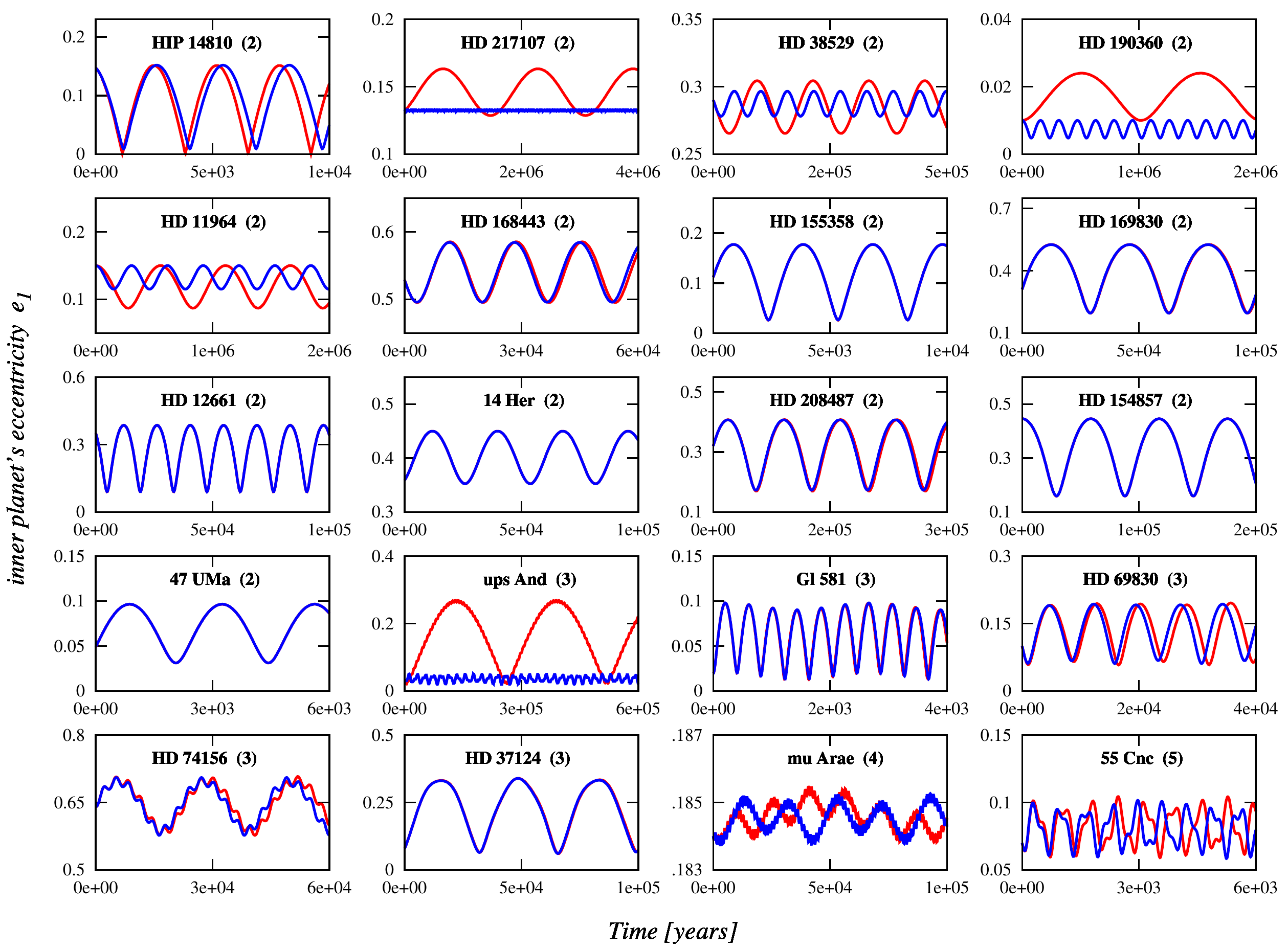}
\caption{
The long-term, secular  evolution of eccentricities of the innermost planets of
known non-resonant extrasolar systems. Red curves are for the point-mass
planets, interacting mutually through Newtonian forces. The blue curves are for 
generalized model of motion, including additional effects  (general relativity
and the quadrupole moment of the star). For all  tested planetary systems, we
fix $T_{\idm{rot}} = 30$~days, $k_L=0.02$. The names of  parent stars are
labeling each panel together with  a number of planets written in brackets.
}
\label{fig:fig1}
\end{figure*}
\else
\begin{figure*}
\centering
\includegraphics[width=17.6cm]{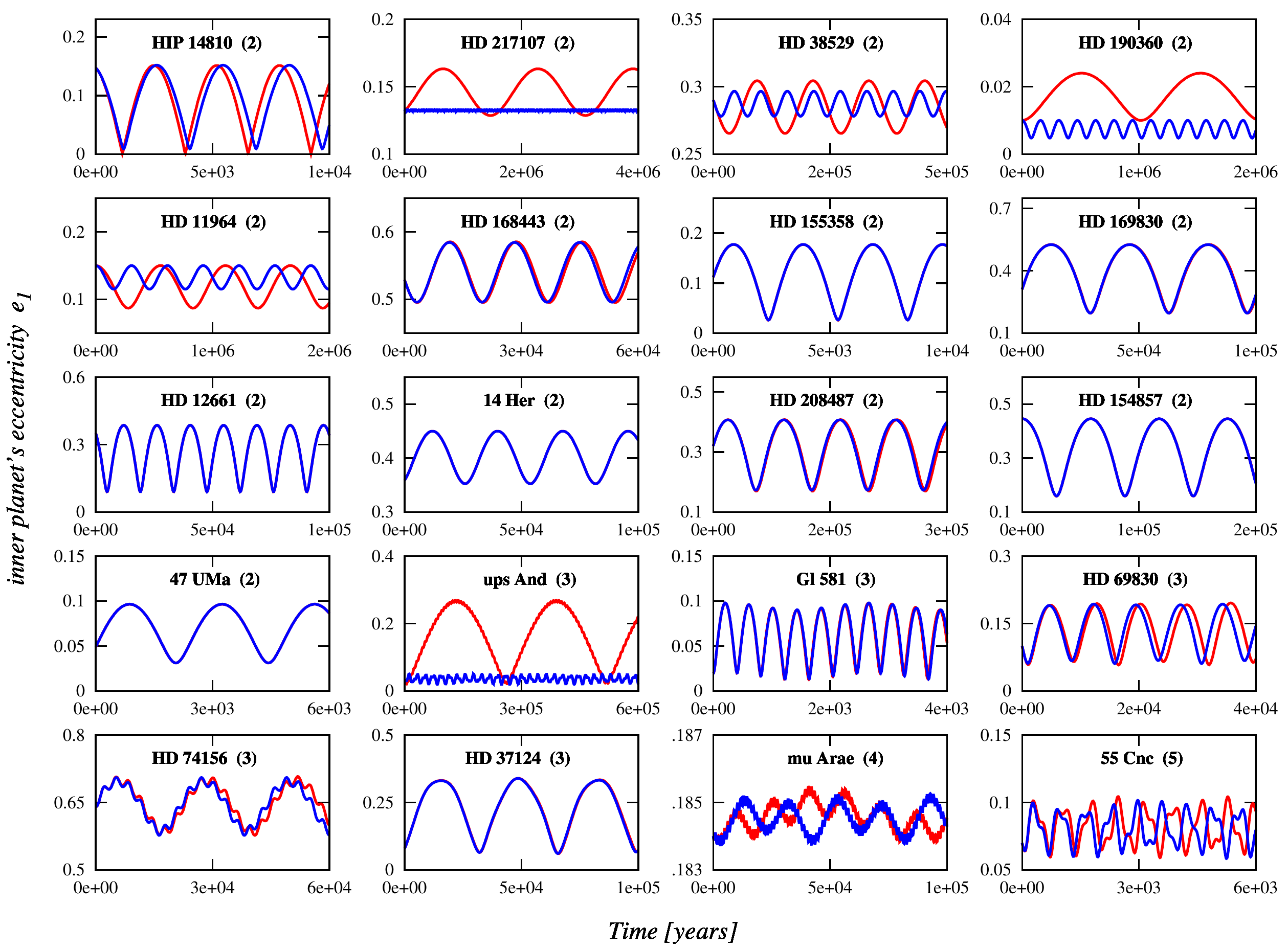}
\caption{
The long-term, secular  evolution of eccentricities of the innermost planets of
known non-resonant extrasolar systems. Red curves are for the point-mass
planets, interacting mutually through Newtonian forces. The blue curves are for 
generalized model of motion, including additional effects  (general relativity
and the quadrupole moment of the star). For all  tested planetary systems, we
fix $T_{\idm{rot}} = 30$~days, $k_L=0.02$. The names of  parent stars are
labeling each panel together with  a number of planets written in brackets.
}
\label{fig:fig1}
\end{figure*}
\fi
%
\subsection{Representative plane of initial conditions and equilibria}
%
The simplest class of solutions which can be studied most effectively, are the
equilibria (or stationary solutions).  In the multi-parameter dynamical systems,
their positions (coordinates), stability, and bifurcations provide  information
on the general structure of the phase space. Hence, we focus on the
stationary solutions emerging in the secular model of a two-planet system with
the GR and QM corrections. 

After the averaging of $\Hfull$ over the mean anomalies (i.e., over
the orbital periods), the secular Hamiltonian ($\Hsec$) does not depend on $l_i
\equiv \mathcal{M}_i$ anymore. Therefore, as we mentioned already,
the conjugate actions $L_i$ become
constants of motion (hence, the mean semi-major axes are also constant).
Moreover, in the coplanar problem the longitudes of nodes are undefined and
irrelevant for the dynamics. The secular energy of two-planet system does not
depend on individual longitudes of pericenters $\varpi_i$, but only on the their
difference  $\Delta\varpi = \varpi_1-\varpi_2$ \citep{Brouwer1961}. These basic
facts can be expressed  with the help of appropriate canonical transformation
\citep{Michtchenko2004}:
\begin{eqnarray}
\label{eqtot1}
\Delta\varpi \equiv \varpi_1 - \varpi_2, \quad &&G_1,\\
\label{eqtot2}
\varpi_2, \quad &&C = G_1 + G_2.
\end{eqnarray}
Because $\varpi_2$ is the cyclic angle, the conjugate momentum equal to the
total angular momentum $C$ is conserved. For a fixed 
angular momentum as a constant parameter, the  phase-space  of reduced system
become two-dimensional,  and  the system is integrable.  We can now choose
$\Delta\varpi$ as the canonical angle and $G_1$ as the
conjugate momentum. Alternatively, the role of the momentum may be attributed to
$e_1$. Obviously,  $G_2$ (or $e_2$) becomes a dependent
variable (through the $C$ integral).  

To study the dynamics of the reduced secular system in a global manner, we
follow a concept of the so called  {\em representative plane} of initial
conditions introduced by \cite{Michtchenko2004}. The representative plane
(\RP{}) comprises of points in the phase space which  lie on a specific plane
crossing all phase trajectories of the system. In the cited paper, it has been
shown that a good choice of the \RP{}-plane flows from the condition of
vanishing derivatives of the secular Hamiltonian over $\Delta{\varpi}$. It is
equivalent to the symmetry of secular interactions with respect to fixed apsidal
line of a selected orbit and it means that time derivatives of the conjugate
momenta (which can be expressed by eccentricities) must vanish.  Indeed, in
accord with the law of conservation of the  total angular momentum,  the
eccentricities must reach at the same time maximal and minimal values along the phase
trajectories, hence $\dot{e_i}=0$, $i=1,2$.  Regarding the classic problem with
Newtonian point-to-point interactions, this condition is satisfied when
$\Delta{\varpi}=0$ or $\Delta{\varpi}=\pi$ \citep{Michtchenko2004} [see also
\citep{Migaszewski2008a}]. If we consider the generalized problem of two-planet
system then the concept of the representative plane is still valid. The {\em
small} perturbations which we introduce do not change the dimension of the phase
space and the motion remains on perturbed Keplerian orbits, only the form of the
secular Hamiltonian is modified. Moreover, we may expect that qualitative
properties of the secular system may be changed because we introduce 
a few new
parameters describing physical setup of the  studied system and interactions
governing its dynamical evolution. To encompass both the specific values of
angle $\Delta{\varpi}=0,\pi$, the representative plane can be defined by the
following set of points:
\[
{\cal S} = \{ e_1 \cos{\Delta{\varpi}} \times e_2;  e_1, e_2 \in [0,1) \}, 
\]
where $\Delta\varpi=\pi$,  $e_1\cos{\Delta{\varpi}}<0$ for the left-hand
half-plane (called the \RPm-plane, from hereafter), and $\Delta\varpi=0$,
$e_1\cos{\Delta{\varpi}}>0$ are for the right-hand half-plane (the \RPp-plane).
Hence, the sign of $e_1 \cos\Delta\varpi$ on the $x$-axis tells us on the value of angle
$\Delta\varpi$.

\subsection{A general view of the representative plane}
To illustrate the influence of additional perturbations on the secular dynamics
of the classic model, we compute a set of the \RP{}-planes for wide ranges of
orbital and physical parameters of the studied two-planet  configurations.
Before we discuss these results, at first  we explain how the \RP-{plane}
illustrates  the dynamical structure of the phase space.  Conveniently,
Figure~\ref{fig:fig2}, which is derived for {\em specific} orbital parameters
(given in the caption), reveals  {\em all} relevant features in a condensed form
and its description can be regarded as a guide useful for the further analysis.

We start from the left-hand panel of Fig.~\ref{fig:fig2}. Smooth half-ellipse
like curves marked with thin lines 
are for the levels of the secular energy of the
generalized problem. The straight lines  are for the collision line of orbits
defined through  $a_1 (1\pm e_1) = e_2 (1-e_2)$.
\ifpdf
\begin{figure*}
\centerline{
\hbox{\includegraphics[width=8.8cm]{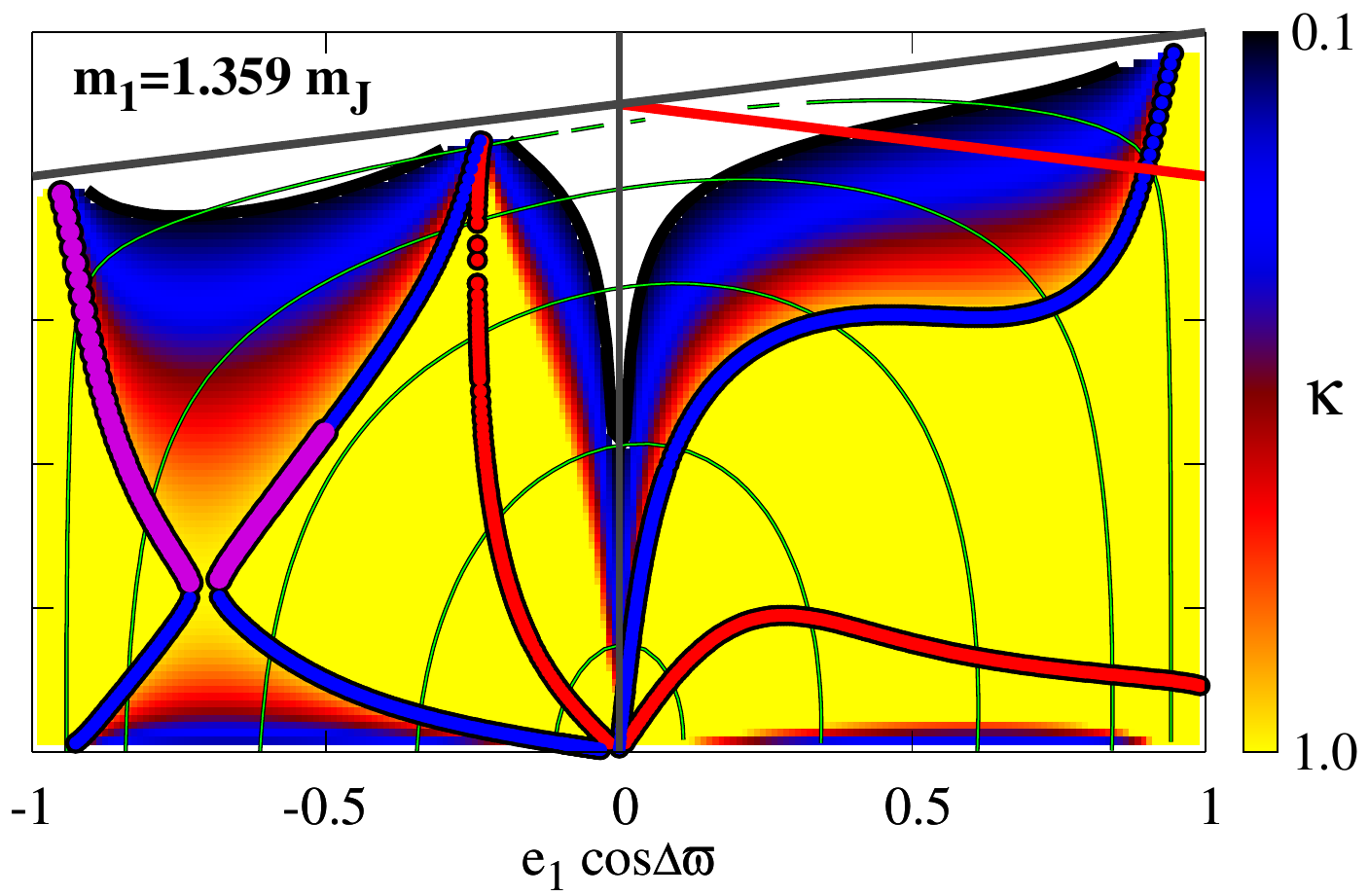}
      \hspace*{0cm}   
      \includegraphics[width=8.8cm]{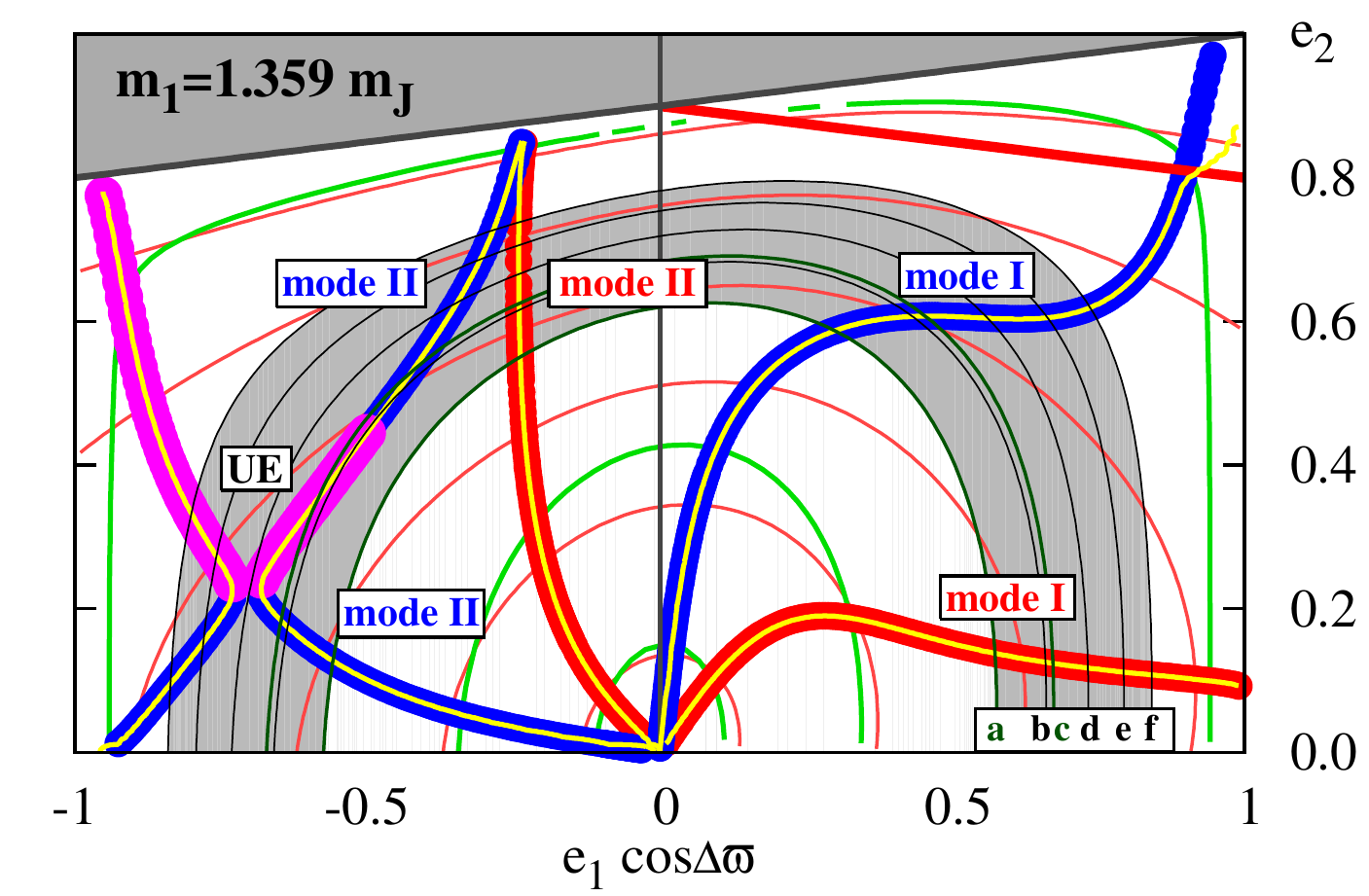}
}
}
\caption{
The representative plane $(e_1 \cos{\Delta{\varpi}}, e_2)$ for the two-planet
system. The elements are $m_1=1.359~\mbox{m}_{\idm{J}}$,
$m_2=0.453~\mbox{m}_{\idm{J}}$, $a_1=0.1$~au, $a_2=1.0$~au. Stellar mass is
1~$M_{\sun}$, rotational period is $T_{\idm{rot}}=30$~days, stellar radius is
1~$R_{\sun}$, $k_L=0.02$. The thick curves are for the stationary solutions: red
curves are for stable equilibria in the classic model, blue and violet curves
are for stable and unstable equilibria of the generalized model, respectively.  
{\em The left panel}:  Color contours are for the ratio of the  apsidal
frequency induced by the general relativity and quadrupole moment  to the 
apsidal frequency caused by mutual interactions between planets, $\kappa$.
\corr{The scale of color code is limited by $\kappa=1$, meaning that the
corrections become more important than the point mass Newtonian gravity. In
general, the yellow colour may refer to $\kappa>1$.} Energy levels of the
generalized model are marked with thin lines.  {\em The right panel}: A few
particular energy levels of the generalized model are labeled with
\textbf{a}---\textbf{f}, accordingly.  See also phase diagrams for these energy
levels  which are shown in Fig.~\ref{fig:fig3} and Fig.~\ref{fig:fig4}. Curves
of equilibria are labeled: mode~I are for aligned apsides ($\Delta\varpi=0$),
mode~II are for anti-aligned apsides ($\Delta\varpi=\pi$),  \UE{} means unstable
equilibria.  The thin half-ellipse like curves are for energy levels computed
for the classic model (red) and generalized model (green). More details can be
found in the text. 
}
\label{fig:fig2}
\end{figure*}
\else 
\begin{figure*}
\centerline{
\hbox{\includegraphics[width=8.8cm]{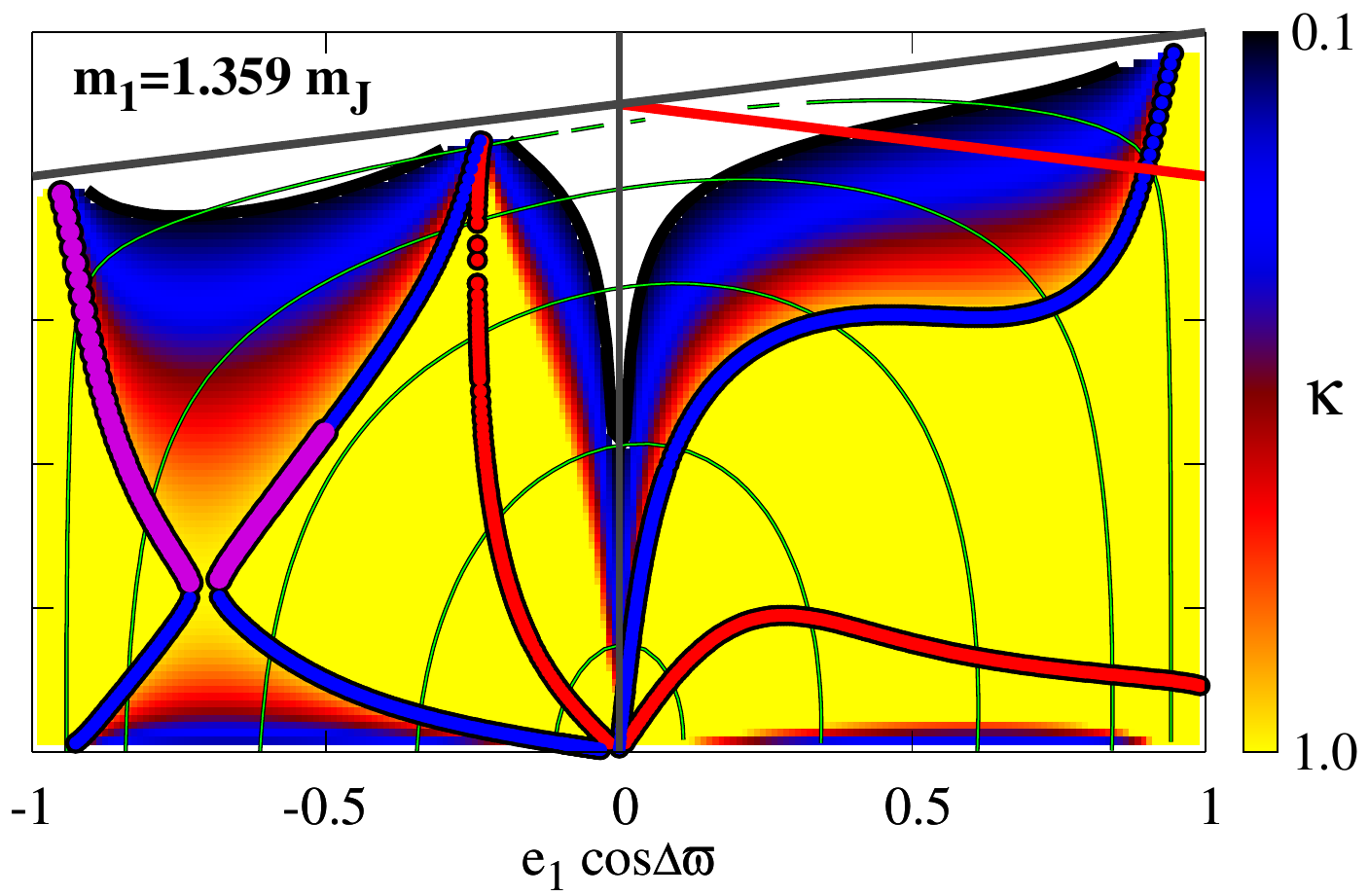}
      \hspace*{0cm}   
      \includegraphics[width=8.8cm]{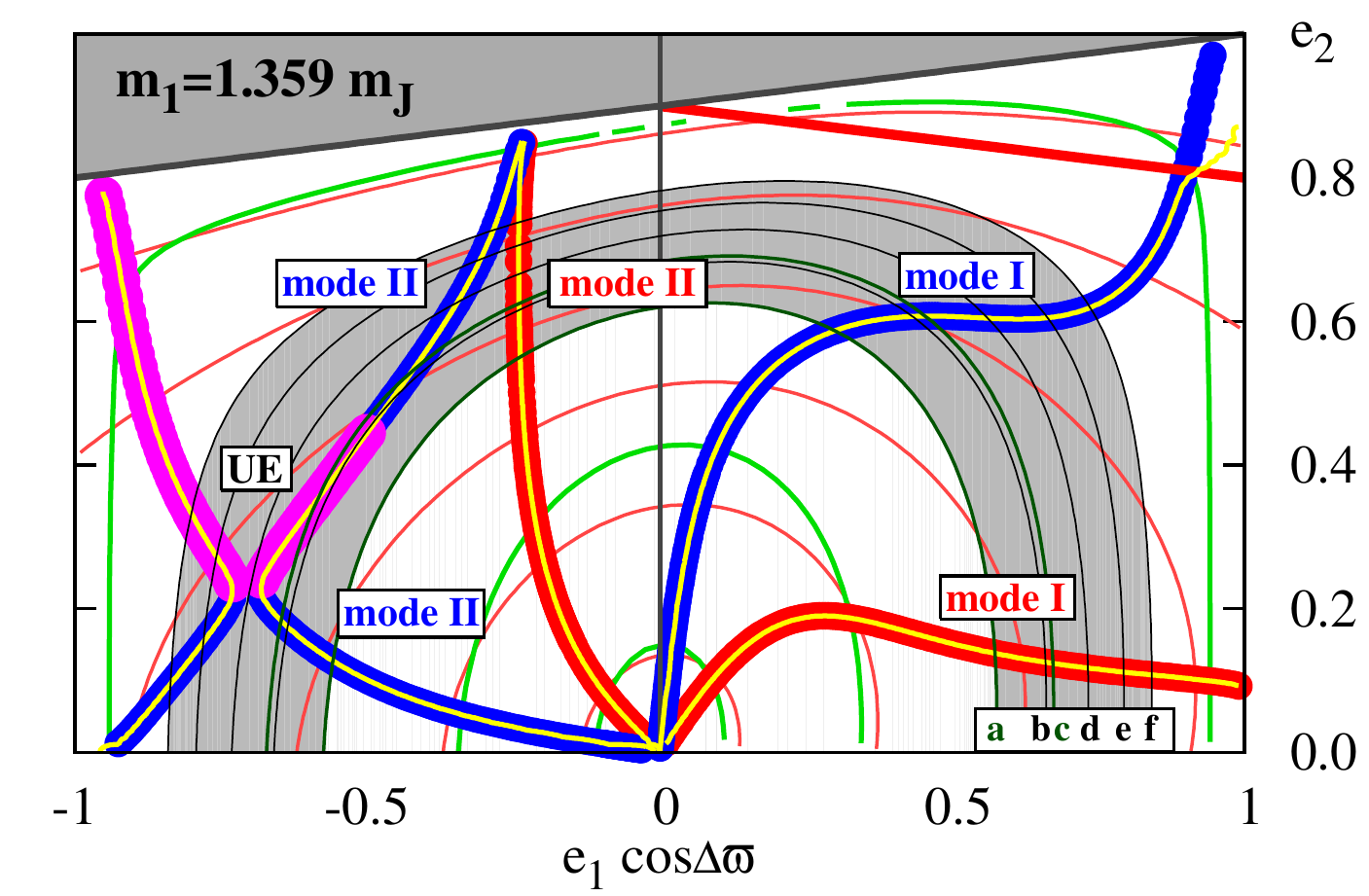}
}
}
\caption{
The representative plane $(e_1 \cos{\Delta{\varpi}}, e_2)$ for the two-planet
system. The elements are $m_1=1.359~\mbox{m}_{\idm{J}}$,
$m_2=0.453~\mbox{m}_{\idm{J}}$, $a_1=0.1$~au, $a_2=1.0$~au. Stellar mass is
1~$M_{\sun}$, rotational period is $T_{\idm{rot}}=30$~days, stellar radius is
1~$R_{\sun}$, $k_L=0.02$. The thick curves are for the stationary solutions: red
curves are for stable equilibria in the classic model, blue and violet curves
are for stable and unstable equilibria of the generalized model, respectively.  
{\em The left panel}:  Color contours are for the ratio of the  apsidal
frequency induced by the general relativity and quadrupole moment  to the 
apsidal frequency caused by mutual interactions between planets, $\kappa$.
\corr{The scale of color code is limited by $\kappa=1$, meaning that the
corrections become more important than the point mass Newtonian gravity. In
general, the yellow colour may refer to $\kappa>1$.} Energy levels of the
generalized model are marked with thin lines.  {\em The right panel}: A few
particular energy levels of the generalized model are labeled with
\textbf{a}---\textbf{f}, accordingly.  See also phase diagrams for these energy
levels  which are shown in Fig.~\ref{fig:fig3} and Fig.~\ref{fig:fig4}. Curves
of equilibria are labeled: mode~I are for aligned apsides ($\Delta\varpi=0$),
mode~II are for anti-aligned apsides ($\Delta\varpi=\pi$),  \UE{} means unstable
equilibria.  The thin half-ellipse like curves are for energy levels computed
for the classic model (red) and generalized model (green). More details can be
found in the text. 
}
\label{fig:fig2}
\end{figure*} 
\fi
The relative magnitude of the corrections to the secular Hamiltonian
is represented by contour levels of the following coefficient:
\begin{equation}
\label{eq:kappa}
\kappa(e_1 \cos \Delta\varpi,e_2)
= \left|\frac{\partial\left<\Hrel\right>/\partial{G_1} +
\partial\left<\Hflat\right>/\partial{G_1}}
{\partial\left<\Hmut\right>/\partial{G_1}}\right|,
\end{equation}
i.e., the ratio of the apsidal frequency of the inner pericenter induced by
$\left<\Hrel + \Hflat\right>$ relative to the  ``natural''  apsidal frequency in
the point mass Newtonian model. Regions of the \RP{}-plane  where $\kappa>0.1$, are
color-coded, according to the levels of constant $\kappa$. Yellow color (light
gray in the printed paper) encodes $\kappa \ge 1$, meaning that
 the apsidal frequency of
the innermost pericenter forced by  {\em perturbing} GR+QM corrections is
larger than the relative pericenter frequency $\dot{\Delta\varpi}$ 
caused by the secular NG interactions between  planets. 

The thick curves defined through:
\begin{equation}
\frac{\partial{\Hsec}}{\partial{G_1}} = 0,
\end{equation}
can be attributed to the stationary solutions of the reduced, two-dimensional system
with $(G_1,\Delta\varpi)$-variables, because  coordinates of all points of 
these curves must also satisfy
\begin{equation}
\frac{\partial{\Hsec}}{\partial{\Delta{\varpi}}} = 0,
\end{equation}
according to the definition of the \RP{}-plane.  Such solutions are periodic
orbits of the full secular system (Eqs.~\ref{eqtot1}--\ref{eqtot2}). We  examine
the Lyapunov stability of these equilibria with the help of the Lyapunov
theorem, adopting  $\Hsec$ as the Lyapunov function in the cases when they 
correspond to its maximum (or a minimum) in the reduced two-dimensional phase
space.  \corr{Because the investigated dynamical system has one degree of
freedom, the stable or unstable equilibria can be easily identified as extrema
or saddles of the secular Hamiltonian in the representative plane,
respectively}. 

The positions of equilibria [or stationary modes \citep{Michtchenko2004}] help
us to distinguish between different types (families) of orbits characterized by
librations of $\Delta\varpi$ around a particular value (libration center).  The
red, thick curves drawn in both panels of Fig.~\ref{fig:fig2} are for stationary
modes in the classic model \citep{Michtchenko2004}, while the equilibria in the
generalized model are drawn with thick, blue curves. In the \RPp{} half-plane,
these stationary solutions are classified by \cite{Michtchenko2004} as mode~I
equilibria, and they are characterized  by librations of $\Delta\varpi$
around~$0$ in the neighboring trajectories.  In the \RPm{} half-plane, we can
find also  Lyapunov stable mode~II solutions related to librations of
$\Delta\varpi$ around~$\pi$ in close trajectories.  Some parts of the equilibria
curves are marked with   violet color (for the generalized model). These points
denote unstable equilibria (\UE{}) accompanied by {\em the true secular
resonance} (the \TSR{} from hereafter). Such unstable equilibria in the
\RPp{}-plane  are discovered  by \cite{Michtchenko2004} in the secular  classic
coplanar model of 2-planets. Obviously, mode~I and mode~II solutions known as
generic features of the classic model, exist also in the generalized problem.
However, their  positions in the phase space may be heavily affected by
apparently subtle GR and QM perturbations.  

Note that along the red thick curves representing equilibria in the classic
model, $\kappa$ is undefined. 

Now, let us examine the right-hand panel of Fig.~\ref{fig:fig2}. In this plot,
besides levels of the secular energy of the classic model (red, thin curves), we
also plot  such levels for the generalized problem (green, thin curves). We can
observe a significant discrepancy between the shapes of contour levels of both
Hamiltonians. 

In this plot we mark also a few specific levels of $\Hsec$: 
$\mathcal{E}_{\idm{\bf a}} = -2.247$,
$\mathcal{E}_{\idm{\bf b}} = -2.24955$,
$\mathcal{E}_{\idm{\bf c}} = -2.25$,
$\mathcal{E}_{\idm{\bf d}} = -2.2525$,
$\mathcal{E}_{\idm{\bf e}} = -2.256$,
$\mathcal{E}_{\idm{\bf f}} = -2.26$,
respectively; here, we skipped all constant terms (the Keplerian part, indirect
term, constant relativistic term) in the full secular Hamiltonian, and  the
energy values are given in terms of $10^{-5}~M_{\sun}\mbox{au}^2\mbox{yr}^{-2}$ 
when 1~$M_{\sun}$, 1~au, and 1~sideral year are taken as units of mass,
distance, and time, respectively (then the Gauss constant $k=2\pi$).  Because
the motion of the secular system  must be confined to fixed energy level, points
at which a particular energy level crosses the curve representing stationary
solutions, tell us on distinct equilibria and their bifurcations. For instance,
following energy level labeled with {\bf a} in the right-hand panel of Fig.~2,
we see that it  crosses the equilibria curves at two points. The first one,
found in the \RPp{} half-plane, corresponds to mode~I solution.  The second
cross-point is found in the \RPm{} half-plane, and it marks the mode~II
equilibrium.

To estimate the precision of the analytical theory, the energy levels  in
Fig.~\ref{fig:fig2} are calculated with exact semi-analytical averaging
\citep{Michtchenko2004} that makes use of precise adaptive Gauss-Legendre
quadratures \citep{Migaszewski2008b}.  For a comparison, the stationary modes
are computed with both methods: we recall that thick curves represent equilibria
calculated with the semi-numerical averaging while the thin curves (over-plotted
on them) are for  stationary solutions calculated with the help of the
analytical theory outlined in Sect.~2 (the right-hand panel of
Fig.~\ref{fig:fig2}). The results of these methods are in excellent accord. The
most significant  discrepancies between the results appear in the \RPp{}
half-plane (for $\Delta\varpi=0$), in the regime of large eccentricity $e_2$. In
such a case, the secular series representing $\Hmut$ diverge over the {\em
anti-collision} line (marked with red, straight line). This problem is discussed
in detail in \citep{Migaszewski2008a}.  Obviously, over this anti-collision
line, $r_2>r_1$ at some parts of orbits, breaking the underlying assumption that
we require to expand the inverse of the mutual distance in convergent series.
Moreover, we demonstrate that the analytical theory reproduces the dynamics of 
hierarchical configurations up to very large $e_2$. Another, direct test of the
precision of the analytic theory is given in Sect.~3.4.
%
\subsection{Phase diagrams and the non-linear secular resonance}
%
To investigate more closely the secular dynamics related to the new mode~II and
\UE{} solutions, and to understand the structure of the phase space in more
details, we compute a number of {\em phase plots} (or {\em phase
diagrams}). The secular energy is kept constant and we draw the phase
trajectories in the ($e_1 \cos{\Delta{\varpi}}, e_1 \sin{\Delta{\varpi}}$)--, as
well as  ($e_2 \cos{\Delta{\varpi}}, e_2 \sin{\Delta{\varpi}}$)--planes,
choosing the initial conditions along the fixed energy level.  The phase plots
are computed with the help of the analytic theory.  For a reference, we recall
the right-hand panel of Fig.~\ref{fig:fig2} which illustrates the  \RP{}-plane
of $(e_1 \cos{\Delta{\varpi}}, e_2)$.   We recall that the \UE{} of the
generalized model are marked with violet curves.

The phase diagrams are illustrated in Fig.~\ref{fig:fig3}. The top row is for
the energy level of $\mathcal{E}_{\idm{a}} = -2.247\times10^{-5}$  which is
labeled with {\bf a} in the right-hand panel of Fig.~\ref{fig:fig2}.  This
energy level crosses the equilibria curves in two points which  can be
recognized in the phase diagrams as libration centers of $\Delta\varpi=0$
(labeled with mode~I), and of $\Delta\varpi=\pi$  (labeled with mode~II),
respectively. Overall, the phase diagrams look like qualitatively the same as in
the classic model. Both libration modes are separated from the circulation zone
of $\Delta\varpi$ by {\em false} separatrices
\citep{Pauwels1983,Michtchenko2004}, i.e., the transition between each mode and
the circulation  of $\Delta\varpi$ {\em does not} involve solutions with
infinite period. \corr{ This is illustrated in smaller, bottom plots
accompanying each phase diagram, which show the secular, fundamental frequency
$g$ of the mean system for initial conditions lying on the $x$ axis of the
respective phase diagram.   This frequency has been determined through 
solutions of the secular equations of motion. Clearly, when the true separatrix
crosses the $e_{1,2}\cos\Delta\varpi$-axis, $g$ decreases to 0. Crossings of
false separatrices do not lead to any discontinuity of smooth plots of $g$. }
\ifpdf
\begin{figure*}
\centerline{
\vbox{
\hbox{\includegraphics[width=68mm]{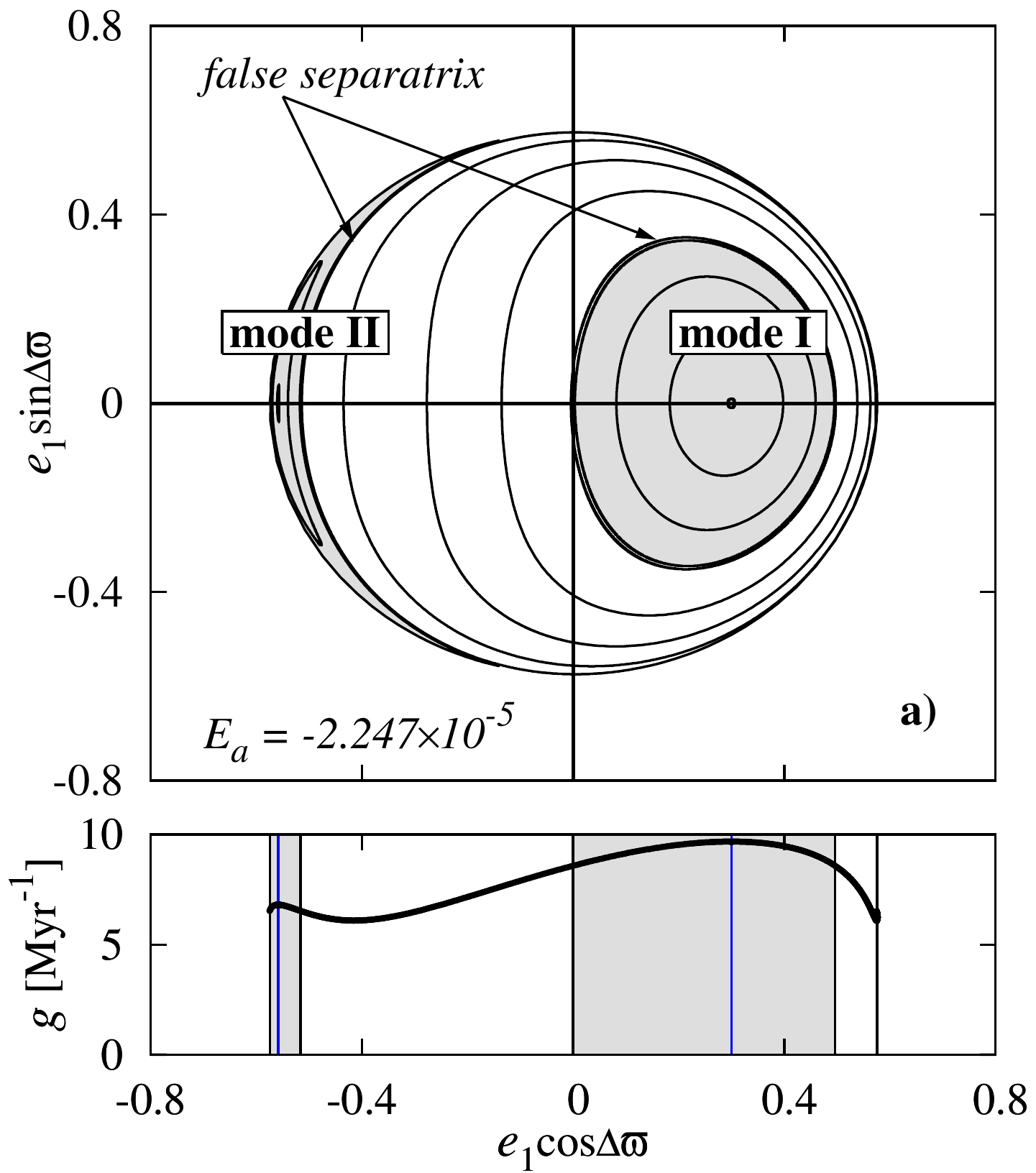}
\hspace*{6mm}
      \includegraphics[width=68mm]{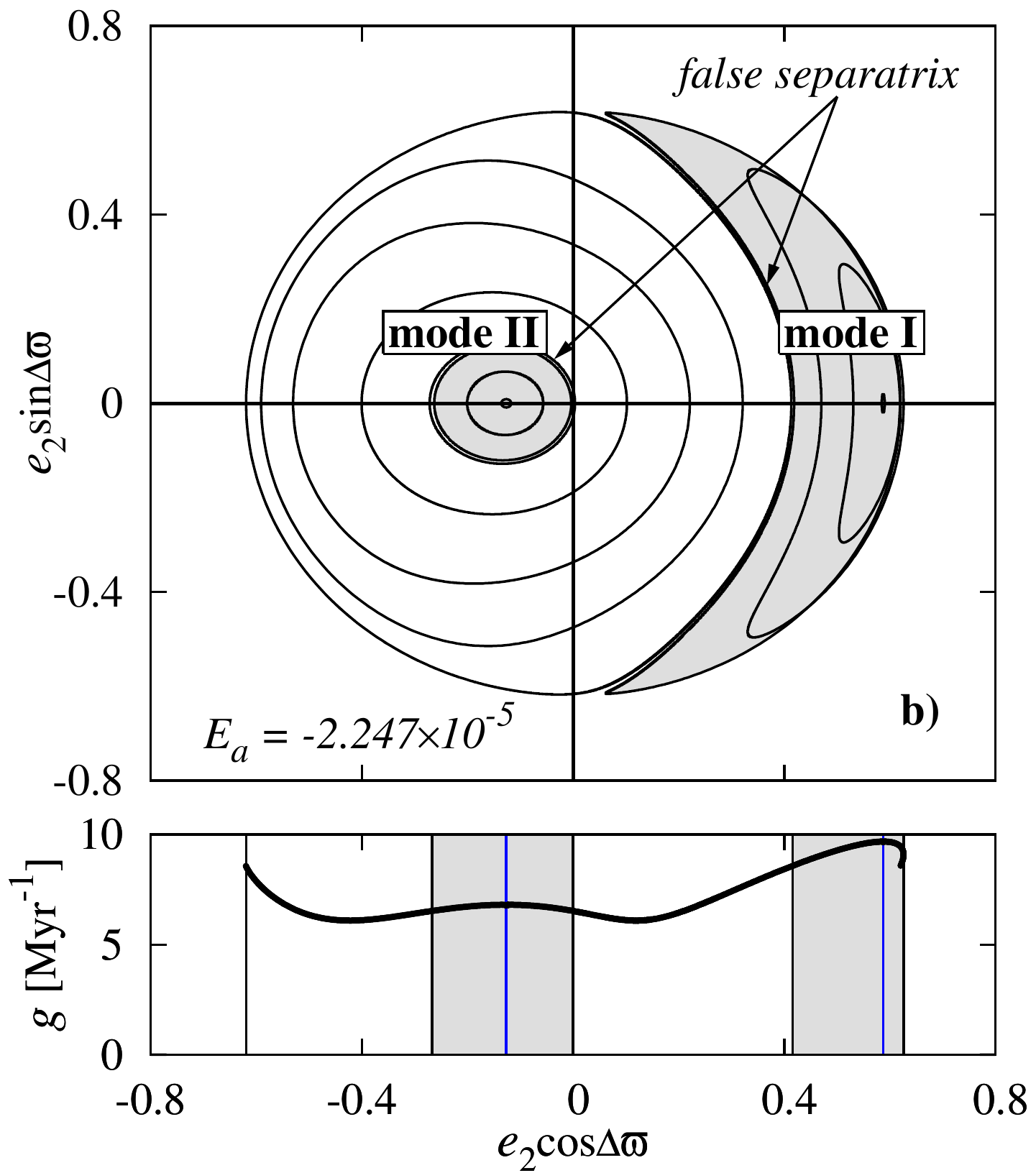}}    
\hbox{\includegraphics[width=68mm]{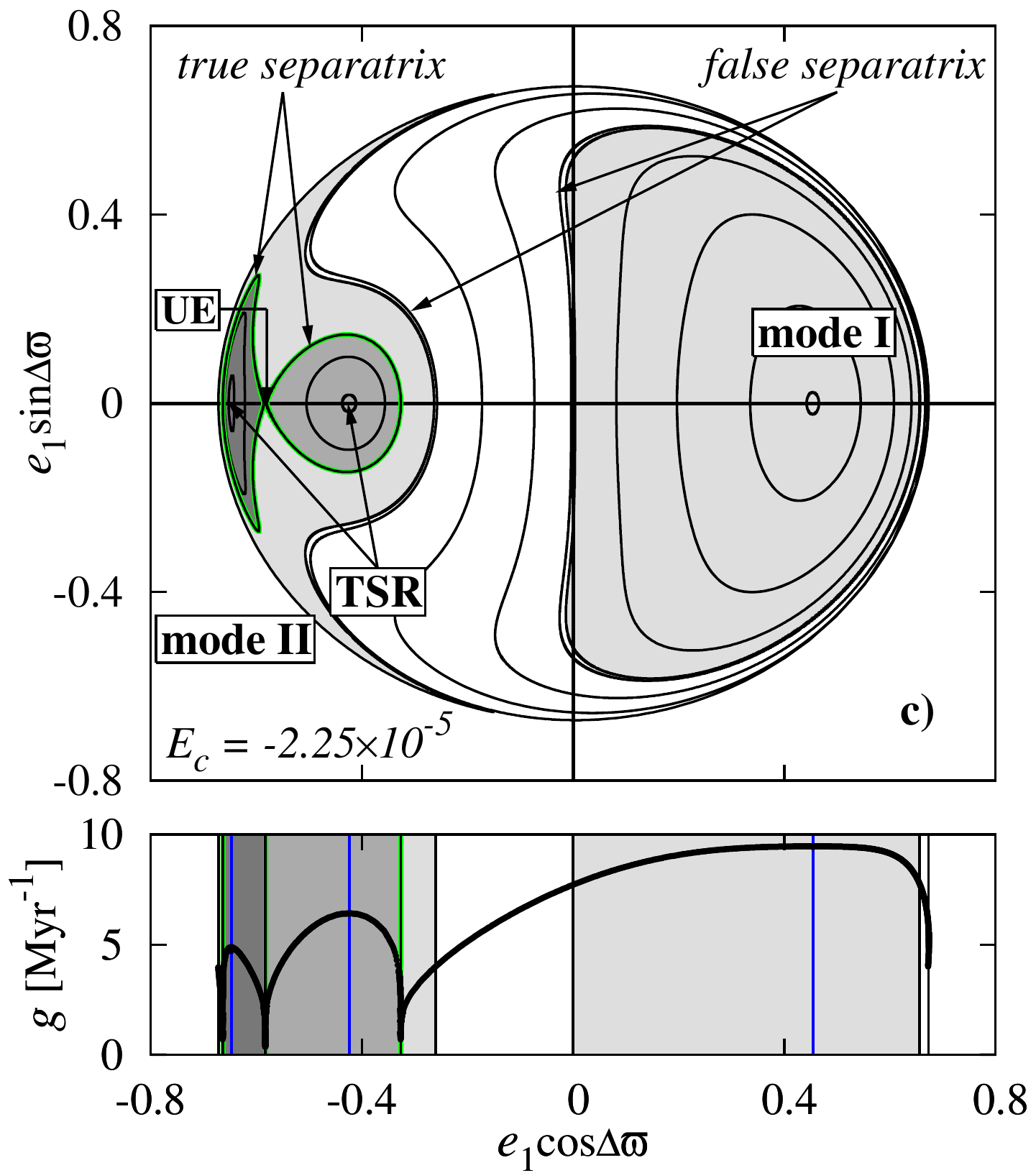}
\hspace*{6mm}
      \includegraphics[width=68mm]{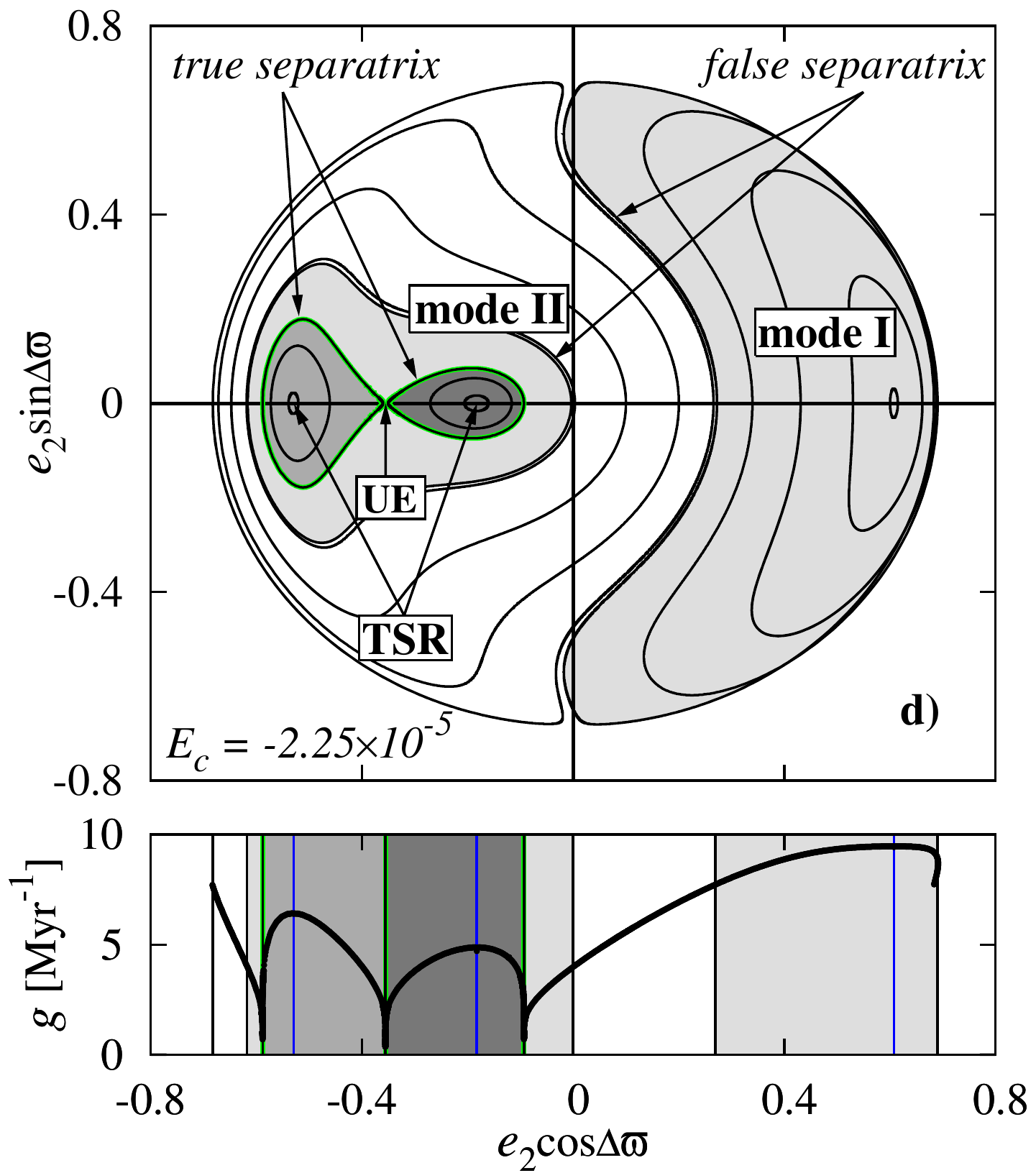}}
}
}      
\caption{
Phase diagrams computed for two-planet secular system described with the same
parameters as used for the construction of Fig. \ref{fig:fig2}. \corr{ Each
phase diagram is accompanied by a smaller  plot of the 
fundamental frequency $g$ of
the secular solutions  calculated for initial conditions lying on  the $y \equiv
e_{1,2}\cos\Delta\varpi$-axis. } Panels in the top row are for the secular
energy $E_{\bf a} = -2.247\times10^{-5}$, panels in the bottom row are for
$E_{\bf c} = -2.25\times10^{-5}$. These levels are labeled in Fig.
\ref{fig:fig2} with ({\bf a}) and ({\bf c}), respectively.  The left-hand panels
are for the ($e_1 \cos{\Delta{\varpi}}, e_1 \sin{\Delta{\varpi}}$)-plane, the
right-hand panels are  for ($e_2 \cos{\Delta{\varpi}}, e_2
\sin{\Delta{\varpi}}$)-plane. Shaded regions mark different  libration zones
around mode~I ($\Delta\varpi=0$) and mode~II ($\Delta\varpi=\pi$), respectively.
\UE{} marks  unstable equilibria  accompanied by libration zones around
$\Delta\varpi=\pi$ and centered at the \TSR{} solutions.  \corr{The true
separatrices are indicated by $g \rightarrow 0$}. See the text for more details.
}
\label{fig:fig3}
\end{figure*}
\else
\begin{figure*}
\centerline{
\vbox{
\hbox{\includegraphics[width=68mm]{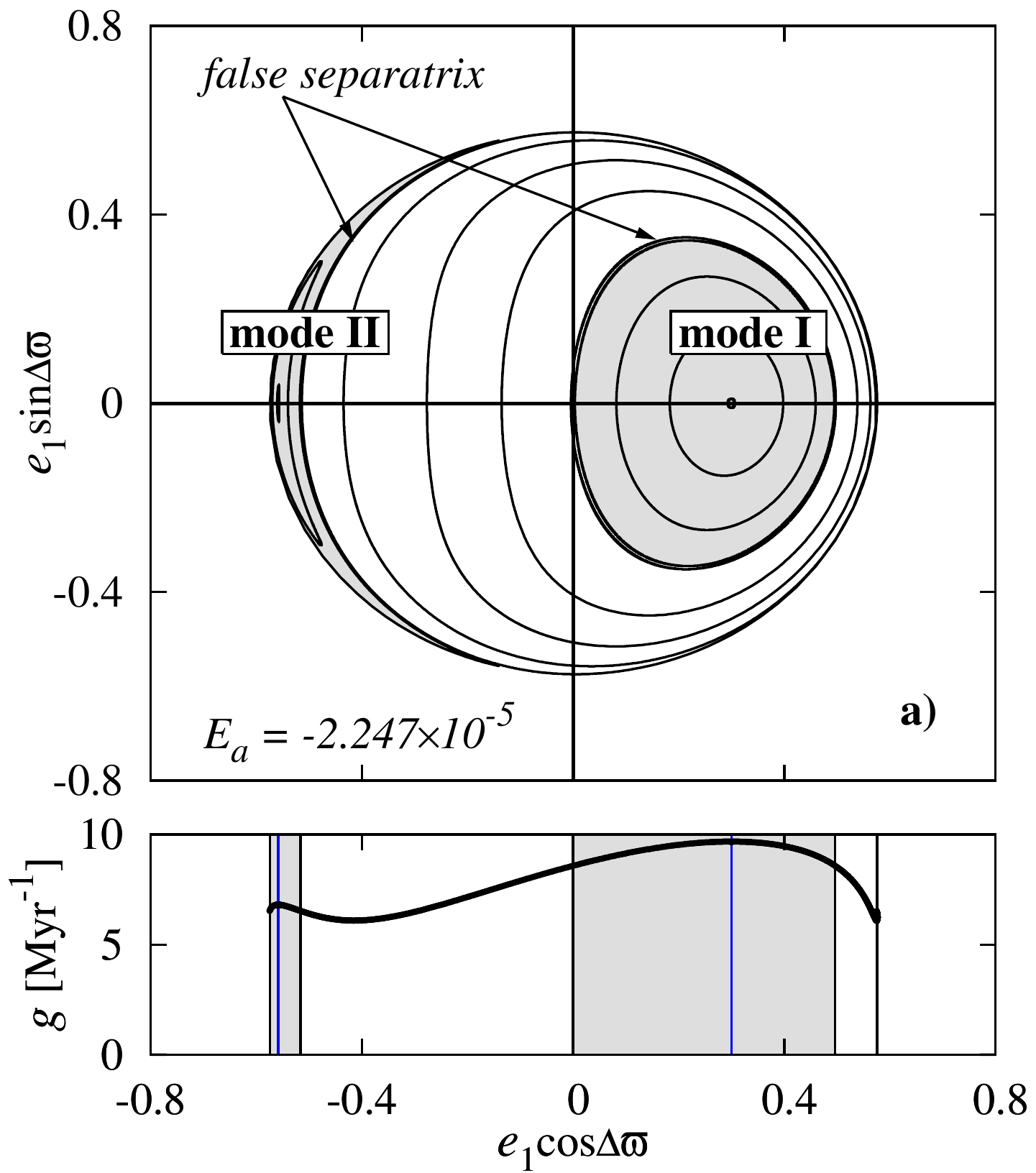}
\hspace*{6mm}
      \includegraphics[width=68mm]{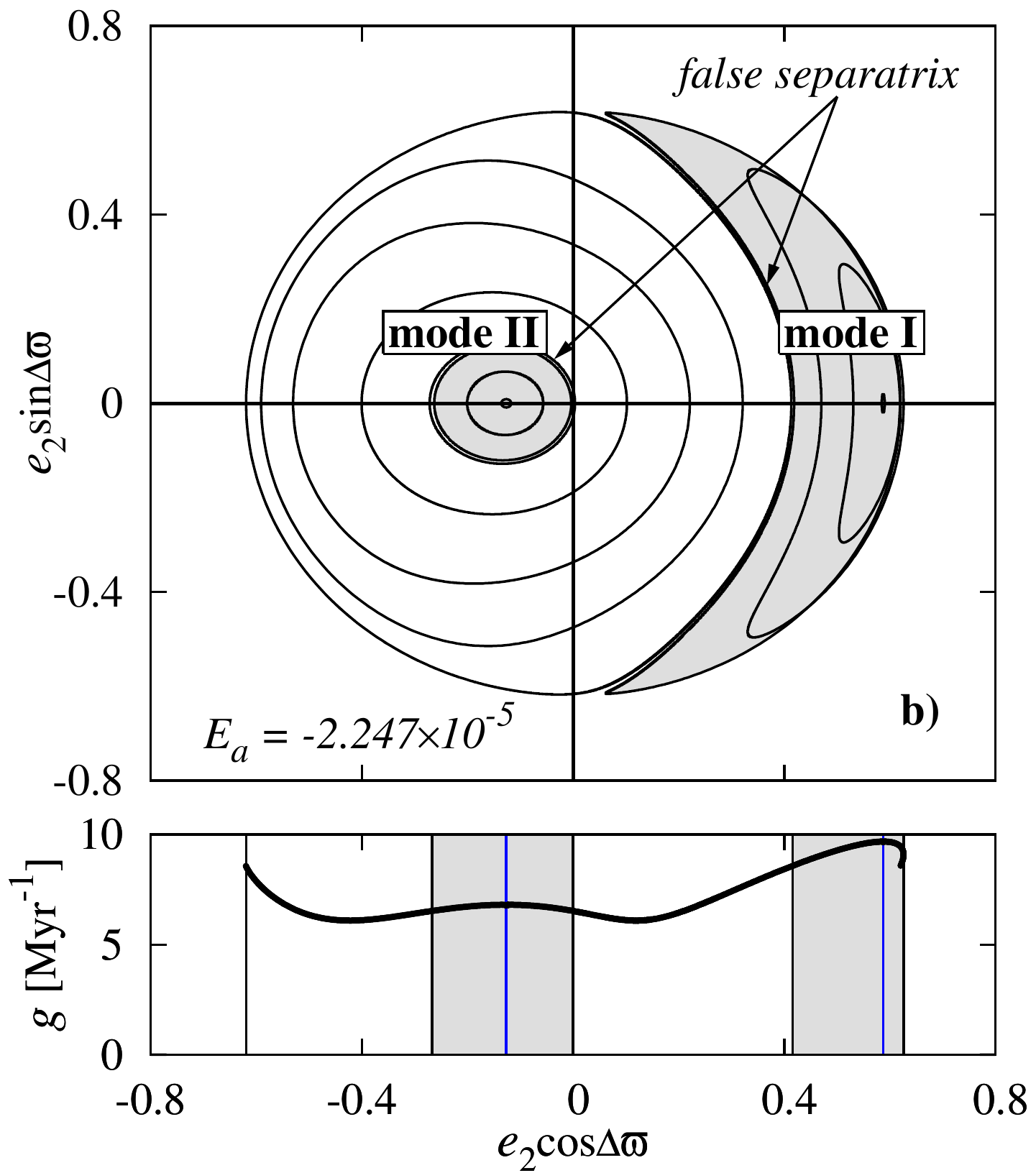}}    
\hbox{\includegraphics[width=68mm]{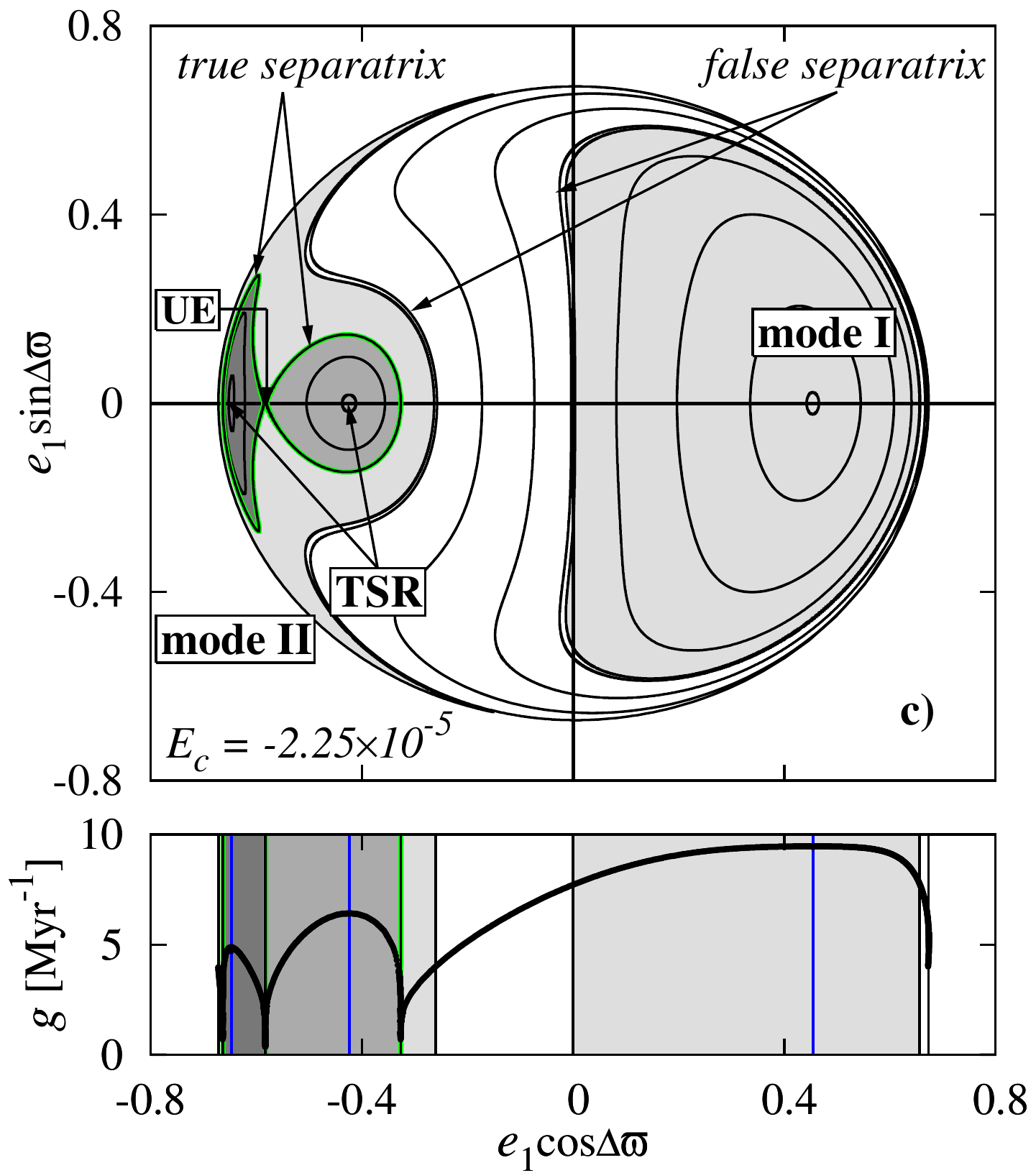}
\hspace*{6mm}
      \includegraphics[width=68mm]{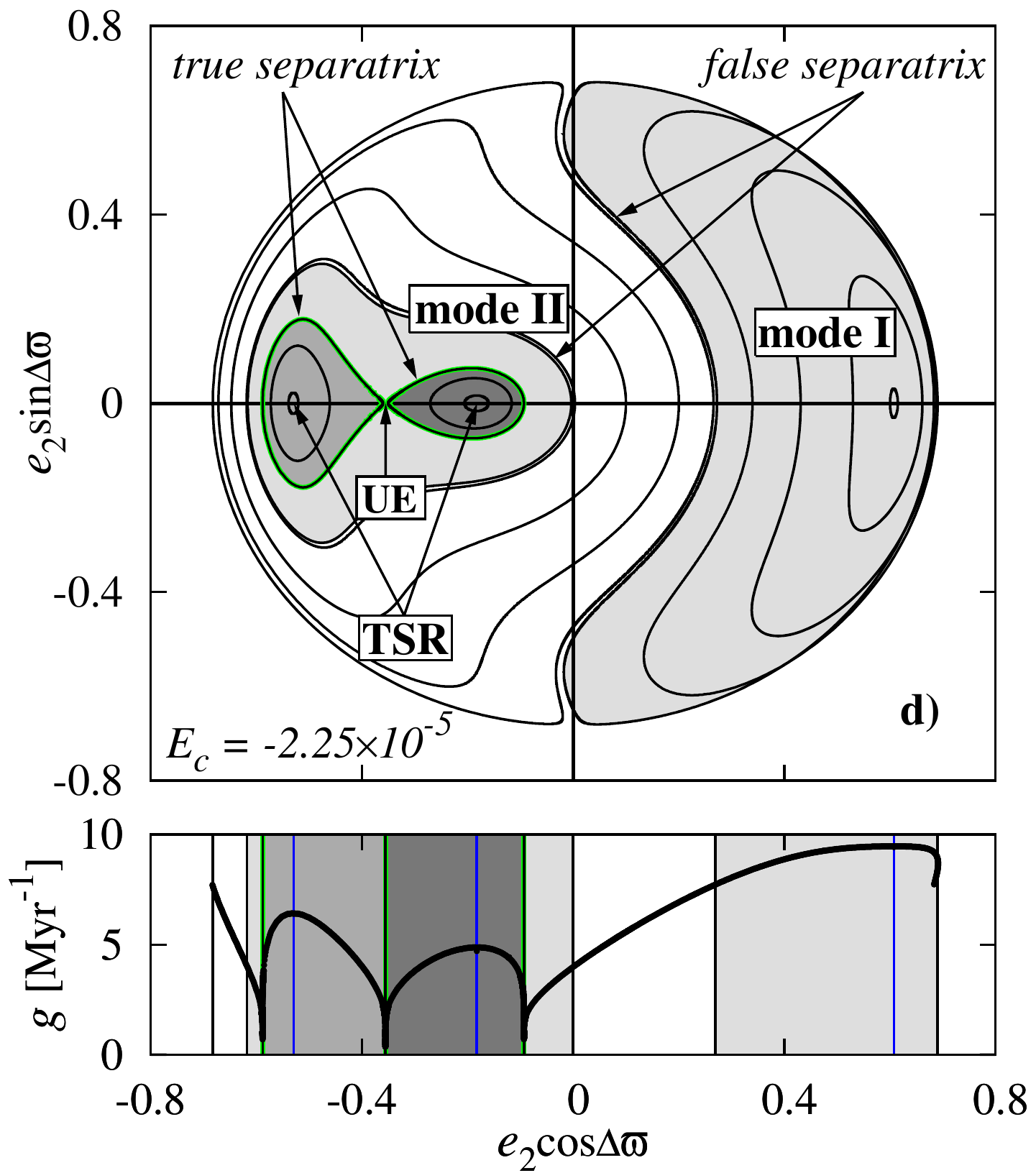}}
}
}      
\caption{
Phase diagrams computed for two-planet secular system described with the same
parameters as used for the construction of Fig. \ref{fig:fig2}. \corr{ Each
phase diagram is accompanied by a smaller  plot of the 
fundamental frequency $g$ of
the secular solutions  calculated for initial conditions lying on  the $y \equiv
e_{1,2}\cos\Delta\varpi$-axis. } Panels in the top row are for the secular
energy $E_{\bf a} = -2.247\times10^{-5}$, panels in the bottom row are for
$E_{\bf c} = -2.25\times10^{-5}$. These levels are labeled in Fig.
\ref{fig:fig2} with ({\bf a}) and ({\bf c}), respectively.  The left-hand panels
are for the ($e_1 \cos{\Delta{\varpi}}, e_1 \sin{\Delta{\varpi}}$)-plane, the
right-hand panels are  for ($e_2 \cos{\Delta{\varpi}}, e_2
\sin{\Delta{\varpi}}$)-plane. Shaded regions mark different  libration zones
around mode~I ($\Delta\varpi=0$) and mode~II ($\Delta\varpi=\pi$), respectively.
\UE{} marks  unstable equilibria  accompanied by libration zones around
$\Delta\varpi=\pi$ and centered at the \TSR{} solutions.  \corr{The true
separatrices are indicated by $g \rightarrow 0$}. See the text for more details.
}
\label{fig:fig3}
\end{figure*}
\fi
The phase diagrams are much more complicated for energy levels labeled with {\bf
b}--{\bf f}, which cross the equilibria curves at more than two points. We
analyze in detail the energy level ({\bf c})  which intersects the curve of
stationary solutions in {\em four} points. These points can be recognized in the
phase diagrams shown in the bottom panels of Fig.~\ref{fig:fig3}. We can
identify them easily in the $x$-axis of these diagrams  (because,  such points
of the \RP{}-plane have $y \equiv  e_{1,2}\sin \Delta\varpi=0$). Starting at the
\RPp{}-plane and following the energy level counter-clockwise, we have a stable
mode~I equilibrium surrounded by large zone of librations of $\Delta\varpi$
around~$0$ which corresponds to single crossing point of the energy level and
the equilibria curve in the \RPp{}-plane. In the \RPm{}-plane, we have {\em
three} such points, two of them are Lyapunov stable, and one point in the middle
is an unstable equilibrium. This part of the phase space, as seen in
Fig.~\ref{fig:fig3}, encompass figure-eight shaded area, involving two islands
of librations (\TSR{}s, or elliptic points) around $\Delta\varpi=\pi$,  and the
hyperbolic \UE{} point lying in the middle between them. Both libration centers
are characterized by $\Delta\varpi=\pi$. Because they are related to stable
equilibria separated by hyperbolic structure of the \UE{}, two parts of  the
phase curve surrounding the islands of the \TSR{}s and that meet in the \UE{},
must form a real separatrix. The whole structure may be  still surrounded by a
zone of librations of $\Delta\varpi$ around $\pi$. It is shaded in light-gray.
Let us note, that this mode~II libration area is confined to the \RPm{}-plane
(hence, in this particular case, angle $\Delta\varpi$  does not pass through
$0$).

A sequence of ($e_2 \cos{\Delta{\varpi}}, e_2 \sin{\Delta{\varpi}}$)-diagrams
for the secular energy levels {\bf a}--{\bf f} are shown in Fig.~\ref{fig:fig4}.
Looking at these levels plotted in the \RP{}-plane, we can now follow a
development of dynamical structures related to the different modes of motion. In
particular, phase diagrams Fig.~\ref{fig:fig4}b and Fig.~\ref{fig:fig4}d reveal
bifurcations of mode~II which emerge the \UE{} and  \TSR{} solutions. Clearly,
the bifurcations may  be identified with points at which the given energy level
is tangent to the equilibria curves in the \RP{}-plane. 

The phase diagrams assure us that the secular dynamics of the generalized model
can be much more complex and rich due to the GR and QM corrections to the NG
Hamiltonian than the secular dynamics in the classic model.
%
\subsection{Numerical test of the analytic secular theory}
%
Finally, we illustrate limitations of  the secular theory and we compare its
results with the outcome of the direct numerical integrations. This comparison
is also directly related to the dynamical stability of the planetary system.

First, we constructed the numerical model of the generalized system
independently on the analytical model. We wrote the equations of motion with
respect to the Jacobi reference frame, using formulation of \cite{Mardling2002} 
who call it {\em the direct code}. In this code, the GR acceleration is modeled 
with the PPN formulae given in \citep{Kidder1995}. In that way we have a
possibility to check the analytic theory in completely independent way, which
also prevents copying logical errors which could be done during the averaging.
After some experiments, we also found that the choice of the reference frame
(e.g., related to  Jacobi, Poincar\'e, or classic-astrocentric coordinates) is
in fact irrelevant for the results of this test. We also do not
account for the difference between the {\em osculating} and the {\em mean}
elements. 

Using the direct code, we integrated numerically a few phase diagrams 
for  nominally {\em
the same initial conditions and parameters} used to draw plots in
Fig.~\ref{fig:fig4}a--f with the help of the analytic, secular model. We computed osculating elements related
to the Jacobi reference  frame over a few secular cycles. The numerically
derived phase curves are drawn with black, filled circles in  respective panels
of Fig.~\ref{fig:fig4}. The solutions obtained with the analytic theory are
over-plotted on these numerical solutions with thiner, green curves. Subsequent
five panels  of Fig.~\ref{fig:fig4}a--e reveal that the agreement of both sets
of solutions is excellent. The analytic theory reproduces qualitative features
seen in the phase plots, and their structure with great accuracy. We find that
both solutions coincide even in the regime of large eccentricities.
Remarkably, the direct code integrations last over CPU time which is by a few
orders of magnitude longer than the calculations carried out with the help of
the analytical theory.

However, in the last panel of Fig.~\ref{fig:fig4}f we can observe significant
deviations of the analytic solutions from the numerical theory, particularly in
the outer parts of the phase diagram.  In fact, in this case $e_2$ is so large that
the assumptions of the secular theory are broken. After examining the
\RP{}-plane (Fig.~\ref{fig:fig2}), we can see that the energy level
corresponding to the last panel passes close to the collision line. To 
illustrate the real border of the dynamical stability, we examined the dynamical
character of solutions in the \RP{}-plane with the help of the Spectral Number
technique \citep{Michtchenko2001}. This simple FFT-based algorithm makes it
possible to distinguish between chaotic and regular solutions. The dynamical
maps shown in Fig.~\ref{fig:fig5} are constructed by counting the number of
frequencies in the FFT-spectrum of the time series, $\{ \sigma(t) = a(t) \exp
\mbox{i}\lambda_i(t) \}$, where $a_i(t)$ and $\lambda_i(t)$ are temporal {\em
canonical} semi-major axes and mean longitude of each planet. The number of
peaks (the Spectral Number, SN from hereafter)  in the spectrum over some noise
level tells us on the character of orbit. Orbits with large SN (grater than
1000) are very chaotic, while the SN $\sim 1$ means small number of frequencies
and a regular, quasi-periodic phase trajectory. Each point in the dynamical maps
represents a phase trajectory  that was integrated over $\sim 10^4~P_{\idm{2}}$.
Although such time span is relevant for the short-term dynamics only, 
calculations took a very long CPU time (a few days on 24 AMD-CPU cores). The
results are shown in two panels of Fig.~\ref{fig:fig5}. The left panel is for
the classic model (only NG interactions are included), while the right panel is
constructed for the generalized model.

Let us analyze the left-hand panel of Fig.~\ref{fig:fig5}
 for the classic model. Solutions, which appear
strongly chaotic  are marked with colors (darker point means larger SN and more
chaotic system).  Clearly, the border of stable motions is irregular and is
shifted towards small $e_2$ by $0.1$--$0.2$ with respect to the formal,
geometrical collision line bordering the triangular region (it is drawn in both
panels of Fig.~\ref{fig:fig5}). The thick red curves mark the equilibria of mode
I and mode~II, respectively. Shaded regions are for the initial conditions in
the \RP{}-plane corresponding to orbital configurations with librating
$\Delta\varpi$.  In the right-hand panel of Fig.~\ref{fig:fig5}, we show the
equilibria curves of  the generalized model and  libration zones of
$\Delta\varpi$ associated with these equilibria. We mark again the SN signature
of  the short-term dynamics. We note that the border of stability is
quite different from that ones of the classic model (compare with the
left-hand panel of Fig.~\ref{fig:fig5}).  This is a very clear example showing
that apparently subtle GR+QM effects may affect the {\em short-term} stability
of the system in a significant way. 

The dynamical map for the generalized model 
(Fig.~\ref{fig:fig5}, the right hand panel) helps us to identify the source of
unstable behavior seen in Fig.~\ref{fig:fig4}f,  revealing that some initial
conditions lead to erratic and irregular behavior. In the dynamical map, we mark
two levels of $\Hsec$ corresponding to phase diagrams drawn in
Fig.~\ref{fig:fig4}e,f, respectively. The energy level ({\bf e}) lies entirely
in the regular region, in spite that $e_2$ may reach values as large as 0.8. The
neighboring level ({\bf f}) may touch the unstable zone, and that is why the
orbital evolution at this energy level  may become very unstable.  Because the
motion must be confined to the fixed energy level, due to the secular evolution,
some initial conditions may be transported to the chaotic zone  (by excitation
of the eccentricity) during  roughly a half of the secular period. Then the
short-term, strong chaos can destabilize the system immediately.  Following the
fixed energy curve, we can also identify three islands of stable motions. The
first one lies in the right half-plane of the phase diagram and is associated
with librations of $\Delta\varpi$ around $0$, see the neighborhood of the
corresponding  cross point in the \RPp-plane, Fig.~\ref{fig:fig2}.  In the
\RPm-plane, we can find corresponding unstable equilibrium and two stable
solutions with associated libration islands shown in Fig.~\ref{fig:fig4}f (see
the left-hand half-plane of the phase diagram) around $(0,-0.1)$ and
$(0,-0.78)$, respectively. 
\ifpdf
\begin{figure*}
\centerline{
\vbox{
\hbox{\includegraphics[width=58mm]{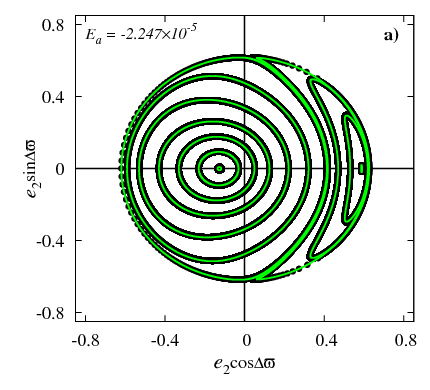}
      \includegraphics[width=58mm]{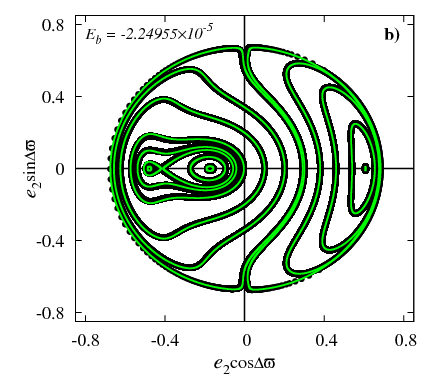}
      \includegraphics[width=58mm]{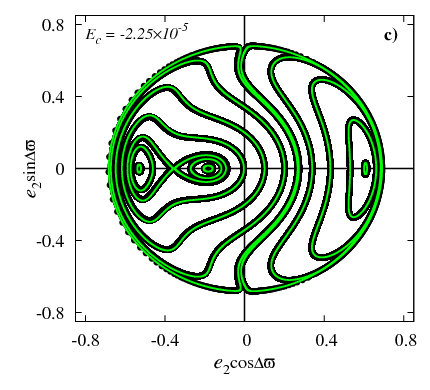}
      }
\hbox{\includegraphics[width=58mm]{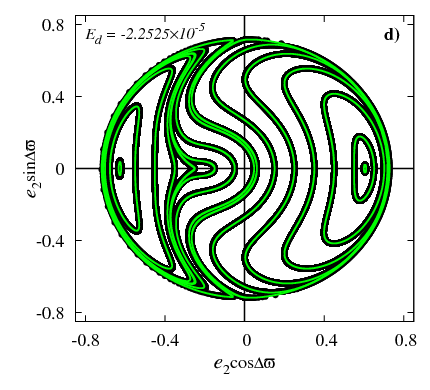}
      \includegraphics[width=58mm]{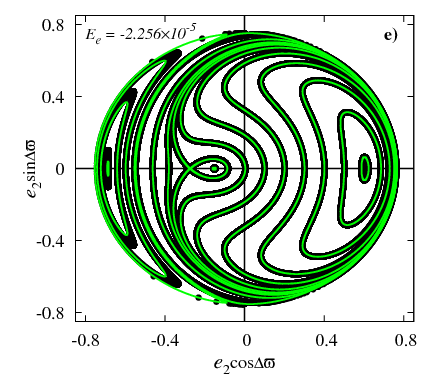}
      \includegraphics[width=58mm]{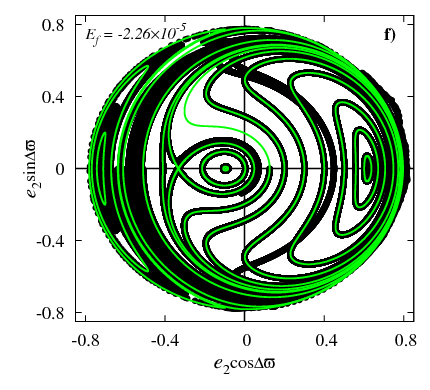}
      }
}
}      
\caption{
Phase diagrams in the $(e_2\cos{\Delta{\varpi}}, e_2
\sin{\Delta{\varpi}})$-plane drawn  to illustrate a comparison of the secular
evolution of the mean system (green, thin curves) with the numerical solutions
of full,  unaveraged system (larger, filled black circles). The orbital
parameters are the same as in Fig. \ref{fig:fig2}. Subsequent panels labeled
with \textbf{a}-\textbf{f}  correspond to  relevant energy levels marked and
labeled in Fig.~\ref{fig:fig2}, accordingly.
}
\label{fig:fig4}
\end{figure*}
\else
\begin{figure*}
\centerline{
\vbox{
\hbox{\includegraphics[width=58mm]{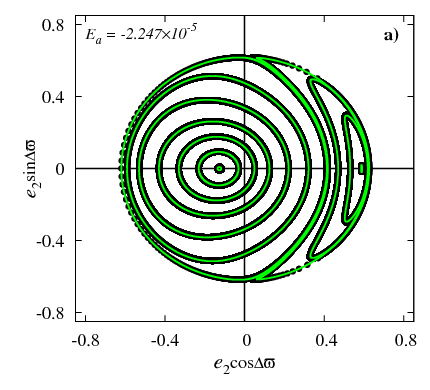}
      \includegraphics[width=58mm]{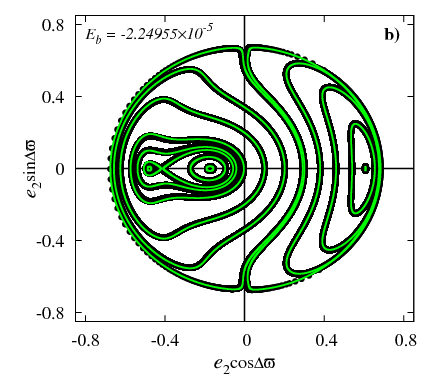}
      \includegraphics[width=58mm]{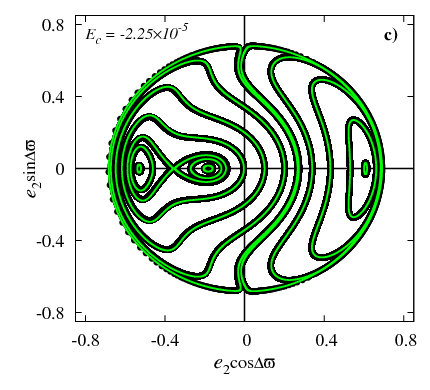}
      }
\hbox{\includegraphics[width=58mm]{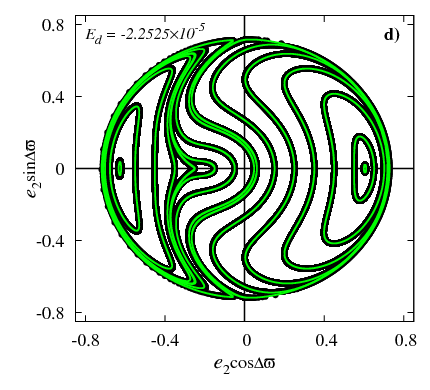}
      \includegraphics[width=58mm]{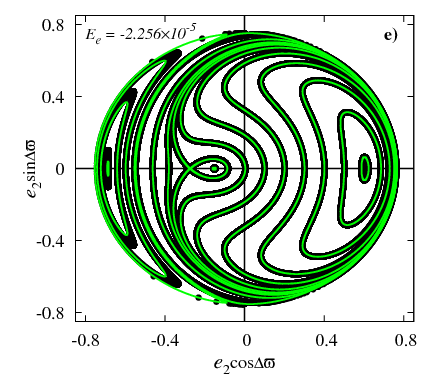}
      \includegraphics[width=58mm]{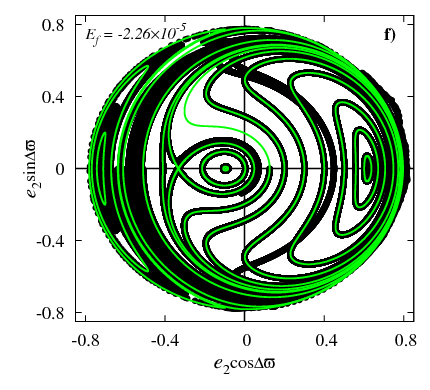}
      }
}
}      
\caption{
Phase diagrams in the $(e_2\cos{\Delta{\varpi}}, e_2
\sin{\Delta{\varpi}})$-plane drawn  to illustrate a comparison of the secular
evolution of the mean system (green, thin curves) with the numerical solutions
of full,  unaveraged system (larger, filled black circles). The orbital
parameters are the same as in Fig. \ref{fig:fig2}. Subsequent panels labeled
with \textbf{a}-\textbf{f}  correspond to  relevant energy levels marked and
labeled in Fig.~\ref{fig:fig2}, accordingly.
}
\label{fig:fig4}
\end{figure*}
\fi
\ifpdf
\begin{figure*}
\centerline{
\hbox{\includegraphics[width=8.7cm]{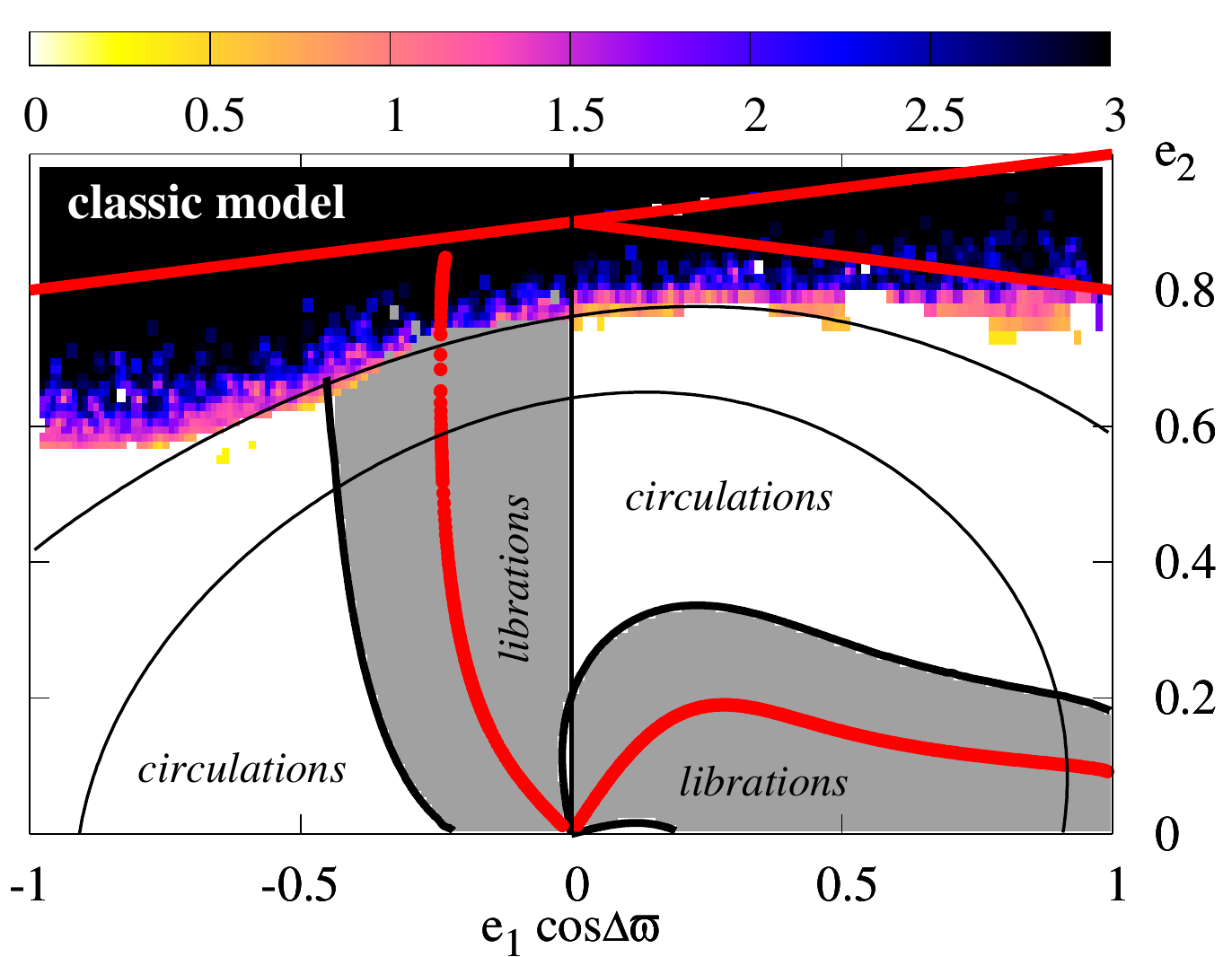}
   \hspace*{0.4cm}   
   \includegraphics[width=8.7cm]{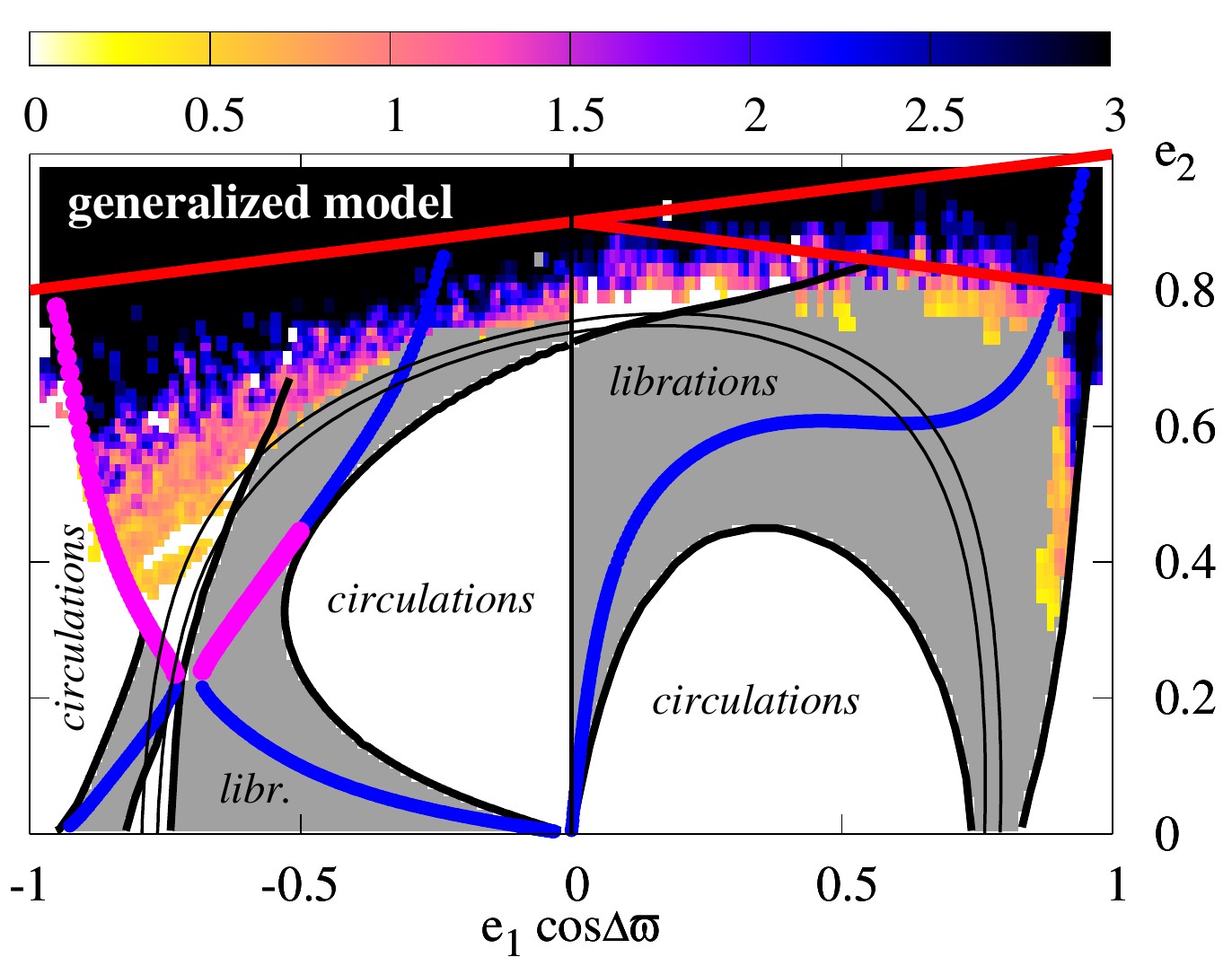}
}
}
\caption{
The \RP{}-planes for the classic model (the left-hand panel) and for the
generalized model (the right-hand panel) for the two-planet system analyzed in
Fig.~\ref{fig:fig2}. The thick, red curves mark stationary modes in  the classic
model, the thick blue curves are for the equilibria in the generalized model.
Shaded areas indicate zones of $\Delta\varpi$ librations. The thick red lines
are for the collision line of orbits. Colors code $\log$~SN of the {\em outer
orbit}, characterizing  solutions derived numerically with the help of the
direct code. Black points are for strongly chaotic solutions with the
$\log\mbox{SN} \sim 3$, white color  is for regular solutions with
$\log\mbox{SN}\sim 0$, intermediate values are marked with the color scale above
the panels, accordingly (see the text for more details).
}
\label{fig:fig5}
\end{figure*}
\else
\begin{figure*}
\centerline{
\hbox{\includegraphics[width=8.7cm]{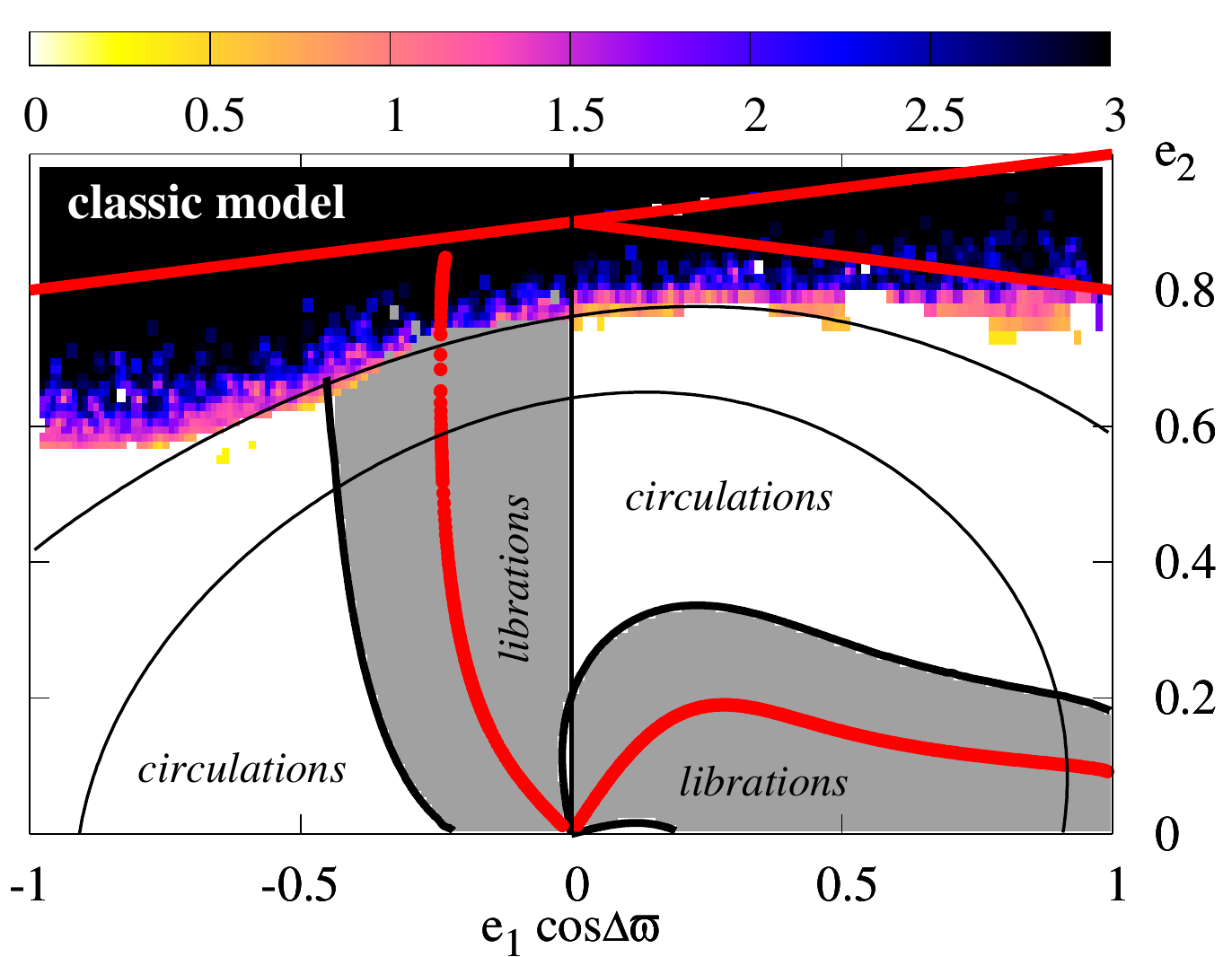}
   \hspace*{0.4cm}   
   \includegraphics[width=8.7cm]{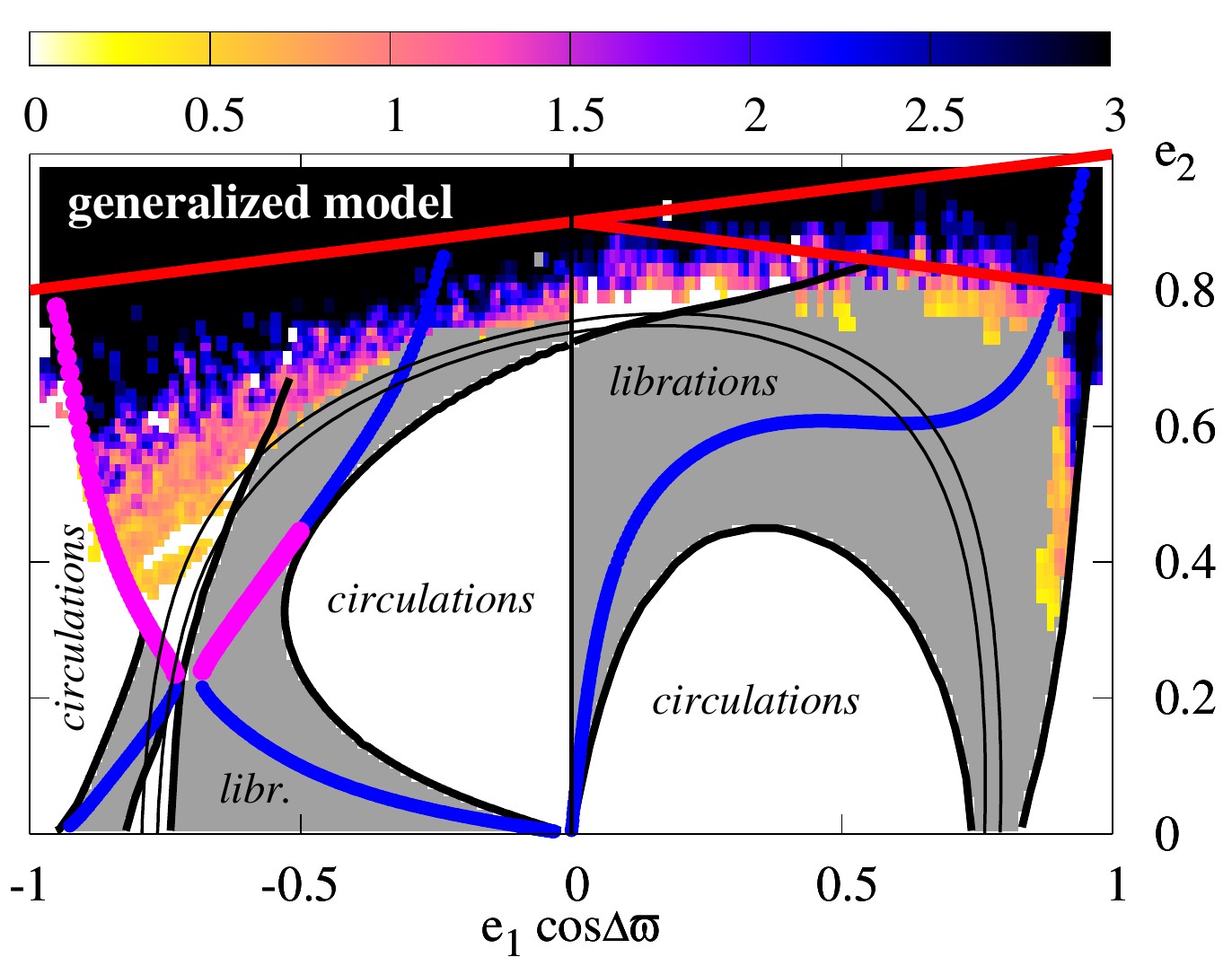}
}
}
\caption{
The \RP{}-planes for the classic model (the left-hand panel) and for the
generalized model (the right-hand panel) for the two-planet system analyzed in
Fig.~\ref{fig:fig2}. The thick, red curves mark stationary modes in  the classic
model, the thick blue curves are for the equilibria in the generalized model.
Shaded areas indicate zones of $\Delta\varpi$ librations. The thick red lines
are for the collision line of orbits. Colors code $\log$~SN of the {\em outer
orbit}, characterizing  solutions derived numerically with the help of the
direct code. Black points are for strongly chaotic solutions with the
$\log\mbox{SN} \sim 3$, white color  is for regular solutions with
$\log\mbox{SN}\sim 0$, intermediate values are marked with the color scale above
the panels, accordingly (see the text for more details).
}
\label{fig:fig5}
\end{figure*} 
\fi
\corr{
Finally, to illustrate the development of the secular instability, we solved the
equations of motion of the full system, starting very close to the UE lying on
the energy level between levels {\bf e} and {\bf f}, as marked in
Fig.~\ref{fig:fig2}, and Fig.~\ref{fig:fig4}e,f. The solution is illustrated in
the phase diagram in the left-hand panel of Fig.~\ref{fig:fig6}. For a reference, the
numerical solution is over-plotted on the analytically derived separatrices of
the UE. The corresponding time evolution of the orbital elements are illustrated
in the right-hand panels of Fig.~\ref{fig:fig6}. Clearly, during quite a long
time the full system stays close to the UE, but after $\sim 10^5$~yrs it follows
a trajectory  close to the inner separatrix, and finally begins to move close to
the outer separatrix, approaching large $e_2$. During this evolution, we observe
not only very irregular behavior of $a_2$ and both  eccentricities, but also
$\Delta\varpi$ changing from large amplitude  librations around $0$ to
circulations. Although the configuration seems bounded during many secular
periods, such behavior may be classified as strongly chaotic.
}

\ifpdf
\begin{figure*}
\centerline{
\vbox{
\hbox{
\includegraphics[height=61mm]{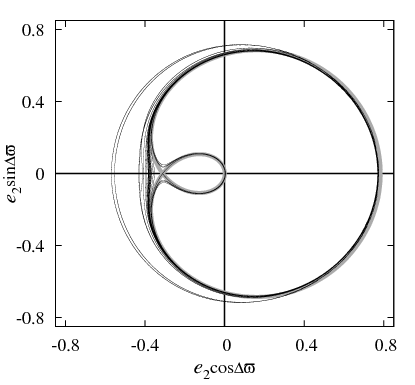}\hskip5mm
\includegraphics[height=61mm]{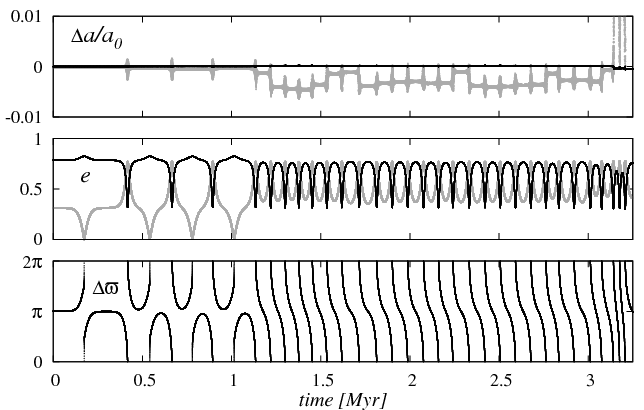}
}
}
}
\caption{
The phase diagram in the $(e_2\cos{\Delta{\varpi}}, e_2
\sin{\Delta{\varpi}})$-plane (the left-hand panel) drawn to illustrate a development
of the secularly unstable behavior of the full system.  The numerical solution
of the full system (thin, black curve) is over-plotted on the analytical,
secular solution (gray, thicker curve). The right-hand panels illustrate the
relative changes of semi-major axes (top panel), eccentricities (middle panel)
[grey curves are for outer orbit, black curves are for the inner orbit], and the
apsidal angle $\Delta\varpi$ (bottom panel).  The initial orbital parameters are
the same as in Fig. \ref{fig:fig2}.  The energy level of this solution lies
between levels \textbf{e}-\textbf{f} marked in Fig.~\ref{fig:fig2}, accordingly,
compare it also with phase diagrams Fig.~\ref{fig:fig4}e,f.
}
\label{fig:fig6}
\end{figure*}
\else
\begin{figure*}
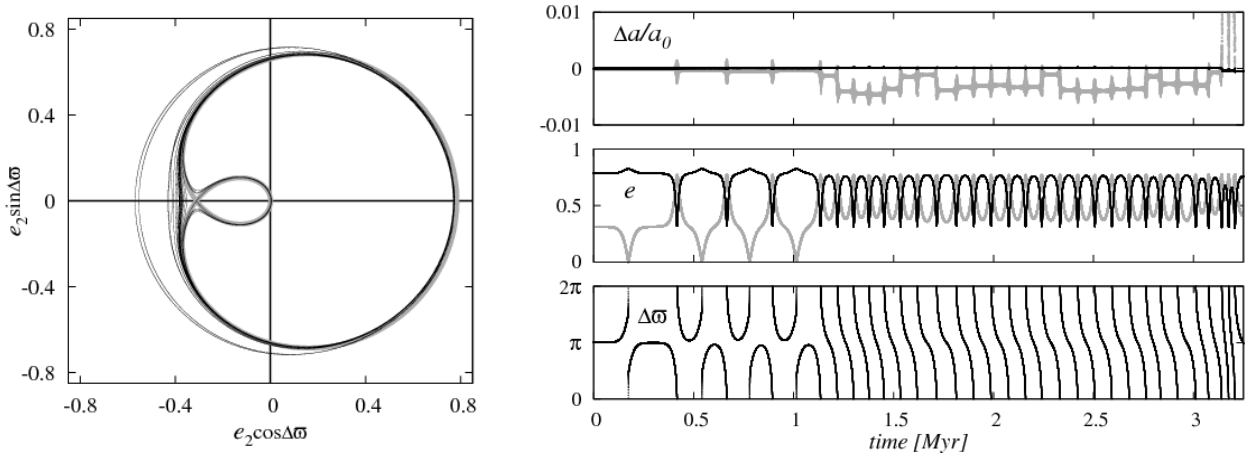

\centerline{
\vbox{
\hbox{
\includegraphics[height=61mm]{fig6a.eps}\hskip5mm
\includegraphics[height=61mm]{fig6b.eps}
}
}
}
\caption{
The phase diagram in the $(e_2\cos{\Delta{\varpi}}, e_2
\sin{\Delta{\varpi}})$-plane (the left-hand panel) drawn to illustrate a development
of the secularly unstable behavior of the full system.  The numerical solution
of the full system (thin, black curve) is over-plotted on the analytical,
secular solution (gray, thicker curve). The right-hand panels illustrate the
relative changes of semi-major axes (top panel), eccentricities (middle panel)
[grey curves are for outer orbit, black curves are for the inner orbit], and the
apsidal angle $\Delta\varpi$ (bottom panel).  The initial orbital parameters are
the same as in Fig. \ref{fig:fig2}.  The energy level of this solution lies
between levels \textbf{e}-\textbf{f} marked in Fig.~\ref{fig:fig2}, accordingly,
compare it also with phase diagrams Fig.~\ref{fig:fig4}e,f.
}
\label{fig:fig6}
\end{figure*}
\fi
%
\section{Parametric survey of two-planet systems}
%
The characterization of the phase space with the help of the representative
plane  can be  very useful to  conduct a survey of the basic features of the
secular dynamics. In particular, we want to understand how it depends on the 
physical and orbital parameters governing the magnitude of the GR and QM
interactions. In subsequent diagrams of the \RP{}-plane, we will always mark the
collision and anti-collision lines. In this way, we can determine the border of
validity of the analytic approach. Yet to derive the stationary modes possibly
exactly, in the whole permitted range of eccentricity,  we compute their
locations with the help of the semi-analytical averaging algorithm.
%
\subsection{Dependence of the secular dynamics on the masses}
\label{mass_dependence}
%
The results of the survey of the secular dynamics of two-planet systems,
including GR and QM interactions, for varied planetary masses, are illustrated
in   Fig.~\ref{fig:fig7}. 
\ifpdf
\begin{figure*}
\centerline{
\vbox{
\hbox{\includegraphics[width=5.9cm]{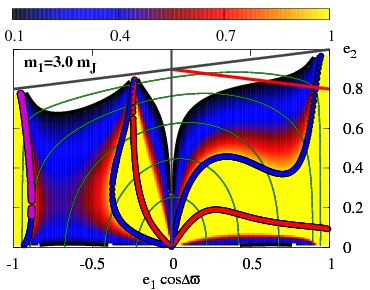}
      \includegraphics[width=5.9cm]{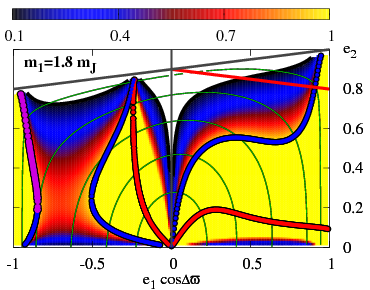}
      \includegraphics[width=5.9cm]{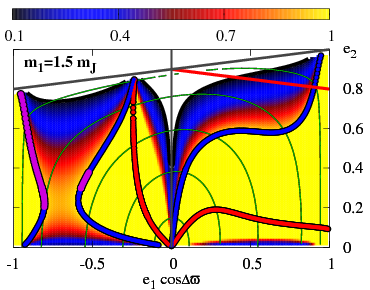}}
\hbox{\includegraphics[width=5.9cm]{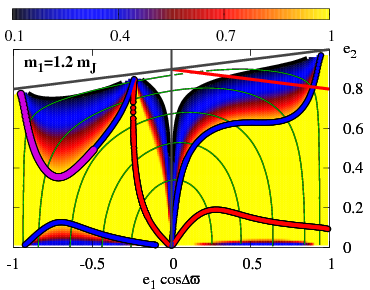}
      \includegraphics[width=5.9cm]{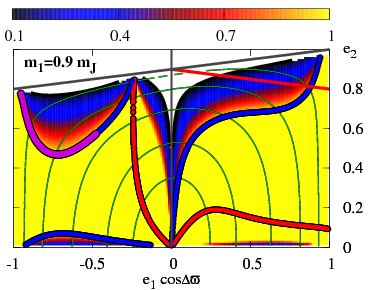}
      \includegraphics[width=5.9cm]{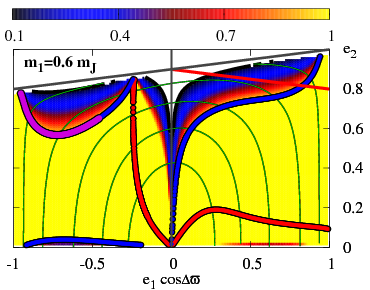}}
}
}      
\caption{
The representative energy planes $(e_1 \cos{\Delta\varpi}, e_2)$ for
$\Delta\varpi=0$ (the right half-planes,\RPp{}) and $\Delta\varpi=\pi$ (the left
half-planes, \RPm{}). 
Color contours are for the ratio of the  apsidal frequency induced by
the general relativity and quadrupole moment  to the  apsidal frequency caused by mutual
interactions between planets. 
Thick curves mark positions of stationary solutions. Red
lines are for stable equilibria in the classic model. Blue and violet 
curves mark
the positions of stable and unstable equilibria of the generalized model. The
thin lines are for the secular energy levels of the generalized model. The thick, skew lines are for the
collision lines defined with $a_1 (1 \pm e_1) = a_2 (1 - e_2)$. Parameters of
these systems are as follows: $m_0 = 1 \mbox{M}_{\sun}$, $a_1=0.1 \mbox{au}$,
$a_2=1 \mbox{au}$, $R_0=1 \mbox{R}_{\sun}$,  $T_{\idm{rot}}=30$~days,
$k_L=0.02$.  Each panel was calculated for varied planetary masses, under the
condition of constant mass ratio $m_1/m_2 = 3$. The mass of the inner planet is
written in the top left corner of each panel.
}
\label{fig:fig7}
\end{figure*}
\else
\begin{figure*}
\centerline{
\vbox{
\hbox{\includegraphics[width=5.9cm]{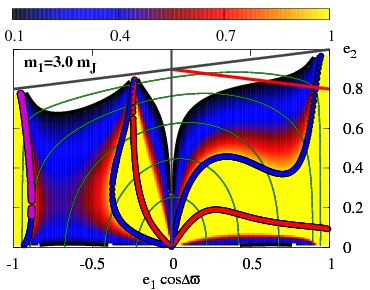}
      \includegraphics[width=5.9cm]{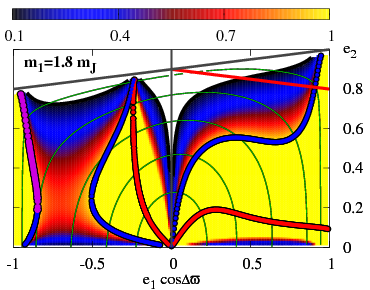}
      \includegraphics[width=5.9cm]{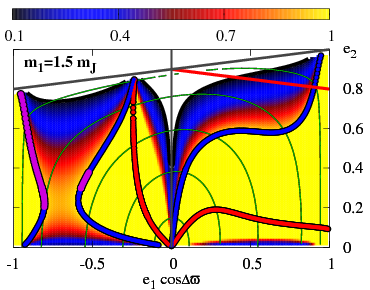}}
\hbox{\includegraphics[width=5.9cm]{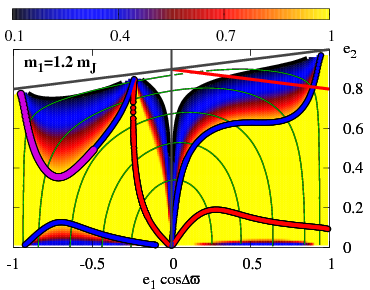}
      \includegraphics[width=5.9cm]{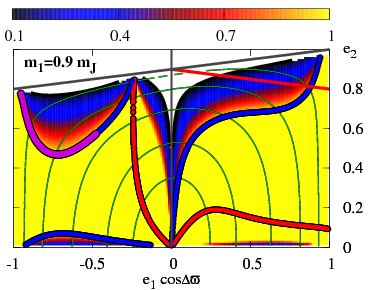}
      \includegraphics[width=5.9cm]{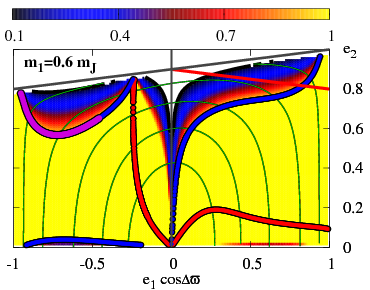}}
}
}      
\caption{
The representative energy planes $(e_1 \cos{\Delta\varpi}, e_2)$ for
$\Delta\varpi=0$ (the right half-planes,\RPp{}) and $\Delta\varpi=\pi$ (the left
half-planes, \RPm{}). 
Color contours are for the ratio of the  apsidal frequency induced by
the general relativity and quadrupole moment  to the  apsidal frequency caused by mutual
interactions between planets. 
Thick curves mark positions of stationary solutions. Red
lines are for stable equilibria in the classic model. Blue and violet 
curves mark
the positions of stable and unstable equilibria of the generalized model. The
thin lines are for the secular energy levels of the generalized model. The thick, skew lines are for the
collision lines defined with $a_1 (1 \pm e_1) = a_2 (1 - e_2)$. Parameters of
these systems are as follows: $m_0 = 1 \mbox{M}_{\sun}$, $a_1=0.1 \mbox{au}$,
$a_2=1 \mbox{au}$, $R_0=1 \mbox{R}_{\sun}$,  $T_{\idm{rot}}=30$~days,
$k_L=0.02$.  Each panel was calculated for varied planetary masses, under the
condition of constant mass ratio $m_1/m_2 = 3$. The mass of the inner planet is
written in the top left corner of each panel.
}
\label{fig:fig7}
\end{figure*}
\fi
We fix the system parameters as follows: the mass of the  parent star is $m_0 =
1~\mbox{M}_{\sun}$,  the equatorial radius of the star $R_0 = 
1~\mbox{R}_{\sun}$, and  $T_{\idm{rot}}=30$~days, $k_L=0.02$ (then $J_2 \sim
10^{-7}$),  the semi-major axes of the planets are $a_1 = 0.1~\mbox{au}$,
and $a_2 = 1.0~\mbox{au}$, respectively. Hence, we consider a typical
hierarchical configuration with the orbital periods ratio $\sim 30$. In this
experiment, the planetary masses are varied, but their ratio is kept constant,
$m_1/m_2=3$. For a reference, the mass of the inner planet is labeled in the
top-left corner of each respective panel in Fig.~\ref{fig:fig7}. Basically, in
this test we can also analyze the effect of unknown inclination of the co-planar
system on the long-term dynamics and stability. However, as we show in a recent
work regarding the 14~Herculis planetary system \citep{Gozdziewski2008},
planetary masses  derived from observations do not necessarily always scale
according to the law of the mass-factor $1/\sin~i$. If the minimal masses are
large then the mutual interactions in low-inclination configurations can
strongly modify the RV signal and even {\em the mass hierarchy} may be reversed 
in the orbital fits.

Figure~\ref{fig:fig7} reveals that curves representing stationary modes,  which
are known in the classic problem, are usually significantly shifted and/or
distorted.  Also new features of the \RP{}-plane appear and  it can be seen in
the top-left panel of Fig.~\ref{fig:fig7}. In general,  the distortions of
equilibria curves are the more stronger when the masses are smaller. It is quite
straightforward to explain this effect. When the planetary masses decrease, also
their mutual interactions (scaled by $m_1 m_2$) are decreasing.  Yet the
pericenter frequency induced by $\left<\Hmut\right>$ is scaled by the mass
product. Simultaneously, the GR and the spin-induced apsidal frequencies do not depend on the
planetary masses directly (see Eq.~\ref{orel} and  Eq.~\ref{oqm}, respectively),
and they can be regarded as approximately constant in the given mass range.
Therefore  $\kappa$ increases with decreasing $m_1$ and $m_2$. 
\corr{Then, also the assumptions of the secular theory are better
fulfilled.}

Some parts of the stationary curves in the \RPm{} half-plane (for
$\Delta{\varpi}=\pi$) comprise of unstable equilibria (they are marked with
violet color). As we mentioned already, to the best of our knowledge, such
solutions are yet unknown in the literature. Similarly to the non-classic
equilibria discovered by \cite{Michtchenko2004}, these solutions are accompanied
by the \TSR{} solutions and correspond to saddles of the secular
Hamiltonian.
The behavior of neighboring solutions tells us that they are Lyapunov unstable.
This has been analyzed  in Sect.~3.

Actually, the sequence of panels in Fig.~\ref{fig:fig7} illustrates a
characteristic development of curves representing the  equilibria, including the
\UE{} solutions. When the masses are relatively large (see the top-left panel of
Fig.~\ref{fig:fig7}), the equilibria  curves are distorted and the unstable
equilibria appear at the very edge of the \RPm{}-plane, in the range of moderate
and large values of eccentricity. On contrary, in the classic model, the
\UE{} solutions
can appear (in fact, they were found) only for $\Delta\varpi=0$
\citep{Michtchenko2004}.
%
%
Seemingly, the new \UE{} branch located in the \RPm{}-plane is
specific only for this model. When the masses decrease then $\kappa$ grows (so
the GR+QM effects become comparable in magnitude to the NG interactions). This
leads to further distortion of mode~II curves and to expanding the \UE{} part
towards moderate $e_1$. At some point (between $m_1\sim 1.8~\mbox{m}_{\idm{J}}$
and $m_1\sim 1.2~\mbox{m}_{\idm{J}}$) both  stationary curves meet in a
bifurcation point. Here, we can explain the particular choice of parameters used
to construct  Fig.~\ref{fig:fig2}. When the masses become smaller, the 
equilibria curves separate along $e_2$. We note that already for  $m_1\sim
1.2~\mbox{m}_{\idm{J}}$,  the \RP{}-plane is dominated by the GR+QM corrections.
We recall that in the classic case, the qualitative features of the \RP{}-plane
do not depend on the masses individually \citep{Michtchenko2004}, only on their
ratio in the approximation of small values (see also the sequence of plots in
Fig.~\ref{fig:fig7}). This conclusion is not true
anymore in the
realm of the generalized model.
%
\subsection{Dependence of the secular dynamics on semi-major axes}
%
In the next experiment, we investigate the dependence of the  secular dynamics
of the generalized model on individual semi-major axes;
note that the dynamics of the classic model depend only
on their ratio, $\alpha$. The results are illustrated in
Fig.~\ref{fig:fig8}. We proceed in the same manner as to draw 
Figs.~\ref{fig:fig2} and \ref{fig:fig7}. We seek for stationary solutions, and we
overplot the found equilibria on color-coded  contour levels of coefficient
$\kappa$. The primary parameters of the tested configurations are the
following:  $m_0 = 1~\mbox{M}_{\sun}$, $R = 1~\mbox{R}_{\sun}$,
$T_{\idm{rot}}=30$~days, $k_L=0.02$, $m_1 = 0.4~\mbox{m}_{\idm{J}}$,  $m_2 =
0.2~\mbox{m}_{\idm{J}}$.  The ratio of semi-major axes is kept constant, $\alpha
\equiv a_1/a_2 = 0.1$, while the individual $a_1, a_2$ are varied.   For a
reference, the nominal value of $a_1$ is labeled in the top-left corner in each
respective panel.  
\ifpdf
\begin{figure*}
\centerline{
\vbox{
\hbox{\includegraphics[width=5.9cm]{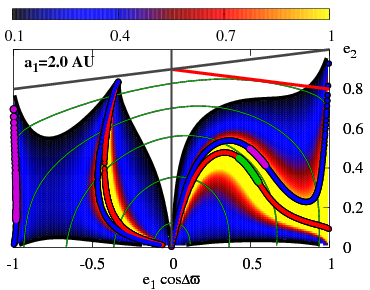}
      \includegraphics[width=5.9cm]{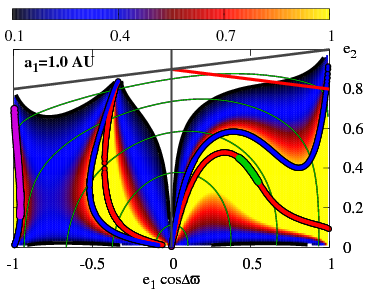}
      \includegraphics[width=5.9cm]{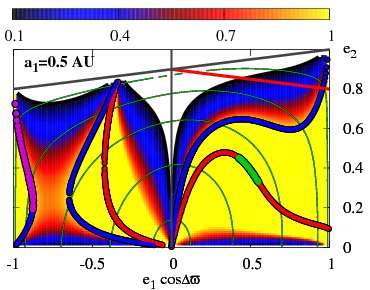}}
\hbox{\includegraphics[width=5.9cm]{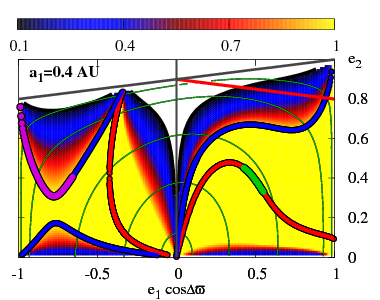}
      \includegraphics[width=5.9cm]{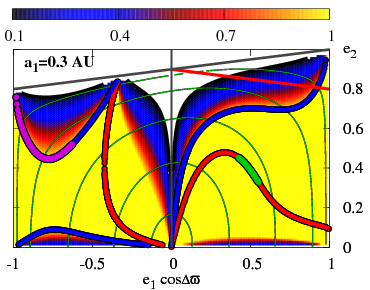}
      \includegraphics[width=5.9cm]{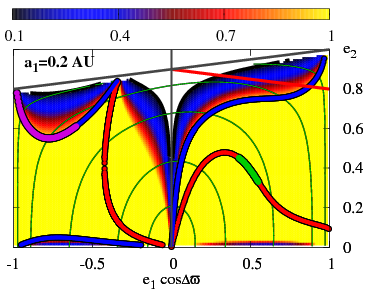}}
      }
      }
\caption{
The survey of the secular dynamics of two-planet system, when the semi-major
axes are varied. Panels are constructed in the same way,  as in Fig.
\ref{fig:fig7}.  
Color contours are for the ratio of the  apsidal frequency induced by
the general relativity and quadrupole moment  to the  apsidal frequency caused by mutual
interactions between planets. 
The thick skew
line is for collisions line defined with $a_1 (1 \pm e_1) = a_2 (1 - e_2)$. The
masses are $m_0 = 1 \mbox{M}_{\sun}$, $m_1=0.4 \mbox{m}_{\idm{J}}$, $m_2=0.2
\mbox{m}_{\idm{J}}$, respectively, the characteristic radius of the star  $R_0=1
\mbox{R}_{\sun}$, $T_{\idm{rot}}=30$~days, $k_L=0.02$. Each panel is for
different semi-major axes, fulfilling the condition of constant $\alpha \equiv
a_1/a_2 = 0.1$.  The semi-major axis of the inner planet labels each respective
panel. Compared to  Fig. \ref{fig:fig7}, an additional  unstable equilibrium for
the classic model appears and it is marked with thick green lines.
}
\label{fig:fig8}
\end{figure*}
\else
\begin{figure*}
\centerline{
\vbox{
\hbox{\includegraphics[width=5.9cm]{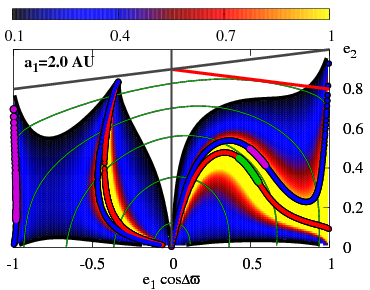}
      \includegraphics[width=5.9cm]{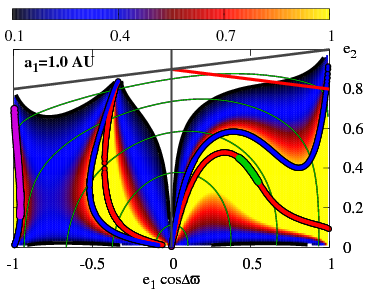}
      \includegraphics[width=5.9cm]{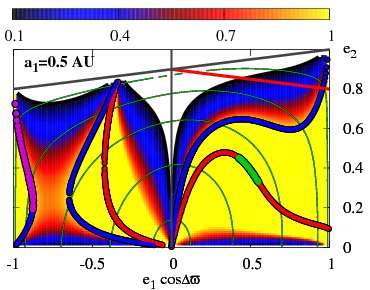}}
\hbox{\includegraphics[width=5.9cm]{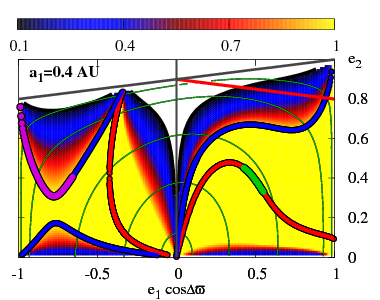}
      \includegraphics[width=5.9cm]{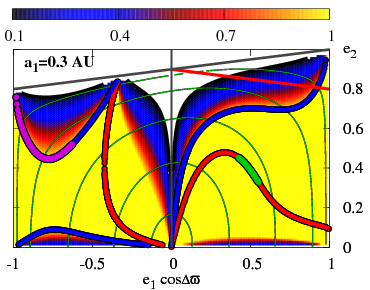}
      \includegraphics[width=5.9cm]{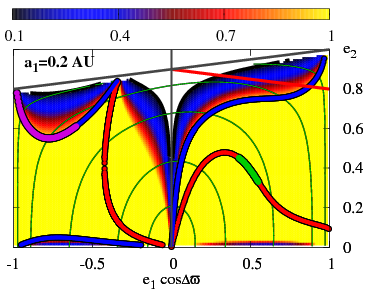}}
      }
      }
\caption{
The survey of the secular dynamics of two-planet system, when the semi-major
axes are varied. Panels are constructed in the same way,  as in Fig.
\ref{fig:fig7}.  
Color contours are for the ratio of the  apsidal frequency induced by
the general relativity and quadrupole moment  to the  apsidal frequency caused by mutual
interactions between planets. 
The thick skew
line is for collisions line defined with $a_1 (1 \pm e_1) = a_2 (1 - e_2)$. The
masses are $m_0 = 1 \mbox{M}_{\sun}$, $m_1=0.4 \mbox{m}_{\idm{J}}$, $m_2=0.2
\mbox{m}_{\idm{J}}$, respectively, the characteristic radius of the star  $R_0=1
\mbox{R}_{\sun}$, $T_{\idm{rot}}=30$~days, $k_L=0.02$. Each panel is for
different semi-major axes, fulfilling the condition of constant $\alpha \equiv
a_1/a_2 = 0.1$.  The semi-major axis of the inner planet labels each respective
panel. Compared to  Fig. \ref{fig:fig7}, an additional  unstable equilibrium for
the classic model appears and it is marked with thick green lines.
}
\label{fig:fig8}
\end{figure*}
\fi
For decreasing $a_1$, the derivatives of $\Hmut, \Hrel, \Hflat$ over $G_1$ 
increase, hence the magnitude of the respective correction to 
the apsidal  frequency grows.
Moreover, the GR and QM induced apsidal frequency increase faster than the rate
of the pericenter advance forced by the NG interactions. In the region of
\RP{}-plane
painted in yellow (light gray), the GR and QM perturbations dominate over the NG
interactions.  This region expands quickly with decreasing semi-major axis of
the inner planet, $a_1$. Already for $a_1 \sim 0.4$~au (which is similar to the
semi-major axis of Mercury in the Solar system), the apparent {\em corrections}
to the classic model, may contribute much more to the pericenter frequency  of
the innermost planet than the point mass Newtonian interactions.

The top-left panel in Fig. \ref{fig:fig8} is for  $a_1 = 2$~au and $a_2 =
20$~au, respectively. For these parameters, a shift of the curve of stationary
solutions, when compared to the ones in the classic model, is already
significant. In the next  panel ($a_1=1$~au, $a_2=10$~au) the distortion of
curves representing stationary modes is even stronger. Moreover, the \UE{} mode
present in the classic model in the \RPp{}-plane, cannot be found  in that
half-plane anymore. Simultaneously, new solutions appear at the very edge of
the  \RPm-plane, in the range of large $e_1$. For $a_1=0.5$~au, this branch of
stationary modes is even more extended. Starting with this value of  $a_1$, the
structure of the \RP{}-plane with respect to the generalized  model is very
different from that ones in the classic case. For smaller $a_1$, the curves of 
stationary modes are still more distorted.  Clearly, these distortions cannot be
regarded as small. This result is  quite unexpected, recalling that the
semi-major axes and the planetary masses by no means  are ``extreme''. In spite
of these ``typical'' parameters, the secular theories of the classic and
generalized models lead to qualitatively different view of the phase space. We
stress again that the notion of the GR and QM effects as corrections (or small
perturbations) to the  secular Hamiltonian should be understood in quite a new
light.
%
\subsection{Dependence of the secular dynamics on the stellar spin}
%
In the last parametric survey,  we study the dependence of the secular dynamics
in the realm of the generalized model on the stellar spin (or, effectively, on
the second zonal harmonic $J_2$). Figure \ref{fig:fig9} illustrates the
\RP{}-plane  computed for  following  parameters: $m_0 = 1~\mbox{M}_{\sun}$, \
$m_1=1.25~\mbox{m}_{\idm{J}}$, $m_2=0.25~\mbox{m}_{\idm{J}}$, $a_1=0.1$~au,
$a_2=1.0$~au,  $R_0=1~\mbox{R}_{\sun}$, $k_L=0.02$. The top-left panel in
Fig.~\ref{fig:fig9} is for the GR correction only and subsequent plots are for
decreasing rotation period (generalized model)  $T_{\idm{rot}}$ of the star (its particular values 
label the respective plots). This sequence corresponds to increasing $J_2$.
\ifpdf
\begin{figure*}
\centerline{
\vbox{
\hbox{\includegraphics[width=5.9cm]{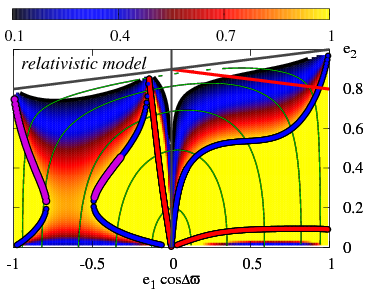}
      \includegraphics[width=5.9cm]{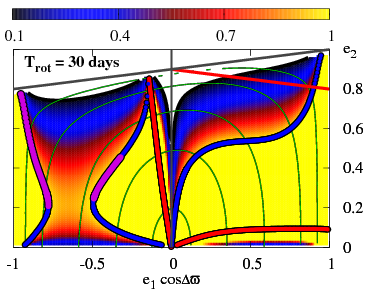}
      \includegraphics[width=5.9cm]{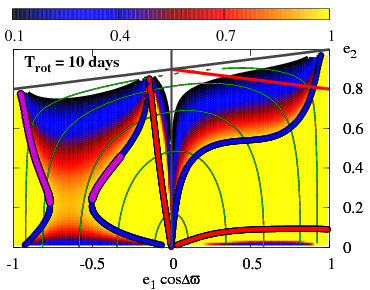}
      }
\hbox{\includegraphics[width=5.9cm]{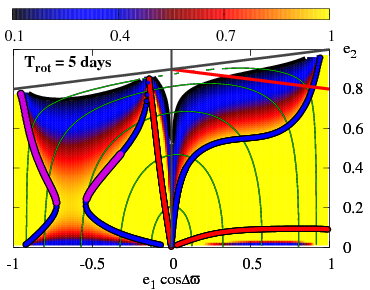}
      \includegraphics[width=5.9cm]{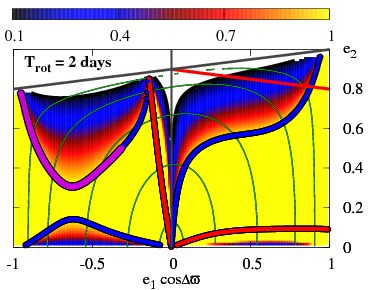}
      \includegraphics[width=5.9cm]{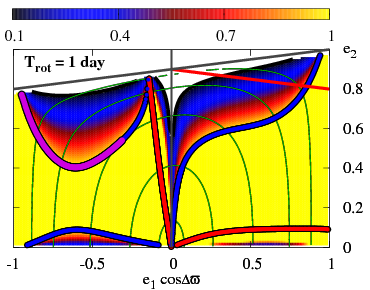}
      }
}}      
\caption{
The representative energy planes $(e_1 \cos{\Delta\varpi}, e_2)$ for 
$\Delta\varpi=0$ (the right half-planes, \RPp{}) and $\Delta\varpi=\pi$  (the
left half-planes, \RPm{}). Colors are for the ratio of the  apsidal frequency
induced by the general relativity and quadrupole moment  to the  apsidal
frequency caused by mutual interactions between planets.  The thick lines mark
the stationary solutions. The thick red curves are for stable equilibria in the
classic model. The blue and violet curves mark  positions of stable and unstable
equilibria in the generalized model, respectively.  Thin lines are for the energy levels of a
generalized model. The thick skew line is for collisions line defined with $a_1
(1 \pm e_1) = a_2 (1 - e_2)$. Parameters of the system are the following: $m_0 =
1.0~\mbox{M}_{\sun}$,  $m_1=1.25~\mbox{m}_{\idm{J}}$,
$m_2=0.25~\mbox{m}_{\idm{J}}$,  $a_1=0.1 \mbox{au}$, $a_2=1.0~\mbox{au}$, 
$R_0=1~\mbox{R}_{\sun}$, $k_L=0.02$.  Subsequent panels are for  different
rotational periods of the star, $T_{\idm{rot}}$ which is labeled in each
respective plot. 
\corr{For a reference,
the top left panel is for the GR correction only.}
}
\label{fig:fig9}
\end{figure*}
\else
\begin{figure*}
\centerline{
\vbox{
\hbox{\includegraphics[width=5.9cm]{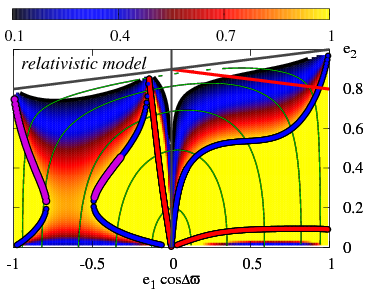}
      \includegraphics[width=5.9cm]{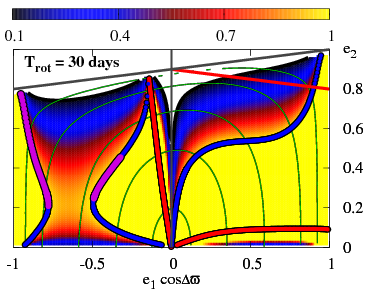}
      \includegraphics[width=5.9cm]{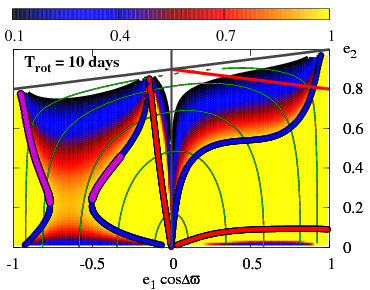}
      }
\hbox{\includegraphics[width=5.9cm]{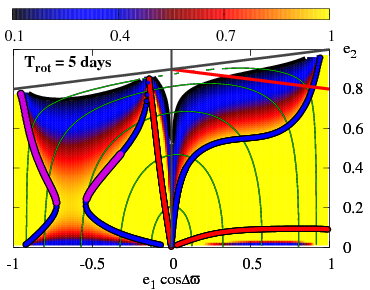}
      \includegraphics[width=5.9cm]{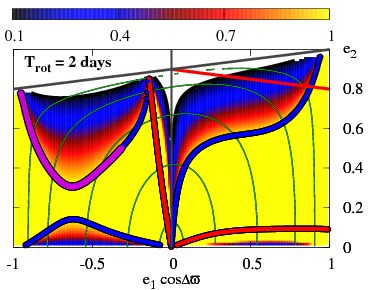}
      \includegraphics[width=5.9cm]{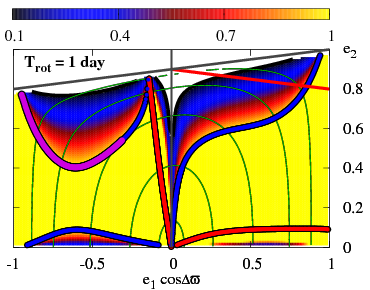}
      }
}}      
\caption{
The representative energy planes $(e_1 \cos{\Delta\varpi}, e_2)$ for 
$\Delta\varpi=0$ (the right half-planes, \RPp{}) and $\Delta\varpi=\pi$  (the
left half-planes, \RPm{}). Colors are for the ratio of the  apsidal frequency
induced by the general relativity and quadrupole moment  to the  apsidal
frequency caused by mutual interactions between planets.  The thick lines mark
the stationary solutions. The thick red curves are for stable equilibria in the
classic model. The blue and violet curves mark  positions of stable and unstable
equilibria in the generalized model, respectively.  Thin lines are for the energy levels of a
generalized model. The thick skew line is for collisions line defined with $a_1
(1 \pm e_1) = a_2 (1 - e_2)$. Parameters of the system are the following: $m_0 =
1.0~\mbox{M}_{\sun}$,  $m_1=1.25~\mbox{m}_{\idm{J}}$,
$m_2=0.25~\mbox{m}_{\idm{J}}$,  $a_1=0.1 \mbox{au}$, $a_2=1.0~\mbox{au}$, 
$R_0=1~\mbox{R}_{\sun}$, $k_L=0.02$.  Subsequent panels are for  different
rotational periods of the star, $T_{\idm{rot}}$ which is labeled in each
respective plot. 
\corr{For a reference,
the top left panel is for the GR correction only.}
}
\label{fig:fig9}
\end{figure*}
\fi
The top-left panel of Fig.~\ref{fig:fig9} is for a spherical,
non-rotating star ($J_2=0$, no tidal bulge),  the bottom-right panel is for 
very fast rotation characteristic for a young object (then
$J_2=1\times10^{-4}$). Curiously,  changes of the spin encompassing that range
are well enough to emerge new families and  stationary solutions which we
described above. They are represented, as before, by thick violet curves drawn in the
\RPm{}-plane.  Actually,  the sequence of plots simulates  variations of
flattening during the lifetime of the star. Hence, after skipping 
dissipative tidal perturbations, we may conclude that the structure of the phase space of
the secular system (and, in general, also its dynamical stability) may depend
not only on the observed or measured  orbital parameters but also on the age and
the spin rate of the host star. 

The structure of the \RP{}-plane and stationary modes which are illustrated
in Fig.~\ref{fig:fig9} remind us closely Fig.~\ref{fig:fig7} and
Fig.~\ref{fig:fig8}. In fact, as we already mentioned,  the dependence of the dynamics on the model
parameters may be described uniformly through  $\kappa$, see Eq.~\ref{eq:kappa}.
Indeed, the increase of the stellar spin leads to increase of the  pericenter
frequency and the nominator of Eq.~\ref{eq:kappa} (note that other terms are
constant). The same behavior of $\kappa$ is caused by decreasing
masses (see the sequence of diagrams in Fig.~\ref{fig:fig7} and also
discussion in Sect.~4.1). In that
case, two terms of the nominator of $\kappa$ do not change, but the denominator
decreases. Finally, if the semi-major axes decrease in constant ratio, $\kappa$
also grows because the GR and QM-induced  correction to the apsidal frequency
grows {\em faster} than the NG-forced apsidal frequency. In that sense, all 
point-mass gravity corrections governed by the described above parameter 
changes, modify similarly the structure of the \RP-plane. 

Considering our simplified secular model, the existence of the branch
of stationary solutions in
the regime of large $e_1$ and small $e_2$ may seem questionable in {\em the real}
planetary configurations. In that regime, the tidal perturbations
may become very significant for the secular evolution.  However,
the position of the bifurcation point 
in the \RPm{}-plane, where the two branches
of equilibria curves meet, and which marks the extent of the branch, is
shifted towards moderate $e_1 \sim 0.6$--$0.7$ when the  mass
ratio $m_1/m_2$  grows.
This effect can be observed in the geometric evolution of
these branches in Figs.~\ref{fig:fig8}, \ref{fig:fig7} and \ref{fig:fig9}
which form a sequence of $m_1/m_2=2,3,5$, respectively.
Hence, planetary configurations in that regime may really be found  
in the Nature.

%
\section{The secular dynamics of the $\upsilon$~And system}
%
Finally, we consider the generalized coplanar problem of three planets.  This
model can be still described by the Hamiltonian written  in the general form of
Eqs.~(\ref{HKepl}--\ref{Hrel}), and (\ref{eq:hrf}--\ref{eq:htb}) with $N=3$. The
averaged perturbing Hamiltonian has the form of Eqs.~(\ref{Hsec}),
(\ref{avHmut}), (\ref{expanssion}), (\ref{avbHrel}), (\ref{avHflat}) and
(\ref{avHflatT}). To study the properties of the secular system, we introduce  
the following set of angle-action variables related to the Poincar\'e canonical
elements
\citep{Migaszewski2008a}:
\begin{eqnarray}
\label{3pl}
&& \sigma_1 \equiv \varpi_3 - \varpi_1,  \quad G'_1, \nonumber\\
&& \sigma_2 \equiv \varpi_3 - \varpi_2 ,  \quad G'_2, \\
&& \sigma_3 \equiv -\varpi_3,  \quad \mbox{AMD} = G'_1 + G'_2 + G'_3, \nonumber
\end{eqnarray}
where $G'_i \equiv L_i-G_i$ (see Eq.~\ref{dvars}). Because $\sigma_3$ is the
cyclic angle, the Angular Momentum Deficit (AMD) is conserved, and the reduced
system has two degrees of freedom. These variables make it possible to construct
the representative energy plane in a similar manner as  for the two-planet
system. Moreover, the choice of such a plane is not unique. One of possible
definitions can be  derived through a direct analogy to the two-planet system
case. The {\em symmetric} representative plane can be defined as the set of
phase-space  points fulfilling the following relation:
\[
\frac{\partial\,\Hsec}{\partial\,\sigma_i} = 0, \quad i=1,2,3,
\]
with the simultaneous conditions that $\sigma_i=0,\pi$. For details, see
\citep{Migaszewski2008a}.

A sequence of symmetric  representative planes shown in Fig.~\ref{fig:fig10}
illustrates the qualitative properties of the generalized secular model of the
$\upsilon$ Andromedae planetary system \citep{Butler1999}. This system comprises
of three planets with masses and semi-major axes  derived through the radial
velocity observations:
$m_0=1.27~\mbox{M}_{\sun}$,
$m_1=0.69~\mbox{m}_{\idm{J}}$, 
$m_2=1.98~\mbox{m}_{\idm{J}}$, 
$m_3=3.95~\mbox{m}_{\idm{J}}$, 
$a_1=0.059~\mbox{au}$, 
$a_2=0.83~\mbox{au}$, 
$a_3=2.51~\mbox{au}$.
The constant of the AMD integral (effectively, the integral
of the total angular momentum) was obtained for the following eccentricities of the
nominal system:  $e_1=0.029$, $e_2=0.254$, $e_3=0.242$. 

We consider the representative plane for varying age of the parent star,
starting with approximately 30~Myr before entering the ZAMS.  Subsequent plots
are labeled by the lifetime $\tau$ relative to this moment taken  as the
zero-point of the time scale, rotation period $T_{\idm{rot}}$ and stellar radius
($R_{0}$) expressed in Sun's radius. For a reference, the  classic model and
with the GR correction  are illustrated in the first two top-left panels of
Fig.~\ref{fig:fig10}. These panels are derived for  non-rotating, non-distorted
spherical star. When the star is spinning, its second zonal harmonic may be as
large $J_2 \sim  10^{-3}$.  The current equatorial  radius 
of $\upsilon$~Andr is approximately 
$R_0=1.26~\mbox{R}_{\sun}$. We note that the stellar radius and the spin period  of
the star at $\tau=-30$~Myr are taken from \citep{Nagasawa2005}, and were
linearly interpolated over $\tau \in [-30,0]$~Myr.

The results are again quite surprising.  After adding the GR corrections to the
Hamiltonian of the classic model, the  overall view of the phase space changes
significantly.  The saddle of the secular Hamiltonian which is present  in the
classic model now vanishes.  At its place, a new {\em maximum} of the secular
Hamiltonian appears. Moreover, at the bottom half-plane of the representative
plane, close to the border of the permitted region of motion,  two new saddle
points appear. 

The next diagrams illustrate changes of the structure of the \RP{}-plane and a
development  of equilibria in a sequence simulating time evolution of the
stellar spin.  At the beginning, before the star enters the ZAMS, the
characteristic plane reveals a sharp maximum and a saddle in the very edge of
the region of permitted motions.  The thin curve surrounding the maximum marks
the energy level of the nominal system. When the rotation period increases
up to $\sim 8$~days, the secondary extremum (the minimum) emerges in the place of the
saddle and it persists shortly before the ZAMS stage and for longer rotational
periods.

Curiously, the only feature seen in the energy diagrams, which survives the spin
variations during the whole lifetime of the star, and persist in  the
generalized model  (with the GR and QM corrections) is the stable equilibrium
point related to the maximum of the secular Hamiltonian, which  is found in the
range of  small $e_1$ and moderate $e_2$. Curiously, the nominal system appears
in the energy level drawn with grey (green), thick line surrounding this
maximum of $\Hsec$. 
\corr{We also may notice that close to
this equilibrium of the generalized model, a
saddle of the classic model appears which is {\em
linearly stable}. 
}
\ifpdf
\begin{figure*}
\centerline{
\vbox{
\hbox{
\includegraphics[width=44mm]{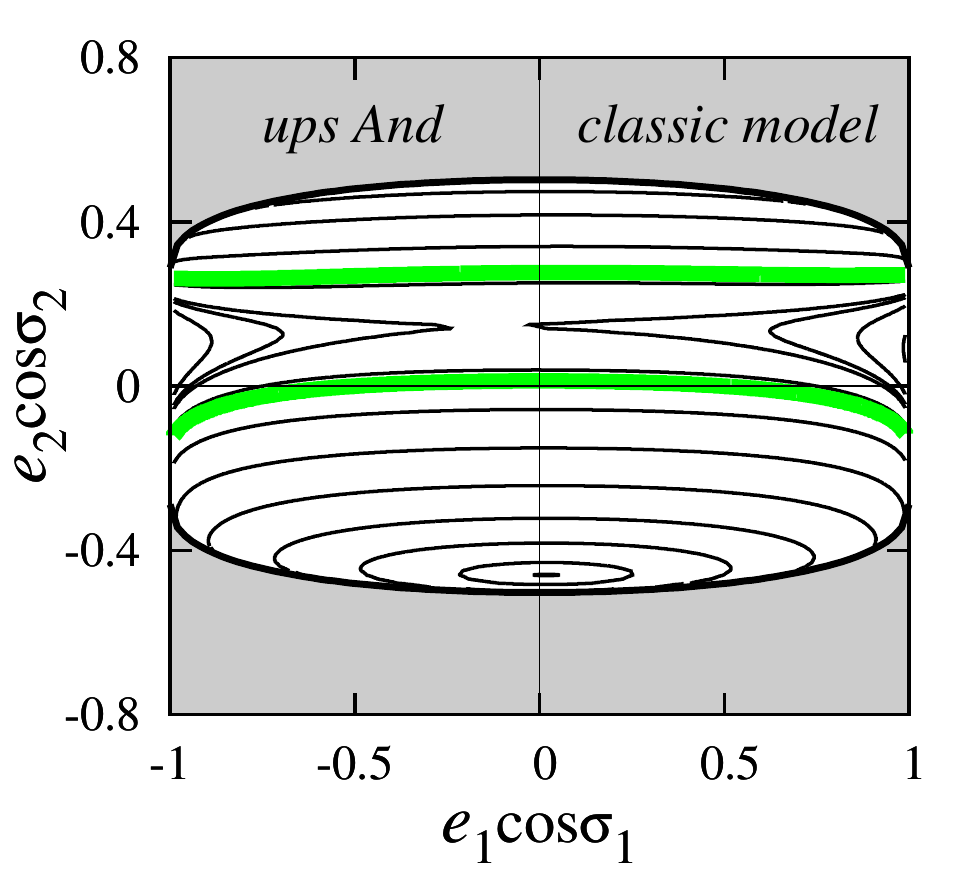}
\includegraphics[width=44mm]{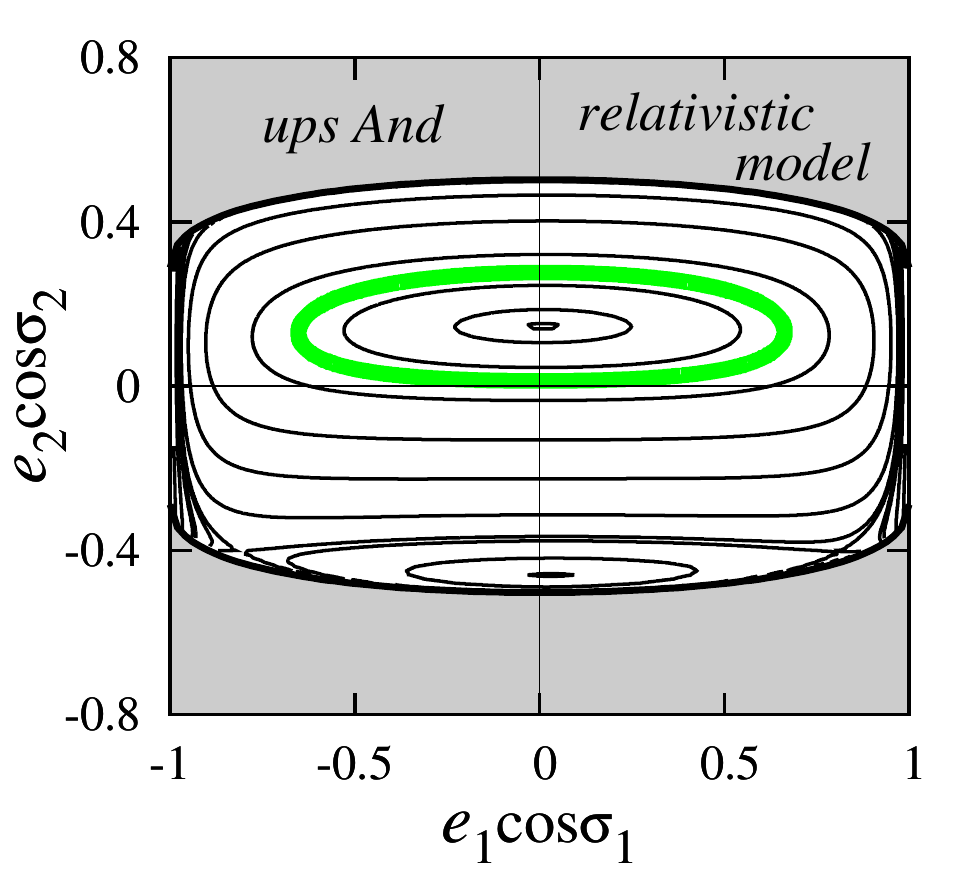}
\includegraphics[width=44mm]{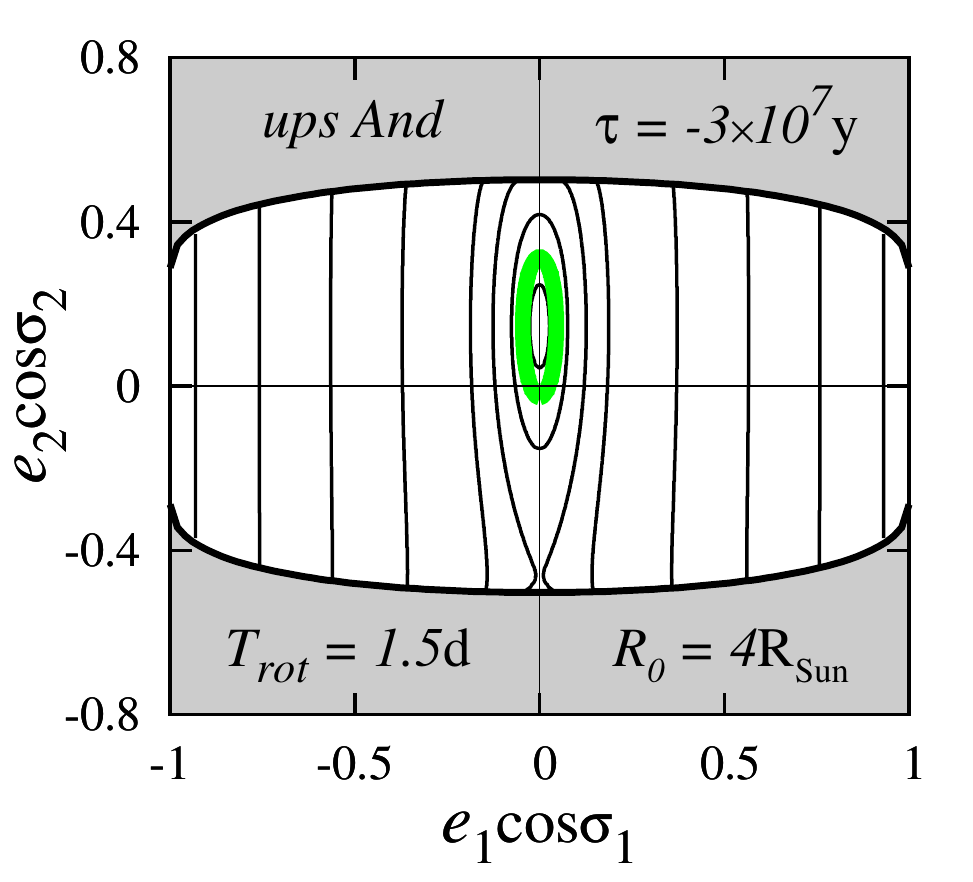}
\includegraphics[width=44mm]{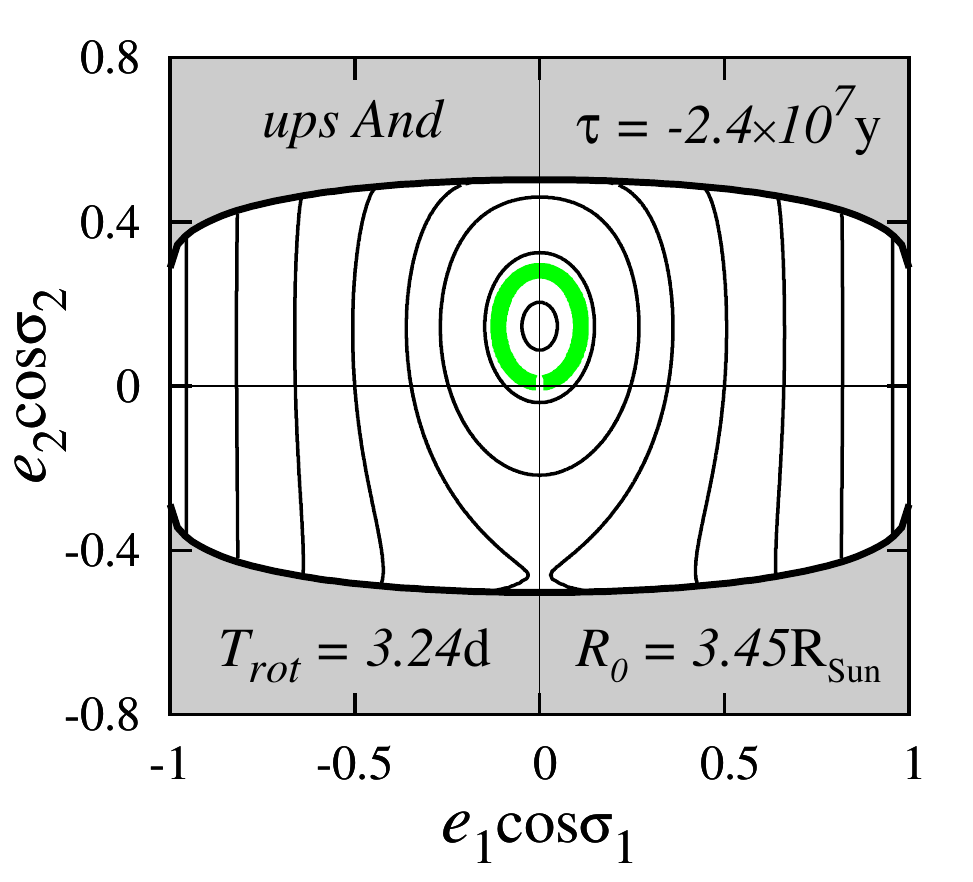}
}
\hbox{
\includegraphics[width=44mm]{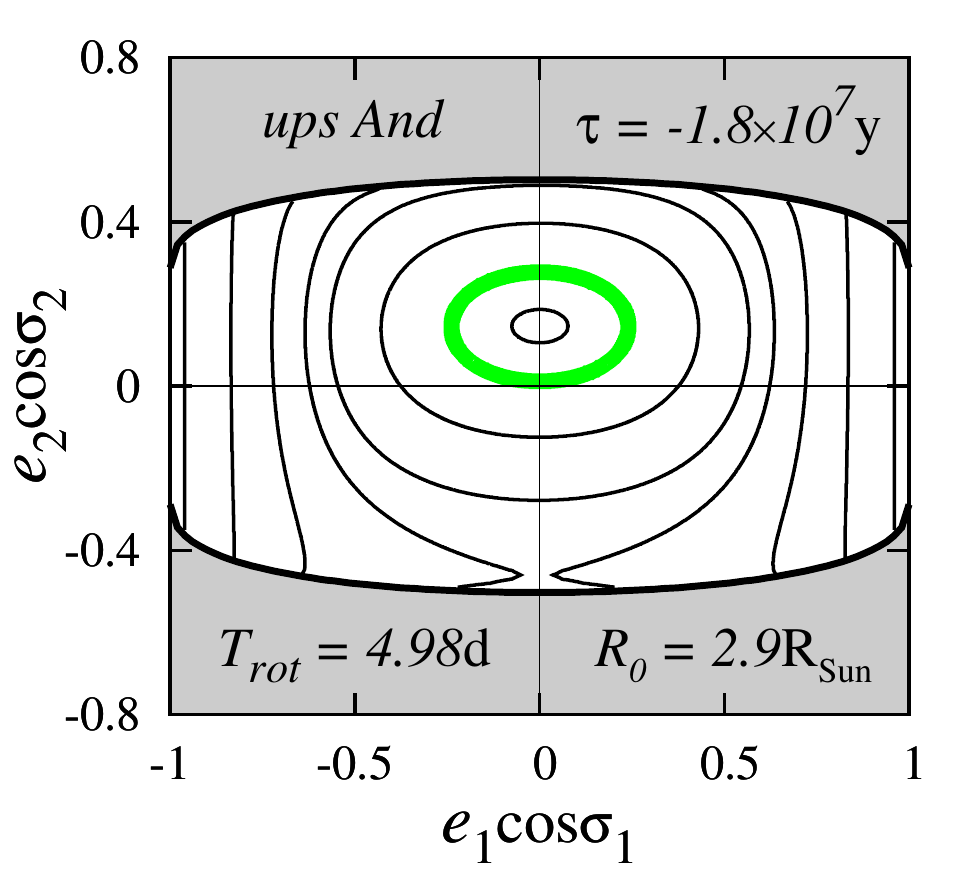}
\includegraphics[width=44mm]{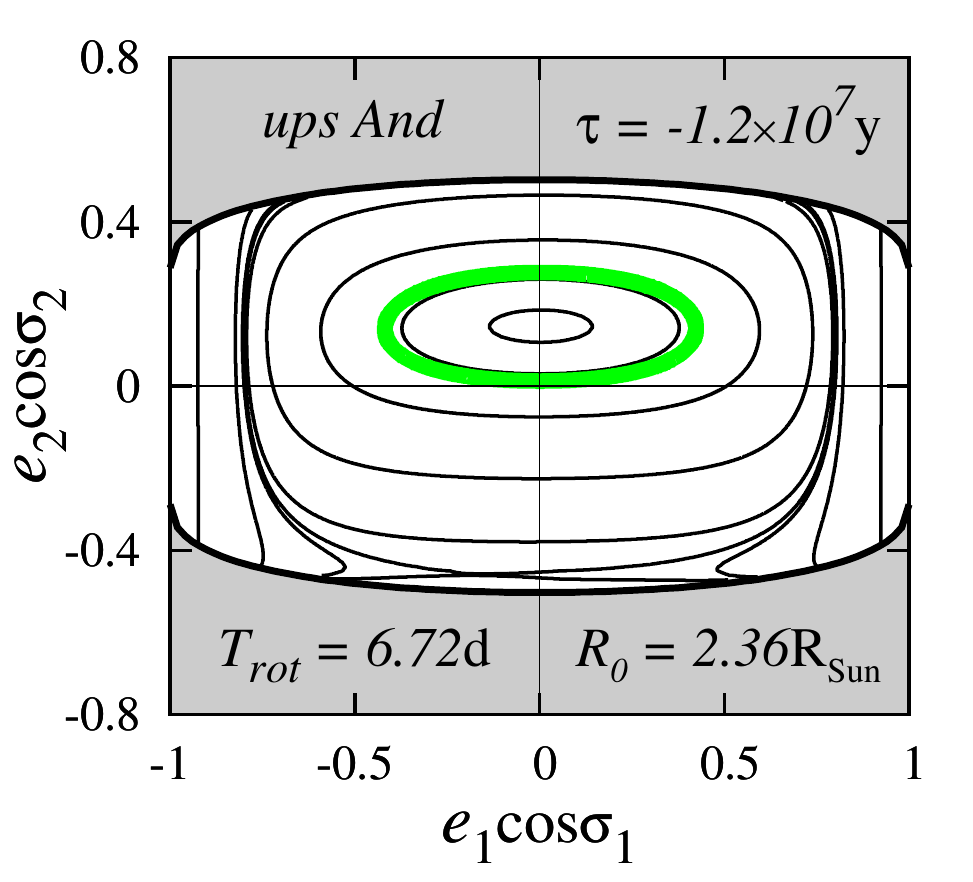}
\includegraphics[width=44mm]{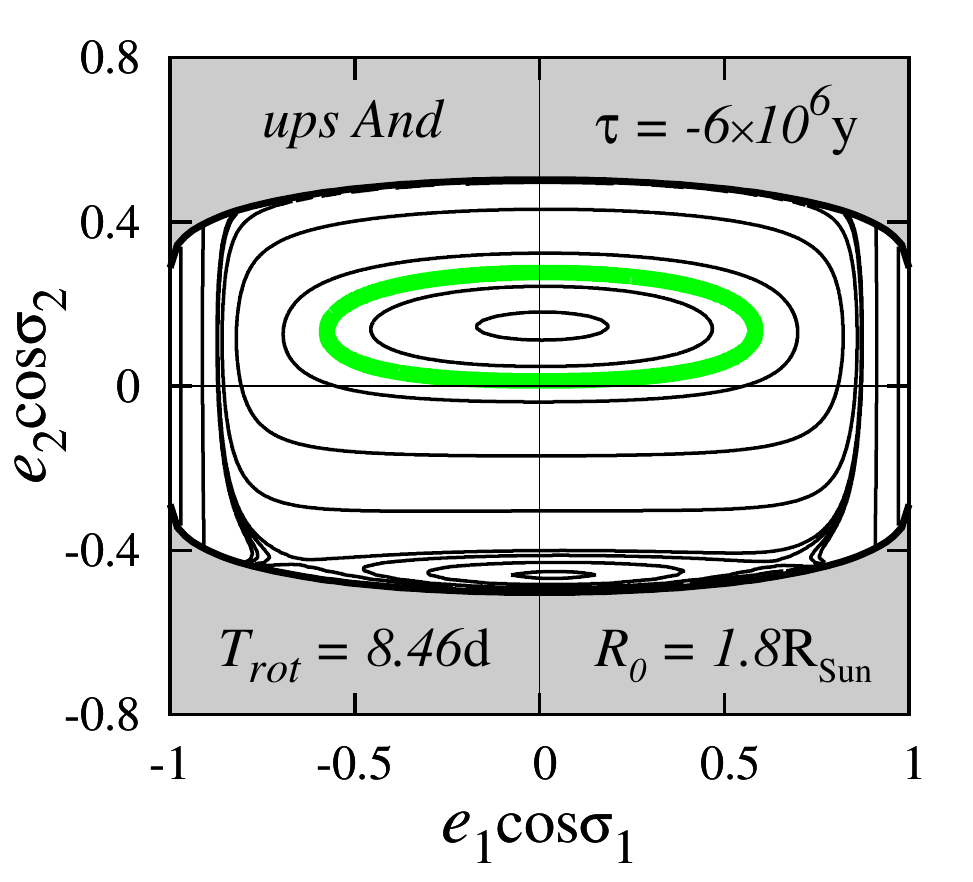}
\includegraphics[width=44mm]{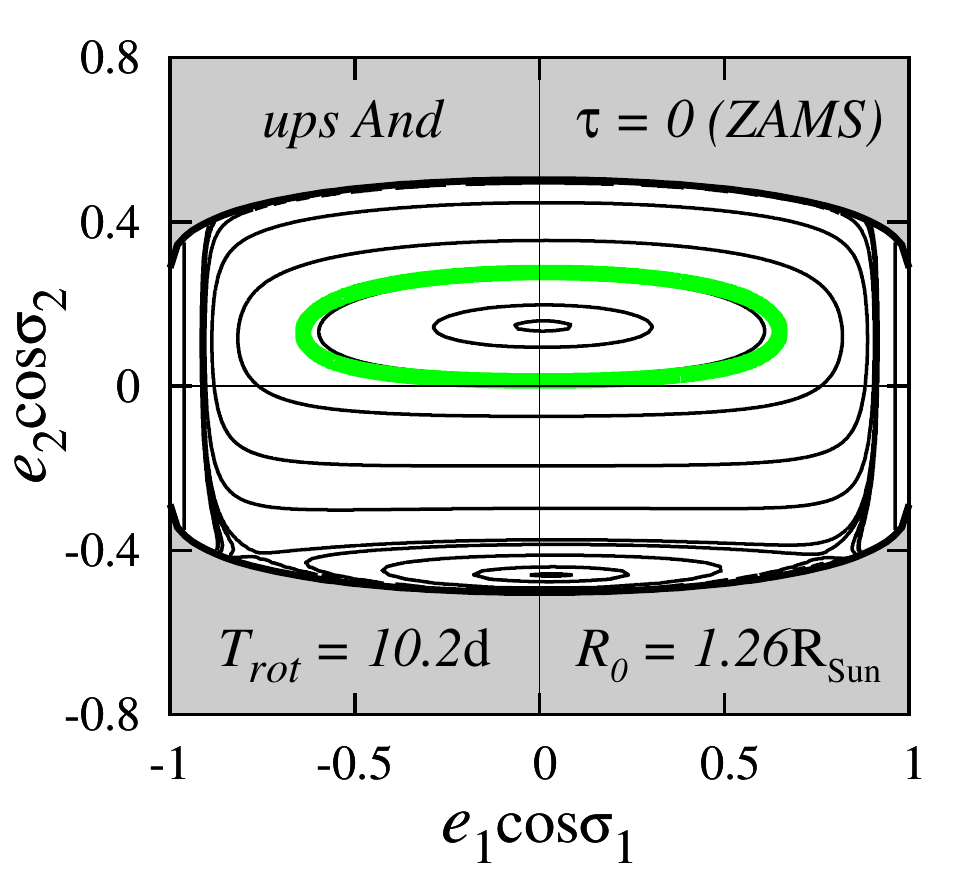}
}
}}
\caption{
The secular energy levels on the \textit{symmetric representative plane} for
three-planet system $\upsilon$ Andromede. The map coordinates are $x \equiv e_1
\cos{\sigma_1}$, $y \equiv e_2 \cos{\sigma_2}$, where $ \sigma_1, \sigma_2$ are
$0$ (positive values of $x$ or $y$) or $\pi$ (negative values of $x$ or $y$).
Grey-colored region means forbidden motions with 
$e_3<0$. Black curves are for the energy levels, green, thick
levels are for the energy of the nominal $\upsilon$~Andr  system. Each panel is for a different
setup of the planetary system model.  
From the {\em left-top panel}: the first panel is for
the classic NG model, the next panel is for the NG+GR model. 
Next panels are for generalized model with the QM corrections parameterized
by the spin rate of the parent star and its lifetime $\tau$ before
the ZAMS stage. 
}
\label{fig:fig10}
\end{figure*}
\else
\begin{figure*}
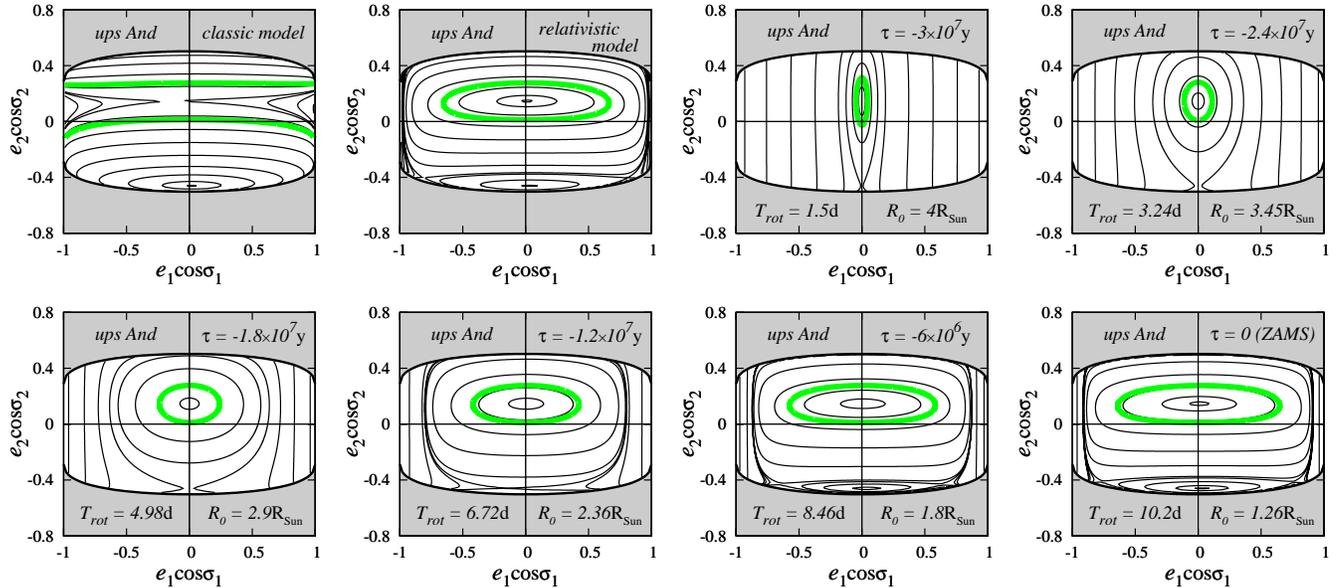

\centerline{
\vbox{
\hbox{
\includegraphics[width=44mm]{fig10a.eps}
\includegraphics[width=44mm]{fig10b.eps}
\includegraphics[width=44mm]{fig10c.eps}
\includegraphics[width=44mm]{fig10d.eps}
}
\hbox{
\includegraphics[width=44mm]{fig10e.eps}
\includegraphics[width=44mm]{fig10f.eps}
\includegraphics[width=44mm]{fig10g.eps}
\includegraphics[width=44mm]{fig10h.eps}
}
}}
\caption{
The secular energy levels on the \textit{symmetric representative plane} for
three-planet system $\upsilon$ Andromede. The map coordinates are $x \equiv e_1
\cos{\sigma_1}$, $y \equiv e_2 \cos{\sigma_2}$, where $ \sigma_1, \sigma_2$ are
$0$ (positive values of $x$ or $y$) or $\pi$ (negative values of $x$ or $y$).
Grey-colored region means forbidden motions with 
$e_3<0$. Black curves are for the energy levels, green, thick
levels are for the energy of the nominal $\upsilon$~Andr  system. Each panel is for a different
setup of the planetary system model.  
From the {\em left-top panel}: the first panel is for
the classic NG model, the next panel is for the NG+GR model. 
Next panels are for generalized model with the QM corrections parameterized
by the spin rate of the parent star and its lifetime $\tau$ before
the ZAMS stage. 
}
\label{fig:fig10}
\end{figure*}
\fi
%
\section{Conclusions}
%
In this work, we consider a generalized secular theory of coplanar, $N$-planet
system. Extending the model analyzed in the recent works devoted 
to the secular planetary dynamics  with mutual Newtonian
point-to-point interactions
\citep[e.g.,][]{Michtchenko2004,Libert2005,Gallardo2005,Migaszewski2008a},  we
consider the influence of the general relativity and quadrupole moment of the
parent star on the secular dynamics of the innermost planet and stability of 
the whole planetary system. In general, these corrections to the classic model
still do not cover all physics governing the dynamics of such systems. In some
cases (for instance, of the short-period hot-Jupiters), the tidal, dissipative
torques acting between the inner planet and the star may be significant for the
orbital evolution. However,  our main  goal is rather to extend the classic
model with the perturbations that are conservative and may be well modeled in
the realm of the Hamiltonian mechanics than to build a complete, general
secular theory. Still, this approach is useful to a wide class of systems, when
the tidal interactions may be regarded as secondary effects, or are acting 
during much longer characteristic time-scale than the GR and QM perturbations.
In reward, for paying the price of less general model, we may investigate the
secular dynamics in a global manner. 

Our analytic model follows assumptions required by the averaging theorem.
Technically, the averaging has been done with  the help of a very simple method.
This algorithm relies on  appropriate change of integration variables. It does
not incorporate any classic Fourier expansion of the perturbing function.  We
obtain a very precise analytic model of the coplanar, $N$-planet system in
terms of the semi-major axes ratio. It can be regarded as a generalization of
the recent analytic secular theories  of the classic model investigated in
many recent papers \citep[e.g.,][]{Ford2000,Lee2003,Michtchenko2004,Libert2005}. On the other
hand,  our work also covers, to some extent, the global dynamics  of the
generalized model studied in \cite{Mardling2002,Nagasawa2005} with the help of
the Gauss/Lagrange planetary equations of motion. We stress, however, that our
investigations are devoted to  more narrow class of systems (regarding the
conservative perturbations).

A general conclusion which can be derived on the basis of the generalized theory
is quite unexpected. Even in a case when the orbital parameters cannot be
regarded as {\em extreme}, the corrections to the classic Hamiltonian stemming
from the general relativity and the quadrupole moment of the star, may affect
the secular dynamics dramatically. Not only  the structure of the phase space of
the secular model changes, and new branches  of stationary solutions appear.
These solutions may bifurcate within small relative ranges of the parameters
(for instance, when the spin of the parent star is changing). We show that there
is no simple and general recipe to predict the behavior of the secular system,
when the perturbations are ``switched on''. The secular dynamics of  the
generalized model becomes extremely complex and rich. For some combinations of
the system parameters, the notion of the GR and QM  effects as {\em
corrections}  to the point-mass  NG interactions does not seem proper anymore.  In some
cases, these effects may be {\em more important} for the secular dynamics than
the mutual, point mass Newtonian interactions between the planets. 

We also show that these effects may be significant for the dynamical stability
of planetary systems both in the short-term and 
in the secular time scales. For instance,
the QM generated perturbations may directly depend on the star age and its
physical parameters ($k_L$, $R_0$). In turn, these effects may strongly
influence the structure of the phase space and can imply short-term, strong
chaotic orbital evolution during a few secular periods.

The direct tests of the analytic theory are very encouraging.  The results
justify its great accuracy. The precision of the analytic calculations is  very
important for studying the global dynamics of hierarchical systems. The
alternative numerical approach would require huge CPU time because the
hierarchical planetary systems evolve during very different time scales. Then
the CPU requirements of the direct numerical integrations are  by orders of
magnitude larger than those ones needed by the analytic formulae. 
%
\section*{Acknowledgments}
%
We are very grateful to Rosemary Mardling for careful reading of the manuscript,
constructive and informative review, many suggestions and invaluable remarks
that greatly improved the work. We would like to thank Tatiana Michchenko for a
discussion and comments on the manuscript. This work is supported by the Polish
Ministry of Science and Education, Grant No. 1P03D-021-29. C.M. is also
supported by Nicolaus Copernicus University Grant No.~408A.

\bibliographystyle{mn2e}
\bibliography{ms}

\begin{thebibliography}{}

\bibitem[\protect\citeauthoryear{{Adams} \& {Laughlin}}{{Adams} \&
  {Laughlin}}{2006a}]{Adams2006b}
{Adams} F.~C.,  {Laughlin} G.,  2006a, ApJ, 649, 992

\bibitem[\protect\citeauthoryear{{Adams} \& {Laughlin}}{{Adams} \&
  {Laughlin}}{2006b}]{Adams2006a}
{Adams} F.~C.,  {Laughlin} G.,  2006b, ArXiv Astrophysics e-prints

\bibitem[\protect\citeauthoryear{{Agol}, {Steffen}, {Sari} \&
  {Clarkson}}{{Agol} et~al.}{2005}]{Agol2005}
{Agol} E.,  {Steffen} J.,  {Sari} R.,    {Clarkson} W.,  2005, MNRAS, 359, 567

\bibitem[\protect\citeauthoryear{{Arnold}, {Kozlov} \& {Neishtadt}}{{Arnold}
  et~al.}{1993}]{Arnold1993}
{Arnold} V.~I.,  {Kozlov} V.~V.,    {Neishtadt} A.~I.,  1993, {Dynamical
  systems III. Mathematical aspects of classical and celestial mechanics}.
Encyclopaedia of mathematical sciences, Springer Verlag

\bibitem[\protect\citeauthoryear{{Benitez} \& {Gallardo}}{{Benitez} \&
  {Gallardo}}{2008}]{Benitez2008}
{Benitez} F.,  {Gallardo} T.,  2008, Celestial Mechanics and Dynamical
  Astronomy, pp 41--+

\bibitem[\protect\citeauthoryear{{Brouwer} \& {Clemence}}{{Brouwer} \&
  {Clemence}}{1961}]{Brouwer1961}
{Brouwer} D.,  {Clemence} G.~M.,  1961, {Methods of celestial mechanics}.
New York: Academic Press, 1961

\bibitem[\protect\citeauthoryear{{Butler}, {Marcy}, {Fischer}, {Brown},
  {Contos}, {Korzennik}, {Nisenson} \& {Noyes}}{{Butler}
  et~al.}{1999}]{Butler1999}
{Butler} R.~P., et al.,
1999, ApJ, 526,
  916

\bibitem[\protect\citeauthoryear{{Butler}, {Wright}, {Marcy}, {Fischer},
  {Vogt}, {Tinney}, {Jones}, {Carter}, {Johnson}, {McCarthy} \&
  {Penny}}{{Butler} et~al.}{2006}]{Butler2006}
{Butler} R.~P., et al., 2006, ApJ, 646, 505

\bibitem[\protect\citeauthoryear{{Charbonneau}, {Brown}, {Latham} \&
  {Mayor}}{{Charbonneau} et~al.}{2000}]{Charbonneau2000}
{Charbonneau} D., et al., 2000, ApJL,
  529, L45

\bibitem[\protect\citeauthoryear{{Ferraz-Mello}, {Michtchenko} \&
  {Beaug{\'e}}}{{Ferraz-Mello} et~al.}{2006}]{FerrazMello2006a}
{Ferraz-Mello} S.,  {Michtchenko} T.~A.,    {Beaug{\'e}} C.,  2006, {Regular
  motions in extra-solar planetary systems}.
Chaotic Worlds: from Order to Disorder in Gravitational N-Body Dynamical
  Systems, pp 255--+

\bibitem[\protect\citeauthoryear{{Fischer}, {Marcy}, {Butler}, {Vogt}, {Frink}
  \& {Apps}}{{Fischer} et~al.}{2001}]{Fischer2001}
{Fischer} D.~A., et al., 2001, ApJ, 551, 1107

\bibitem[\protect\citeauthoryear{{Ford}, {Kozinsky} \& {Rasio}}{{Ford}
  et~al.}{2000}]{Ford2000}
{Ford} E.~B.,  {Kozinsky} B.,    {Rasio} F.~A.,  2000, ApJ, 535, 385

\bibitem[\protect\citeauthoryear{{Godier} \& {Rozelot}}{{Godier} \&
  {Rozelot}}{1999}]{Godier1999}
{Godier} S.,  {Rozelot} J.-P.,  1999, A\&A, 350, 310

\bibitem[\protect\citeauthoryear{{Go{\'z}dziewski}, {Migaszewski} \&
  {Konacki}}{{Go{\'z}dziewski} et~al.}{2008}]{Gozdziewski2008}
{Go{\'z}dziewski} K.,  {Migaszewski} C.,    {Konacki} M.,  2008, MNRAS, 385,
  957

\bibitem[\protect\citeauthoryear{{Iorio}}{{Iorio}}{2006}]{Iorio2006}
{Iorio} L.,  2006, ArXiv General Relativity and Quantum Cosmology e-prints,
  gr-qc/0609112

\bibitem[\protect\citeauthoryear{{Kidder}}{{Kidder}}{1995}]{Kidder1995}
{Kidder} L.~E.,  1995, Phys.~Rev.~D, 52, 821

\bibitem[\protect\citeauthoryear{{Laskar}}{{Laskar}}{2008}]{Laskar2008}
{Laskar} J.,  2008, Icarus, 196, 1

\bibitem[\protect\citeauthoryear{{Laskar} \& {Robutel}}{{Laskar} \&
  {Robutel}}{1995}]{Laskar1995}
{Laskar} J.,  {Robutel} P.,  1995, Celestial Mechanics and Dynamical Astronomy,
  62, 193

\bibitem[\protect\citeauthoryear{{Lee} \& {Peale}}{{Lee} \&
  {Peale}}{2003}]{Lee2003}
{Lee} M.~H.,  {Peale} S.~J.,  2003, ApJ, 592, 1201

\bibitem[\protect\citeauthoryear{{Libert} \& {Henrard}}{{Libert} \&
  {Henrard}}{2005}]{Libert2005}
{Libert} A.-S.,  {Henrard} J.,  2005, Celestial Mechanics and Dynamical
  Astronomy, 93, 187

\bibitem[\protect\citeauthoryear{{Mardling}}{{Mardling}}{2007}]{Mardling2007}
{Mardling} R.~A.,  2007, MNRAS, 382, 1768

\bibitem[\protect\citeauthoryear{{Mardling} \& {Lin}}{{Mardling} \&
  {Lin}}{2002}]{Mardling2002}
{Mardling} R.~A.,  {Lin} D.~N.~C.,  2002, ApJ, 573, 829

\bibitem[\protect\citeauthoryear{{Michtchenko} \& {Ferraz-Mello}}{{Michtchenko}
  \& {Ferraz-Mello}}{2001}]{Michtchenko2001}
{Michtchenko} T.~A.,  {Ferraz-Mello} S.,  2001, AJ, 122, 474

\bibitem[\protect\citeauthoryear{{Michtchenko}, {Ferraz-Mello} \&
  {Beaug{\'e}}}{{Michtchenko} et~al.}{2006}]{Michtchenko2006}
{Michtchenko} T.~A.,  {Ferraz-Mello} S.,    {Beaug{\'e}} C.,  2006, Icarus,
  181, 555

\bibitem[\protect\citeauthoryear{{Michtchenko} \& {Malhotra}}{{Michtchenko} \&
  {Malhotra}}{2004}]{Michtchenko2004}
{Michtchenko} T.~A.,  {Malhotra} R.,  2004, Icarus, 168, 237

\bibitem[\protect\citeauthoryear{{Migaszewski} \&
  {Go{\'z}dziewski}}{{Migaszewski} \&
  {Go{\'z}dziewski}}{2008a}]{Migaszewski2008a}
{Migaszewski} C.,  {Go{\'z}dziewski} K.,  2008a, MNRAS, 388, 789

\bibitem[\protect\citeauthoryear{{Migaszewski} \&
  {Go{\'z}dziewski}}{{Migaszewski} \&
  {Go{\'z}dziewski}}{2008b}]{Migaszewski2008b}
{Migaszewski} C.,  {Go{\'z}dziewski} K.,  2008b, MNRAS, submitted

\bibitem[\protect\citeauthoryear{{Miralda-Escud{\'e}}}{{Miralda-Escud{\'e}}}{2%
002}]{MiraldaEscude2002}
{Miralda-Escud{\'e}} J.,  2002, ApJ, 564, 1019

\bibitem[\protect\citeauthoryear{{Morbidelli}}{{Morbidelli}}{2002}]{Morbidelli%
2003}
{Morbidelli} A.,  2002, {Modern celestial mechanics : aspects of Solar system
  dynamics}.
Taylor {\&} Francis

\bibitem[\protect\citeauthoryear{{Murray} \& {Dermott}}{{Murray} \&
  {Dermott}}{2000}]{Murray2000}
{Murray} C.~D.,  {Dermott} S.~F.,  2000, {Solar System Dynamics}.
Cambridge Univ. Press

\bibitem[\protect\citeauthoryear{{Nagasawa} \& {Lin}}{{Nagasawa} \&
  {Lin}}{2005}]{Nagasawa2005}
{Nagasawa} M.,  {Lin} D.~N.~C.,  2005, ApJ, 632, 1140

\bibitem[\protect\citeauthoryear{{Pauwels}}{{Pauwels}}{1983}]{Pauwels1983}
{Pauwels} T.,  1983, Celestial Mechanics, 30, 229

\bibitem[\protect\citeauthoryear{{Pijpers}}{{Pijpers}}{1998}]{Pijpers1998}
{Pijpers} F.~P.,  1998, MNRAS, 297, L76

\bibitem[\protect\citeauthoryear{{Poincar\'e}}{{Poincar\'e}}{1897}]{Poincare1897}
{Poincar\'e} H.,  1897, Bulletin Astronomique, Serie I, 14, 53

\bibitem[\protect\citeauthoryear{{Richardson} \& {Kelly}}{{Richardson} \&
  {Kelly}}{1988}]{Richardson1988}
{Richardson} D.~L.,  {Kelly} T.~J.,  1988, Celestial Mechanics, 43, 193

\bibitem[\protect\citeauthoryear{{Rodr{\'{\i}}guez} \&
  {Gallardo}}{{Rodr{\'{\i}}guez} \& {Gallardo}}{2005}]{Gallardo2005}
{Rodr{\'{\i}}guez} A.,  {Gallardo} T.,  2005, ApJ, 628, 1006

\bibitem[\protect\citeauthoryear{{Veras} \& {Armitage}}{{Veras} \&
  {Armitage}}{2007}]{Veras2007}
{Veras} D.,  {Armitage} P.~J.,  2007, ApJ, 661, 1311

\bibitem[\protect\citeauthoryear{{Vogt}, {Butler}, {Marcy}, {Fischer}, {Henry},
  {Laughlin}, {Wright} \& {Johnson}}{{Vogt} et~al.}{2005}]{Vogt2005}
{Vogt} S.~S., et al., 2005, ApJ, 632,
  638

\bibitem[\protect\citeauthoryear{{Winn}, {Noyes}, {Holman}, {Charbonneau},
  {Ohta}, {Taruya}, {Suto}, {Narita}, {Turner}, {Johnson}, {Marcy}, {Butler} \&
  {Vogt}}{{Winn} et~al.}{2005}]{Winn2005}
{Winn} J.~N., et al.
2005, ApJ, 631, 1215

\end{thebibliography}
\label{lastpage}
\end{document}